\documentclass[fleqn,usenatbib]{mnras}

\usepackage{newtxtext,newtxmath}

\usepackage[T1]{fontenc}

\DeclareRobustCommand{\VAN}[3]{#2}
\let\VANthebibliography\thebibliography
\def\thebibliography{\DeclareRobustCommand{\VAN}[3]{##3}\VANthebibliography}

\usepackage{graphicx}	
\usepackage{amsmath}	
\usepackage{amssymb}	
\usepackage{subfigure}
\usepackage{float,lscape}

\newcommand{\Msol}{\ensuremath{\mathrm{M}_{\odot}}}

\title[G4Jy: Broadband spectral analysis]{The GLEAM 4-Jy (G4Jy) Sample: V. Broadband spectral analysis }

\author[S. V. White et al.]{Sarah V. White$^{1}$\thanks{E-mail: sarahwhite.astro@gmail.com},
Siegfried A. Gawenda$^{2}$, Giulia Macario$^{3,4}$,
Samuel P. Hansen$^{5}$,
Ross J. Turner$^{5}$, \newauthor
Thomas M.O. Franzen$^{6}$, Joseph R. Callingham$^{7,8}$, Randall Wayth$^{3}$, and Kshitij Thorat$^{9}$
\\
$^{1}$South African Astronomical Observatory (SAAO), PO Box 9, Observatory, 7935, South Africa\\
$^{2}$Department of Physics and Astronomy, The University of Texas, R{\'i}o Grande Valley, Brownsville, TX 78520, USA\\
$^{3}$Square Kilometre Array Observatory (SKAO), 26 Dick Perry Ave, Kensington, WA 6151, Australia\\
$^{4}$ Istituto Nazionale di Astrofisica (INAF), Osservatorio Astrofisico di Arcetri, Largo Enrico Fermi 5, Firenze, Italy, 50125\\
$^{5}$School of Natural Sciences, University of Tasmania, Private Bag 37, Hobart, 7001, Australia\\
$^{6}$SKAO, Jodrell Bank, Lower Withington, Macclesfield, Cheshire, SK11 9FT, UK\\
$^{7}$ASTRON, Netherlands Institute for Radio Astronomy, Oude Hoogeveensedijk 4, 7991 PD, Dwingeloo, The Netherlands\\
$^{8}$Anton Pannekoek Institute for Astronomy, University of Amsterdam, Science Park 904, 1098 XH, Amsterdam, The Netherlands\\
$^{9}$Department of Physics, University of Pretoria, Hatfield, Pretoria, 0028, South Africa\\
}

\date{Accepted XXX. Received YYY; in original form ZZZ}

\pubyear{\the\year{}}

\begin{document}
\label{firstpage}
\pagerange{\pageref{firstpage}--\pageref{lastpage}}
\maketitle

\begin{abstract}
The G4Jy Sample consists of 1,863 of the brightest sources in the southern sky, as selected at 151\,MHz using the GaLactic and Extragalactic All-sky MWA (GLEAM) survey. We supplement existing, low-frequency radio data by completing visual inspection of radio images from the Rapid ASKAP Continuum Survey (RACS) at 887.5, 1367.5 and 1655.5\,MHz. Doing so on a source-by-source basis allows us to assess the quality of the association of the radio emission, which we indicate via a flag system. Having multiple radio flux-densities allows us to construct broadband radio spectral energy distributions (SEDs), and these extend up to 20\,GHz for sources that also have data from the Australia Telescope Compact Array (ATCA). We fit polynomial functions to these SEDs, and determine the distribution of sources that are best-fit by absorption models, these accounting for 11 per cent of the sample. The SEDs are further modelled using the Radio AGN in Semi-analytic Environments (RAiSE) framework, for three assumed density-profiles (set by typical halo masses of 10$^{12.5}$, 10$^{13.5}$, and 10$^{14.5}$\,$\Msol$). The derived source ages show the expected distribution in the radio-power--size ($P$--$D$) diagram, whilst derived jet powers show non-linear behaviour. The latter is believed to be the result of higher densities, associated with higher magnetic fields, restricting the expansion of the radio jets/lobes.   
\end{abstract}

\begin{keywords}
galaxies: active -- galaxies: evolution -- galaxies: high-redshift -- radio continuum: galaxies -- catalogues
\end{keywords}



\section{Introduction}

Active galactic nuclei (AGN) are known to have an important role in galaxy evolution, with the supermassive black-hole at the centre appearing to be correlated with growth of the host galaxy \citep[e.g.][]{Gultekin2009}. For this work we focus on those that are deemed to be `radio loud' AGN, where the radio jets can have both a positive impact on star-formation \citep[e.g.][]{Croft2006} and a negative impact \citep[e.g.][]{Croton2006}. The overall impact of AGN feedback is thus open to debate.

We aim to investigate further by studying the brightest radio galaxies in the southern sky, known as the G4Jy Sample \citep{White2020a, White2020b}. Being selected at low frequencies ($S_{\mathrm{151\,MHz}} > 4$\,Jy), via the GaLactic and Extragalactic All-sky Murchison Widefield Array catalogue \citep[GLEAM;][]{Wayth2015,HurleyWalker2017}, these sources are unbiased with respect to the orientation of the radio-jet axis \citep{Barthel1989} and therefore give us the most complete picture of how AGN interact with their surroundings. For such a study we need to construct broadband radio spectral energy distributions (SEDs) in order to place constraints on the electron energies associated with extended radio emission \citep[e.g.][]{Mack1998}. GLEAM provides excellent constraints at the low-frequency end, and we employ radio data from the Australian Square Kilometre Array Pathfinder (ASKAP) and the Australia Telescope Compact Array (ATCA) to constrain emission at mid-to-high radio frequencies.

As shown by previous work \citep[e.g.][]{Turner2018b}, broadband SEDs allow us to understand intrinsic properties such as the jet power of the sources. This differs from the radio luminosity as it incorporates information on the `work done' by the radio source on its surroundings \citep{Turner2026b}, rather than tracing synchrotron radiation (and inverse-Compton losses). Broadband SEDs are also fundamental for estimating the age of the radio source \citep[e.g.][]{Harwood2017,Quici2022}, with lower frequencies enabling longer AGN lifecycles to be probed. This is particularly interesting as some AGN may have more-recently `switched off', meaning that the radio offers the only window into past black-hole accretion activity \citep{Shulevski2017}. 

The difference with the current work is that past spectral analyses tend to have each revolved around a handful of sources, whereas we complete our analysis en masse for nearly $\sim$2,000 radio galaxies. Doing so allows us to draw out general trends that allow us to put extreme examples (e.g. giant radio galaxies) into context. Furthermore, we emphasise that a significant fraction of sources have spectral curvature that is indicative of restarted AGN activity \citep{White2026b}, which further highlights that we cannot assume that a power-law function ($S_{\nu}\propto\nu^{\alpha}$) accurately describes the SED.

\subsection{Paper outline}

In the next section we outline the radio data that is used for our work, followed by how the Rapid ASKAP Continuum Survey (RACS) allows us to determine mid-frequency flux-densities, extending the work by \citet{Gawenda2026} for the G4Jy-3CRE subset ($S_{{\mathrm{178\, MHz}}}> 9$\,Jy). Then, in Section~\ref{sec:broadband_spectra}, we present the results of fitting the broadband radio SEDs of G4Jy sources, from low to mid-frequencies. For a subset of the sample, we can extend this fitting to higher frequencies, as described in Section~\ref{sec:atca_seds}. Based on this multifrequency analysis, we then use theoretical models in Section~\ref{sec:raise_models} to determine intrinsic properties, namely the source age and jet power. Finally, in Section~\ref{sec:conclusions} we present our conclusions.

\section{Data}

We supplement the multiwavelength G4Jy catalogue \citep{White2026a,White2026b} with radio data from the Rapid ASKAP Continuum Survey (RACS) and the Australia Telescope Compact Array (ATCA). Doing so allows us to study broadband radio spectra from low frequencies via GLEAM (72--231\,MHz) up to frequencies as high as 20\,GHz, with the spectral shape providing insights into the physical processes of these radio-bright sources. 

\subsection{RACS-low images and catalogue}
\label{sec:racs_low} 

RACS is a survey carried out at three frequencies using ASKAP, with the first data release being the Stokes-$I$ images and catalogue at 887.5 MHz \citep{McConnell2021,Hale2021}. Given its similar observing frequency at 843\,MHz, \citet{Hale2021} use the Sydney University Molonglo Sky Survey \citep[SUMSS;][]{Mauch2003,Murphy2007} to assess the flux-density scale of RACS-low and determine that the latter has an error of $\pm16$ per cent. Furthermore, RACS-low has a point-source completeness of $\sim$95 per cent for sources with an integrated flux-density of $\sim$3\,mJy. The images are convolved to a common resolution of 25\,arcsec, which is applied for the Declination range of $-$80 to +30\,deg. (We here note that 15 G4Jy sources are at more-southerly Declinations.) Like for each of the RACS catalogues, source-finding is completed with {\sc PyBDSF} \citep{Mohan2015}, and we make use of the Gaussian-component catalogue over the extragalactic sky (i.e. with a $|b| < 5$\,deg removal of the Galactic Plane).    

\subsection{RACS-mid images and catalogue}
\label{sec:racs_mid} 

The second observing frequency of RACS is at 1367.5\,MHz, with images released by \citet{Duchesne2023} and the corresponding catalogue generated by \citet{Duchesne2024}. The former retain their native, Declination-dependent spatial resolution of 8.1--47.5\,arcsec in order to preserve surface-brightness sensitivity, where the median noise is 198\,$\upmu$Jy\,PSF$^{-1}$. Covering the sky below Dec. = +49\,deg, the median angular resolution is 11.2\,arcsec $\times$ 9.3\,arcsec. Again, we use the original Gaussian-component catalogue (rather than the source catalogue subsequently produced by {\sc PyBDSF}) in order to better-determine the spatial extent of the fitted radio emission. Finally we note that \citet{Duchesne2023,Duchesne2024} estimated a 6-per-cent uncertainty in the brightness scale based upon comparison with the NRAO (National Radio Astronomy Observatory) VLA (Very Large Array) Sky Survey \citep[NVSS;][]{Condon1998} at 1400\,MHz.

\subsection{RACS-high images and catalogue}
\label{sec:racs_high} 

The third and final observing frequency of RACS, corresponding to `RACS-high', is 1655.5\,MHz \citep{Duchesne2025}. The catalogue is constructed from images that have a median spatial resolution of 11.8\,arcsec $\times$ 8.1\,arcsec and a median noise level of $\sim$195\,$\upmu$Jy\,PSF$^{-1}$. These cover the southern sky up to Dec. $\simeq$+48\,deg, with the brightness scale estimated to have an uncertainty of 10 per cent. We acknowledge that a Declination-dependent offset has been discovered in the astrometry \citep{Duchesne2025}, of maximum magnitude 2.5\,arcsec, but we continue with the original Declination determinations in order to maintain consistency between the catalogued components and the imaging from which they are derived.

\subsection{ATCA data}
\label{sec:atca_data}

\begin{figure*}
\centering
\includegraphics[scale=0.98]{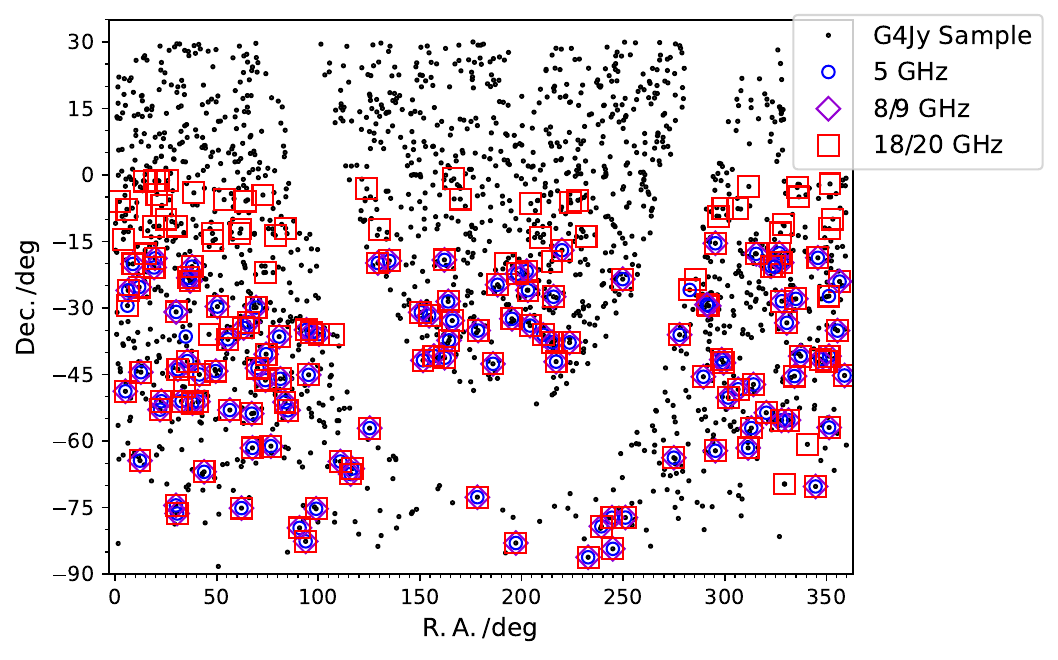} 
\caption{ Sky coverage for the full G4Jy Sample (black dots), with those observed at 5\,GHz (blue circles), 8/9\,GHz (purple diamonds) and 18/20\,GHz (red squares) highlighted. These high-radio-frequency measurements are courtesy of the AT20G Survey and later follow-up using ATCA (Section~\ref{sec:atca_data}). }
\label{fig:atca_coverage}
\end{figure*}

As part of the original visual inspection completed for the G4Jy Sample \citep{White2020a,White2020b}, it was noted which sources coincided with detections from the Australia Telescope 20-GHz (AT20G) Survey \citep[][]{Murphy2010}. Tracing the highest frequencies of the radio emission from G4Jy sources, these detections either pin-pointed the position of the radio core or revealed the location of a hotspot amongst otherwise poor-spatial-resolution imaging ($\sim$45\,arcsec from SUMSS and NVSS). The AT20G detections have a positional accuracy of $\sim$1\,arcsec and the survey reaches an 8-$\sigma$ flux-density limit of 40\,mJy. However, this blind-scan survey has varying completeness \cite[e.g. 92 per cent at 50\,mJy\,beam$^{-1}$;][]{Hancock2011} on account of weather conditions strongly affecting 20-GHz observations. AT20G covers the southern hemisphere (at $|b| > 1.5$\,deg) and includes near-simultaneous follow-up observations at 4.8\,GHz and 8.6\,GHz for detections below Dec. = $-15$\,deg (Figure~\ref{fig:atca_coverage}).

To supplement the AT20G data, ATCA follow-up was conducted (specifically for G4Jy sources) at 5, 8, and 18\,GHz (PI: C. Jackson). We note that the highest frequency was centred on 18\,GHz rather than 20\,GHz in order to take advantage of the better continuum sensitivity available there. These new data are made available in the G4Jy catalogue accompanying this paper, with flags to indicate unreliable flux-density measurements due to poor coverage in the $u-v$ visibility plane. Whilst the AT20G catalogue provides peak flux-density measurements, we provide integrated flux-density measurements based on images generated using {\sc Miriad} \citep{Sault1995}. The result is that 177 G4Jy sources have coverage at 18/20\,GHz for high-frequency spectral-index calculations (Section~\ref{sec:atca_seds}). Furthermore, we treat 18/20-GHz flux-densities as lower limits owing to extended radio emission being resolved out in the ATCA images. 

\section{Determining associated RACS emission}
\label{sec:racs}

The excellent spectral coverage provided by GLEAM \citep{HurleyWalker2017} at low radio frequencies means that broadband radio spectra for the G4Jy Sample \citep{White2020a,White2020b} are very well-constrained below $\sim$230\,MHz. Meanwhile, we take advantage of the multiple RACS data releases (Sections~\ref{sec:racs_low}--\ref{sec:racs_high}) to better-constrain the `mid-frequency' range of the spectrum, $\sim$1\,GHz. This is not as straightforward as it sounds due to the varying sensitivity of the data towards {\it extended} radio emission, and irregular morphologies being approximated by multiple Gaussian components. Consequently, we conduct thorough visual inspection of G4Jy sources, accompanied by hundreds of manual adjustments, as summarised in this section.  

\subsection{Downloading RACS cutouts}

For downloading RACS-low, RACS-mid and RACS-high data, we used a modified version of the Jupyter notebook provided on the CASDA website (\url{https://research.csiro.au/casda/services/}). This uses the {\sc astroquery} Python package \citep{Ginsburg2019} to generate image cutouts via the TAP (Table Access Protocol) URL \url{https://casda.csiro.au/casda\_vo\_tools/tap}. These cutouts are centred on the radio centroid position determined by \citet{White2020a,White2020b} using SUMSS/NVSS, and are large enough to encompass all of the radio emission associated with G4Jy sources. 

\subsection{A flood-fill algorithm as the basis}
\label{sec:floodfill}

\begin{landscape}
\begin{figure}
\centering
\includegraphics[scale=0.16]{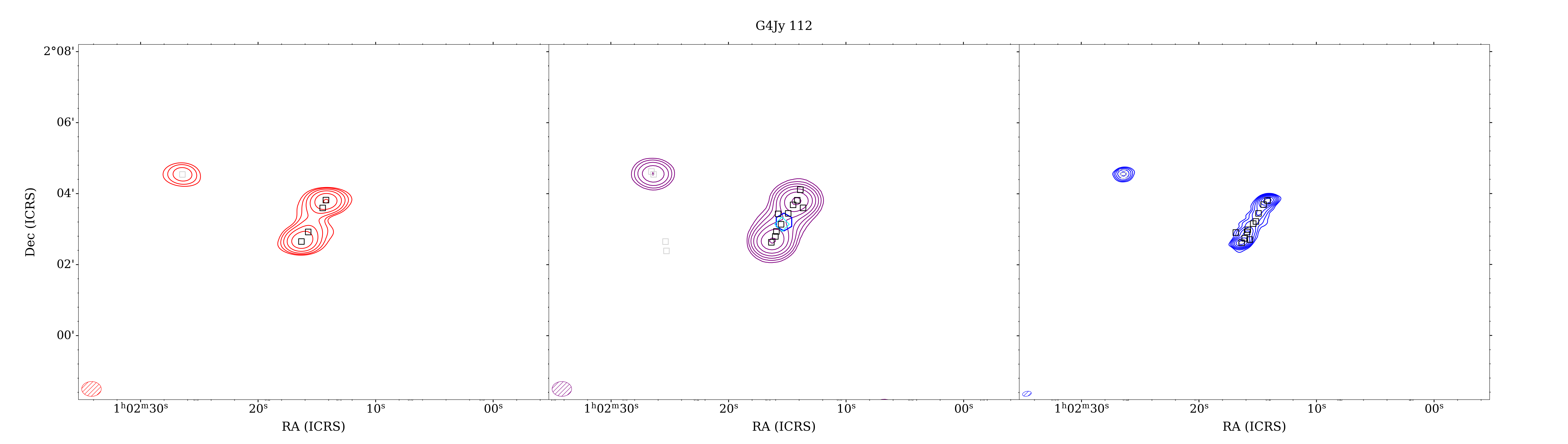} 
\includegraphics[scale=0.16]{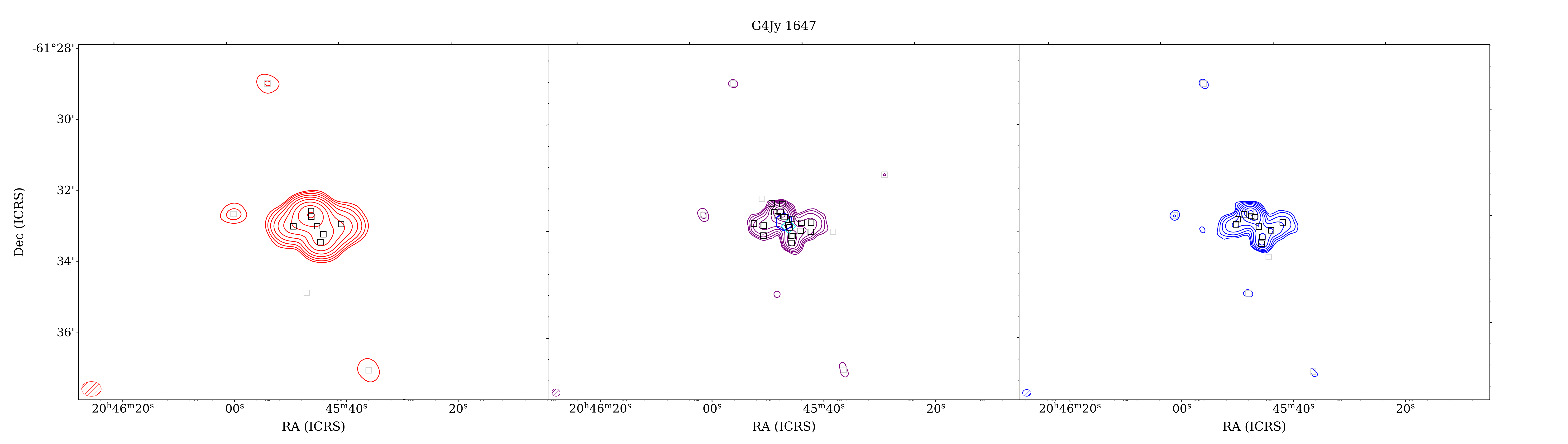} 
\caption{ Examples of the RACS-low (red), RACS-mid (purple), and RACS-high (blue) contour plots that are inspected after flood-filling has been applied. The lowest contour is at 6 times the local rms ($\sigma$), and doubles with each subsequent contour. Meanwhile, the RACS beam is indicated in the lower left-hand corner of each sub-plot. Light-grey squares represent all RACS components that are within 5\,arcmin of the radio centroid position (blue hexagon), whilst black squares represent RACS components that are deemed to be associated with the G4Jy source. The host galaxy position is marked with a cyan diamond and determines the starting point for the flood-fill algorithm (Section~\ref{sec:floodfill}). As shown for G4Jy 1647 (lower panels), the RACS contours are of high-enough resolution to reveal the X-shaped radio morphology of this source. }
\label{fig:first_visual_inspection}
\end{figure}
\end{landscape}

For each of the RACS datasets, we begin by considering all RACS components (light-grey squares in Figure~\ref{fig:first_visual_inspection}) that are within 5\,arcmin of the radio centroid position (blue hexagon) for the G4Jy source. We then apply a flood-filling algorithm that aims to fill to the 6-$\sigma$ contour (where $\sigma$ is the local root-mean-square of the pixel values) and starts from the host-galaxy position (cyan diamond). The latter is available for 93 per cent of the sample \citep{White2026a} and RACS components that lie within the 6-$\sigma$ contour are deemed to be associated with the source (black squares in Figure~\ref{fig:first_visual_inspection}). 

\begin{landscape}
\begin{figure}
\centering
\includegraphics[scale=0.16]{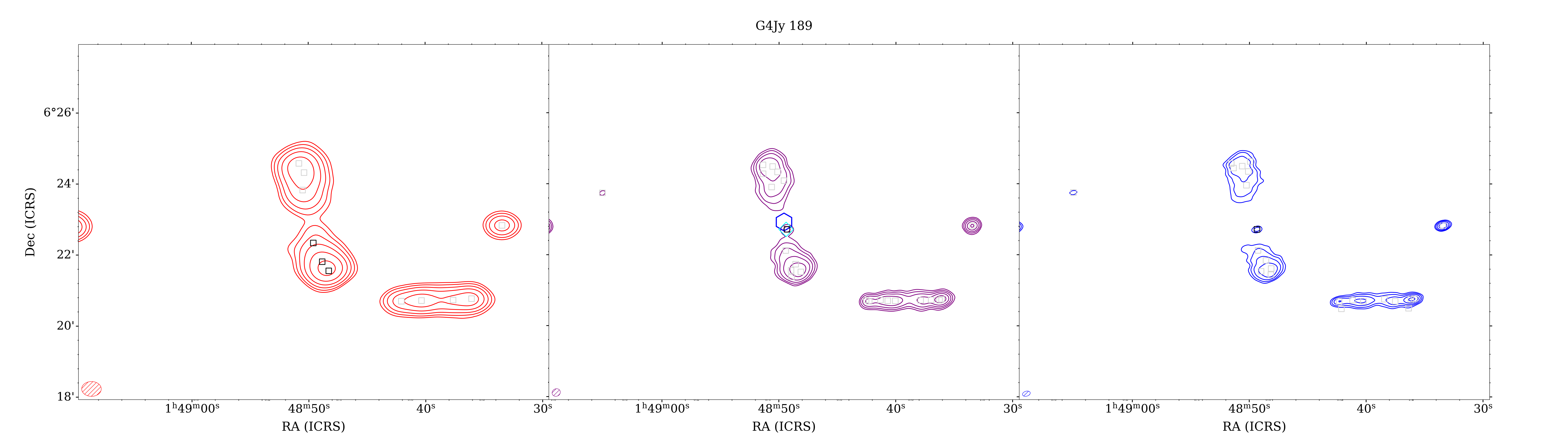} 
\includegraphics[scale=0.16]{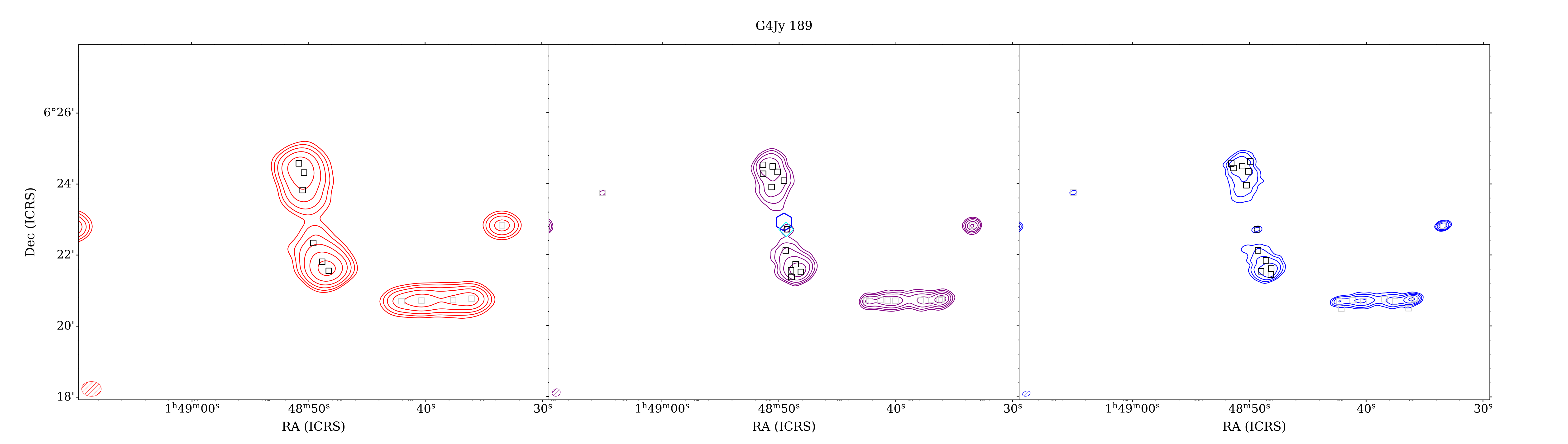} 
\caption{ An example of the RACS-low (red), RACS-mid (purple), and RACS-high (blue) contour plots that are inspected after flood-filling has been applied (upper row) and after manual adjustment of the lists detailing the associated RACS components (lower row). The lowest contour is at 6 times the local rms ($\sigma$), and doubles with each subsequent contour. Meanwhile, the RACS beam is indicated in the lower left-hand corner of each sub-plot. Light-grey squares represent all RACS components that are within 5\,arcmin of the radio centroid position (blue hexagon), whilst black squares represent RACS components that are deemed to be associated with the G4Jy source. The host galaxy position is marked with a cyan diamond and determines the starting point for the flood-fill algorithm (Section~\ref{sec:floodfill}). Unrelated nearby sources remain populated by unassociated RACS components (light-grey squares). }
\label{fig:second_visual_inspection}
\end{figure}
\end{landscape}

Naturally, not all of the components will be properly associated if the source has radio-core emission that is well-separated from the lobes, so (as is the case for G4Jy 189 in Figure~\ref{fig:second_visual_inspection}) manual adjustment is needed. This involves editing a RACS-component list (that is created per G4Jy source) and using {\sc TOPCAT} \citep{Taylor2005} to visually identify which components need to be included (or disassociated). Meanwhile, the criterion that the flood-filled region must encompass the host-galaxy position means that unrelated radio-sources are not included in the association. 

\begin{landscape}
\begin{figure}
\centering
\includegraphics[scale=0.16]{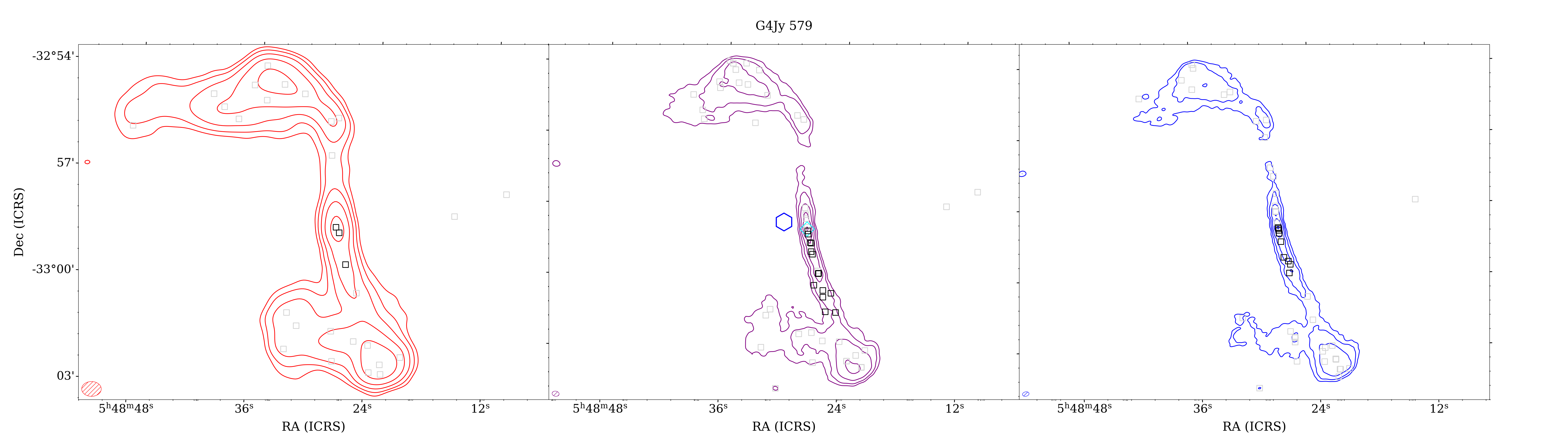} 
\includegraphics[scale=0.16]{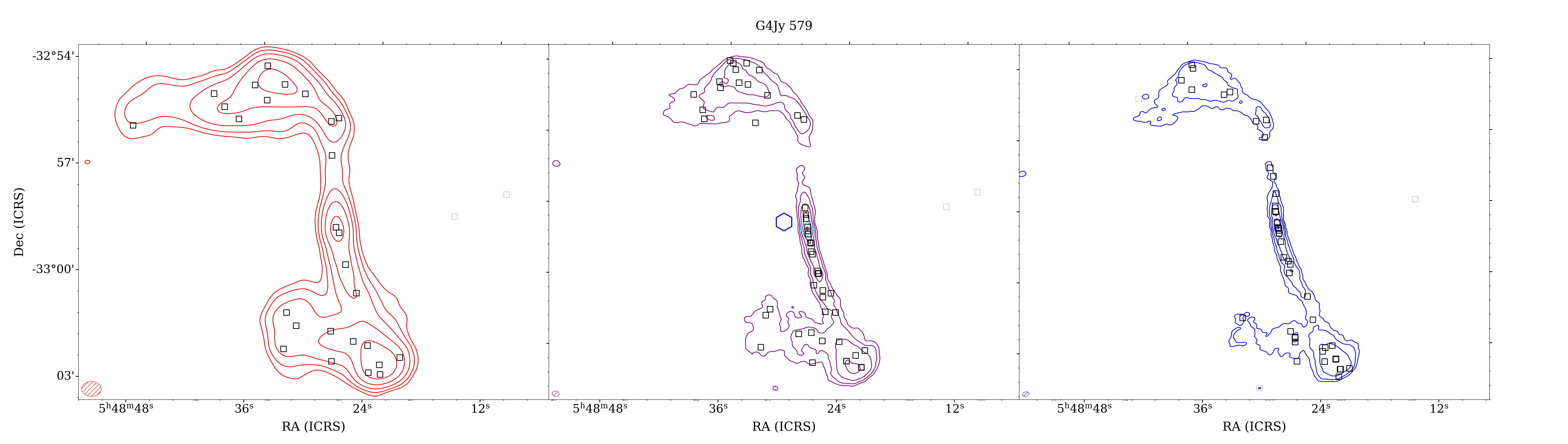} 
\caption{ An example of the RACS-low (red), RACS-mid (purple), and RACS-high (blue) contour plots that are inspected after flood-filling has been applied (upper row) and after manual adjustment of the lists detailing the associated RACS components (lower row). The lowest contour is at 6 times the local rms ($\sigma$), and doubles with each subsequent contour. Meanwhile, the RACS beam is indicated in the lower left-hand corner of each sub-plot. Light-grey squares represent all RACS components that are within 5\,arcmin of the radio centroid position (blue hexagon), whilst black squares represent RACS components that are deemed to be associated with the G4Jy source. The host galaxy position is marked with a cyan diamond and determines the starting point for the flood-fill algorithm (Section~\ref{sec:floodfill}). This radio galaxy, G4Jy 579, is a particularly difficult example where the flood-filling has not extended to the 6-$\sigma$ contour and (in any case) the northern lobe is not smoothly-connected to the radio-core emission.  }
\label{fig:third_visual_inspection}
\end{figure}
\end{landscape}

\begin{landscape}
\begin{figure}
\centering
\includegraphics[scale=0.16]{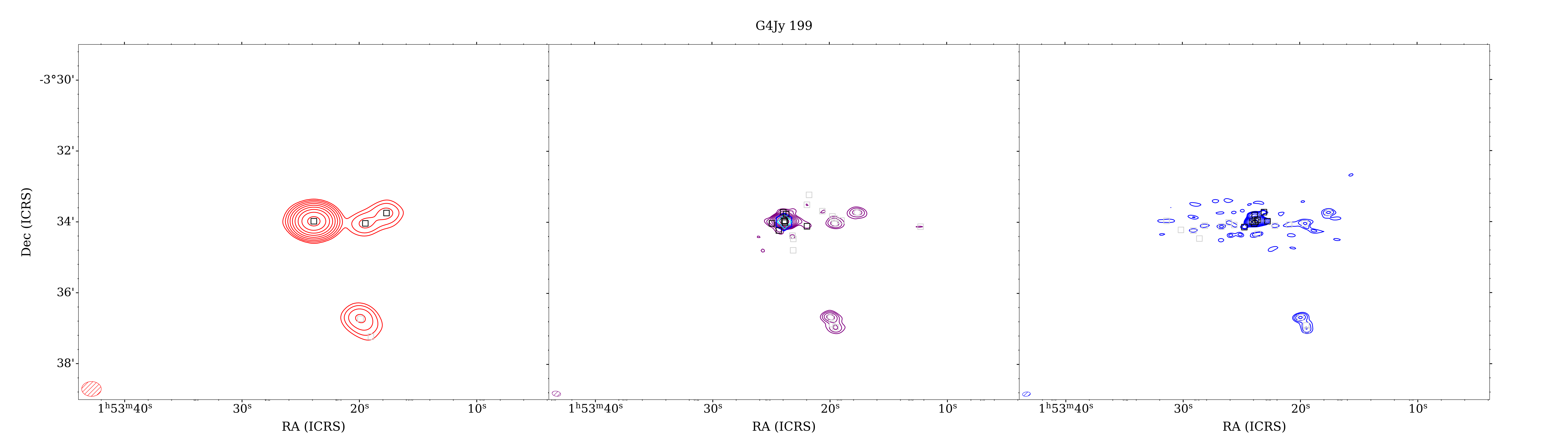} 
\includegraphics[scale=0.16]{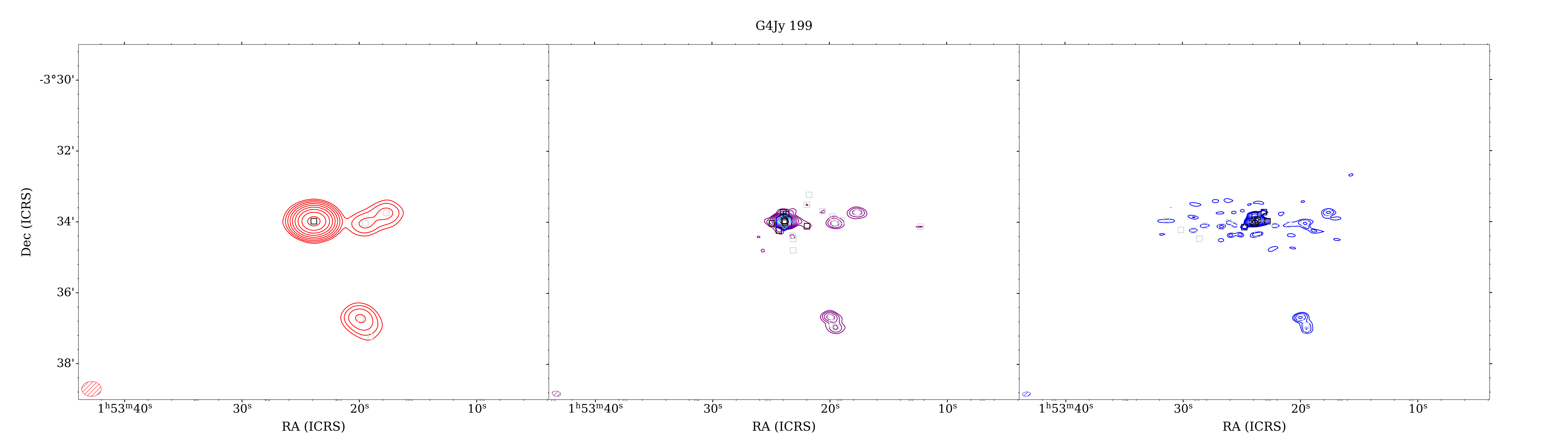} 
\caption{ An example of the RACS-low (red), RACS-mid (purple), and RACS-high (blue) contour plots that are inspected after flood-filling has been applied (upper row) and after manual adjustment of the lists detailing the associated RACS components (lower row). The lowest contour is at 6 times the local rms ($\sigma$), and doubles with each subsequent contour. Meanwhile, the RACS beam is indicated in the lower left-hand corner of each sub-plot. Light-grey squares represent all RACS components that are within 5\,arcmin of the radio centroid position (blue hexagon), whilst black squares represent RACS components that are deemed to be associated with the G4Jy source. The host galaxy position is marked with a cyan diamond and determines the starting point for the flood-fill algorithm (Section~\ref{sec:floodfill}). This is an example of an unrelated `double' radio-galaxy that is originally deemed as associated due to its 6-sigma, RACS-low contour overlapping with that of the G4Jy source in question (G4Jy 199).  }
\label{fig:fourth_visual_inspection}
\end{figure}
\end{landscape}

However, in the case of G4Jy 579 (Figure~\ref{fig:third_visual_inspection}) and hundreds of G4Jy sources, the flood-filling does not work completely owing to discontinuities in the radio emission. To identify and correct for this scenario, we visually inspect intermediate contour plots (Figure~\ref{fig:third_visual_inspection}a) and again use {\sc TOPCAT} to adjust the corresponding RACS-component list. The result is an updated contour plot (Figure~\ref{fig:third_visual_inspection}b) that is then assessed for accuracy. The flags that we assign to the RACS-low (red), RACS-mid (purple), and RACS-high (blue) contours are:

\begin{itemize}
\item 0 -- The correct number of RACS components are associated with the G4Jy source.
\item $-$1 -- The total RACS flux-density will be slightly underestimated due to a RACS component being missed.
\item 1 -- The total RACS flux-density will be slightly overestimated due to a RACS component being associated when it should not be. For example, this flag is assigned when the associated components include artefacts (Figure~\ref{fig:fourth_visual_inspection}).
\end{itemize}

We provide these flags, integrated RACS flux-densities, and number-of-RACS-components in the G4Jy catalogue associated with this paper. In addition, we make the RACS contour-plots available via the G4Jy Zenodo repository: \url{https://zenodo.org/communities/g4jy/records}. 

\subsection{More on visual inspection and manual adjustments}
\label{sec:manual}

In this sub-section we highlight additional limitations of the flood-filling algorithm, {\it which we correct for via manual adjustment}. Firstly, Figure~\ref{fig:fourth_visual_inspection} illustrates a case where the 6-$\sigma$ contour of the G4Jy source intersects with the 6-$\sigma$ contour of a neighbouring source, leading to both being considered for the summed flux-density in RACS-low. 

Secondly, a search radius of 5\,arcmin is used by default, but some G4Jy sources are more extended than 10\,arcmin across. In these cases we extend the search radius based upon the angular size listed in the G4Jy catalogue (courtesy of NVSS/SUMSS imaging) and re-run the flood-fill algorithm and subsequent inspection. (For interest, each of these larger G4Jy sources are presented in Appendix~\ref{app:extended}.)

\begin{landscape}
\begin{figure}
\centering
\includegraphics[scale=0.16]{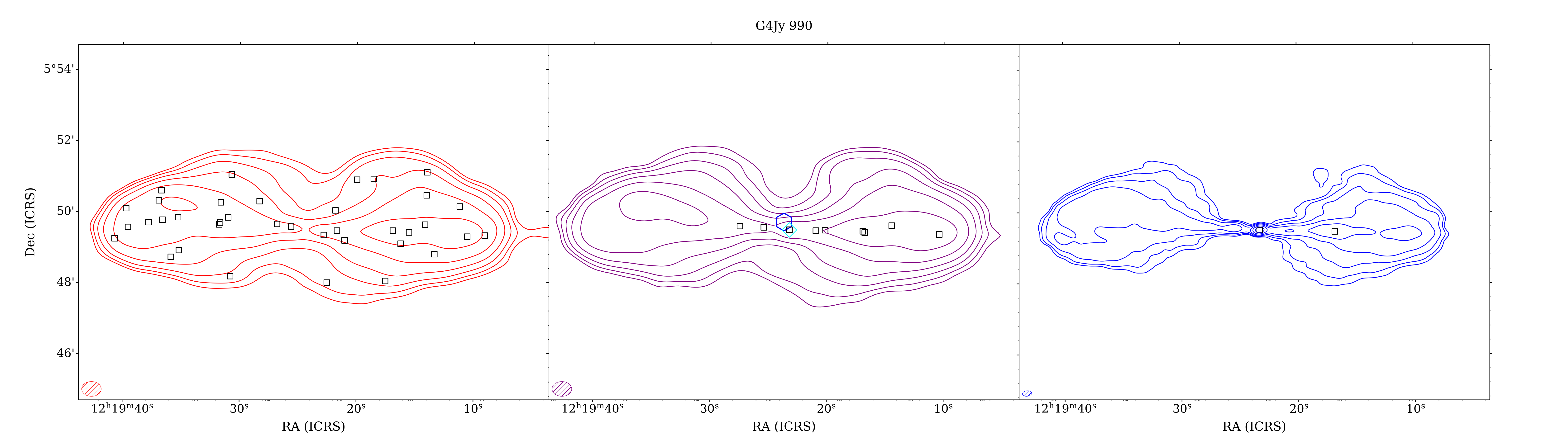} 
\includegraphics[scale=0.16]{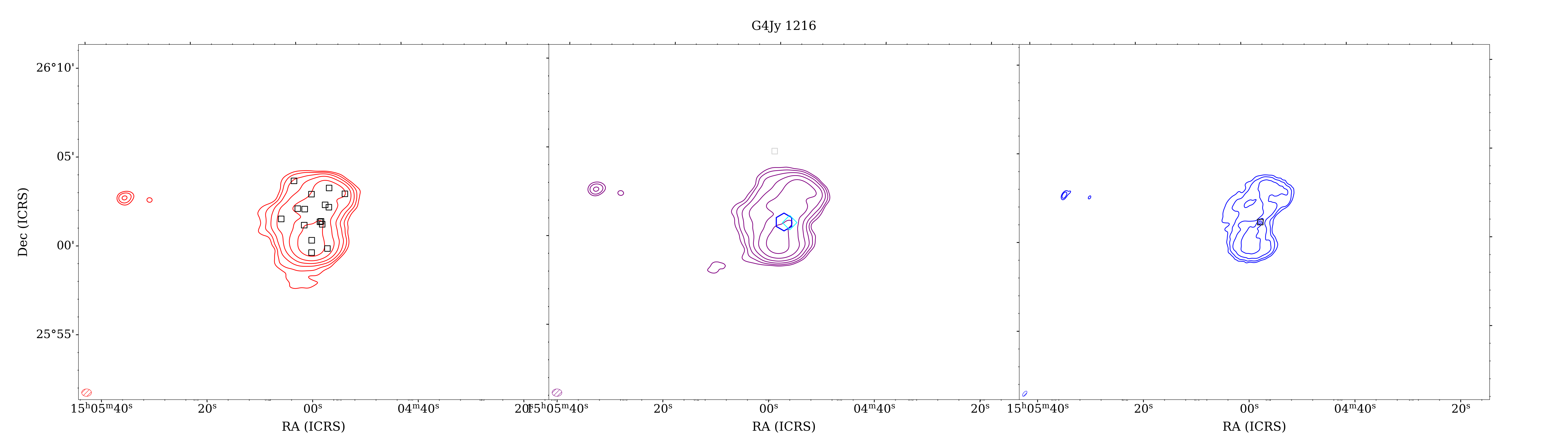} 
\caption{ Two examples of the RACS-low (red), RACS-mid (purple), and RACS-high (blue) contour plots that are inspected after manual adjustment of the lists detailing the associated RACS components. The lowest contour is at 6 times the local rms ($\sigma$), and doubles with each subsequent contour. Meanwhile, the RACS beam is indicated in the lower left-hand corner of each sub-plot. Light-grey squares represent all RACS components that are within 5\,arcmin of the radio centroid position (blue hexagon), whilst black squares represent RACS components that are deemed to be associated with the G4Jy source. The host galaxy position is marked with a cyan diamond and determines the starting point for the flood-fill algorithm (Section~\ref{sec:floodfill}). These sources show diffuse extended emission that {\sc PyBDSF} appears to not be able to characterise (Section~\ref{sec:manual}).  }
\label{fig:fifth_visual_inspection}
\end{figure}
\end{landscape}

Thirdly, we caution that the RACS flux-densities that we determine via contouring and visual-inspection may still be underestimates due to {\sc PyBDSF} not fitting diffuse emission very well. This is because small box-sizes are used by the source-finder to calculate the local rms, and so relatively-uniform emission may be misinterpreted as noise. A couple of such examples are shown in Figure~\ref{fig:fifth_visual_inspection} and should be borne in mind when interpreting the broadband radio SEDs. Specifically, this affects 28 sources (1.6 per cent of the sample), which are listed in Table~\ref{tab:pybdsf_flag}.  

\begin{table}

	\centering
	\caption{The G4Jy sources that most-likely have their RACS-mid or RACS-high flux-density underestimated (Section~\ref{sec:manual}) as a result of {\sc PyBDSF} struggling to fit diffuse/extended radio emission (PyBDSF\_flag = `1').}
	\label{tab:example_table}
	\begin{tabular}{ccccc} 
		\hline
		Source name & & & \\
		\hline
    G4Jy\,74 & G4Jy\,133 & G4Jy\,151 & G4Jy\,171 & G4Jy\,204 \\ 
    G4Jy\,224 & G4Jy\,347 & G4Jy\,381 & G4Jy\,386 & G4Jy\,462 \\
G4Jy\,614 & G4Jy\,619 & G4Jy\,641 & G4Jy\,680 & G4Jy\,886 \\
G4Jy\,949 & G4Jy\,990 & G4Jy\,1034 & G4Jy\,1048 & G4Jy\,1135 \\
G4Jy\,1172 & G4Jy\,1173 & G4Jy\,1216 & G4Jy\,1402 & G4Jy\,1496 \\
          G4Jy\,1499 & G4Jy\,1670 &  G4Jy\,1671 & \\

		\hline
	\end{tabular}

\label{tab:pybdsf_flag}
\end{table}

\section{Broadband radio SEDs (72 -- 1700 MHz)}
\label{sec:broadband_spectra}

Having determined integrated flux-densities at RACS frequencies (Section~\ref{sec:racs}), we revisit the spectral-index calculation of G4Jy Paper IV \citep{White2026b} before further analysing the broadband radio spectra. 

\subsection{Spectral indices between 151\,MHz and 1.4\,GHz}
\label{sec:alpha_mid}

In previous work for the G4Jy Sample we studied the degree of spectral curvature in the radio by combining G4Jy flux-densities \citep{White2020a,White2020b} with those from NVSS \citep{Condon1998}. The NVSS coverage (at 1.4\,GHz) accounts for 77 per cent of the G4Jy Sample, whereas RACS-mid (at 1.3675\,GHz) provides us with (1,728/1,733) $\sim$100 per cent coverage (Table~\ref{tab:coverage}). As such, we recalculate the two-point spectral index, $\alpha_{\mathrm{mid}} = \alpha_{151\,\mathrm{MHz}}^{1368\,\mathrm{MHz}}$, assuming a power-law description of the radio emission ($S_{\nu} \propto \nu^\alpha$) and using 1.3675\,GHz rather than 1.4\,GHz for the calculation. Another difference is that we define the low-frequency spectral-index, $\alpha_{\mathrm{low}}$, to also be between two frequencies (76 and 227\,MHz), rather than a power-law fitted to 20 low-frequency flux-density measurements. Such a relaxation of the spectral-index calculation means that we can also consider G4Jy sources that show significant spectral curvature {\it within} the GLEAM band (spanning 72 to 231\,MHz).

\begin{table}
	\centering
	\caption{Coverage of the G4Jy Sample at different radio frequencies (Section~\ref{sec:broadband_spectra}), and how many sources have flagged data (via visual inspection).}
	\label{tab:example_table}
	\begin{tabular}{lr} 
		\hline
		Characteristic & No. sources\\
		\hline
		The full G4Jy Sample \citep{White2020a,White2020b} & 1,863\\
        Sources that suffer from confusion at 151\,MHz & 386 \\
		Sources with host-galaxy identifications & 1,733\\
		Coverage provided by NVSS (Dec.\,$>-39.5$\,deg)  & 1,437 \\
        Coverage provided by RACS-low (at 887.5\,MHz)  & 1,705 \\
        Sources with underestimated RACS-low flux-density & 15 \\
        Sources with overestimated RACS-low flux-density & 22 \\
        Coverage provided by RACS-mid (at 1367.5\,MHz)  & 1,728 \\
        Sources with underestimated RACS-mid flux-density & 15 \\
        Sources with overestimated RACS-mid flux-density & 65 \\
        Coverage provided by RACS-high (at 1655.5\,MHz)  & 1,727 \\
        Sources with underestimated RACS-high flux-density & 30 \\
        Sources with overestimated RACS-high flux-density & 114 \\
		\hline
	\end{tabular}
    \label{tab:coverage}
\end{table}

\begin{figure*}
\centering
\subfigure[Best-fitted spectra for the full sample]{
\includegraphics[scale=0.52]{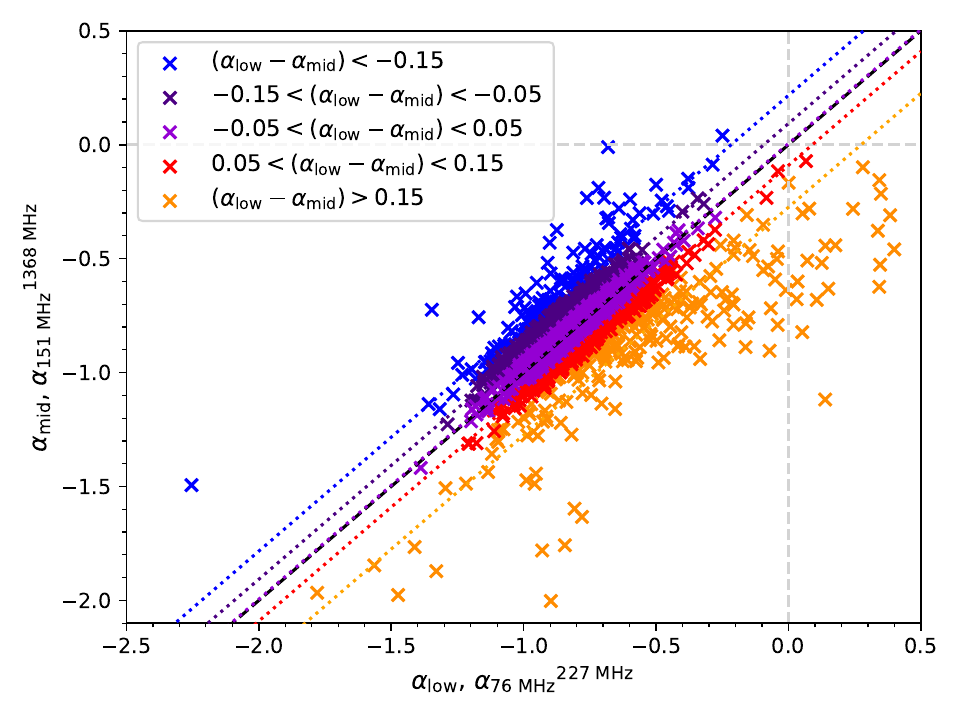} }
\subfigure[Sources with SCI $< -0.15$]{
\includegraphics[scale=0.52]{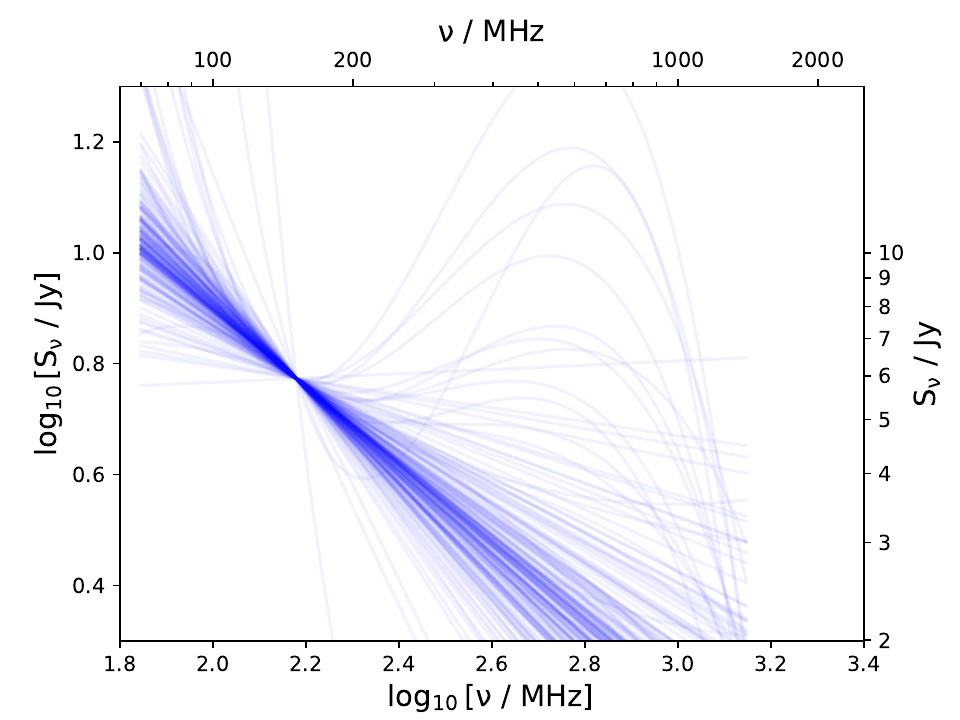} }
\subfigure[Sources with $-0.15 <$ SCI $< -0.05$]{
\includegraphics[scale=0.52]{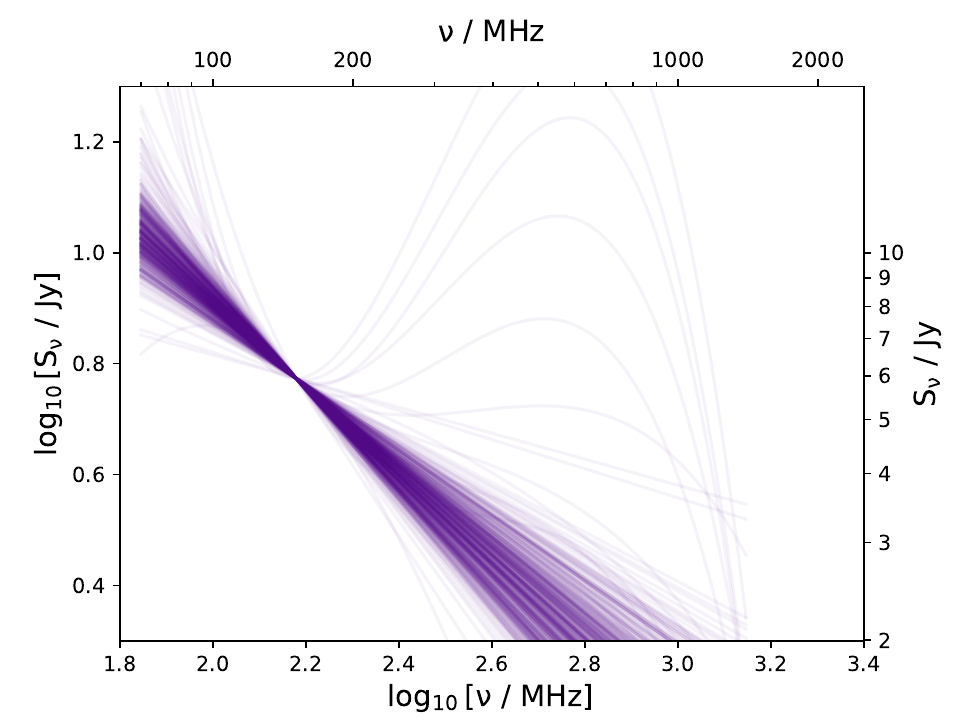} }
\subfigure[Sources with $-0.05 <$ SCI $< 0.05$]{
\includegraphics[scale=0.52]{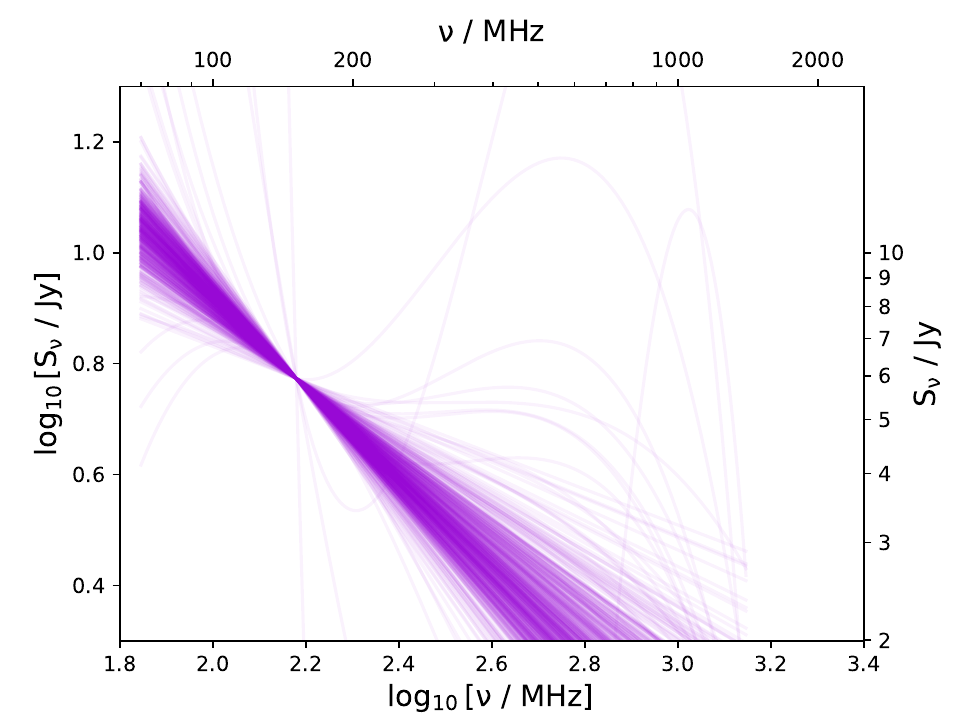} }
\subfigure[Sources with $0.05 <$ SCI $< 0.15$]{
\includegraphics[scale=0.52]{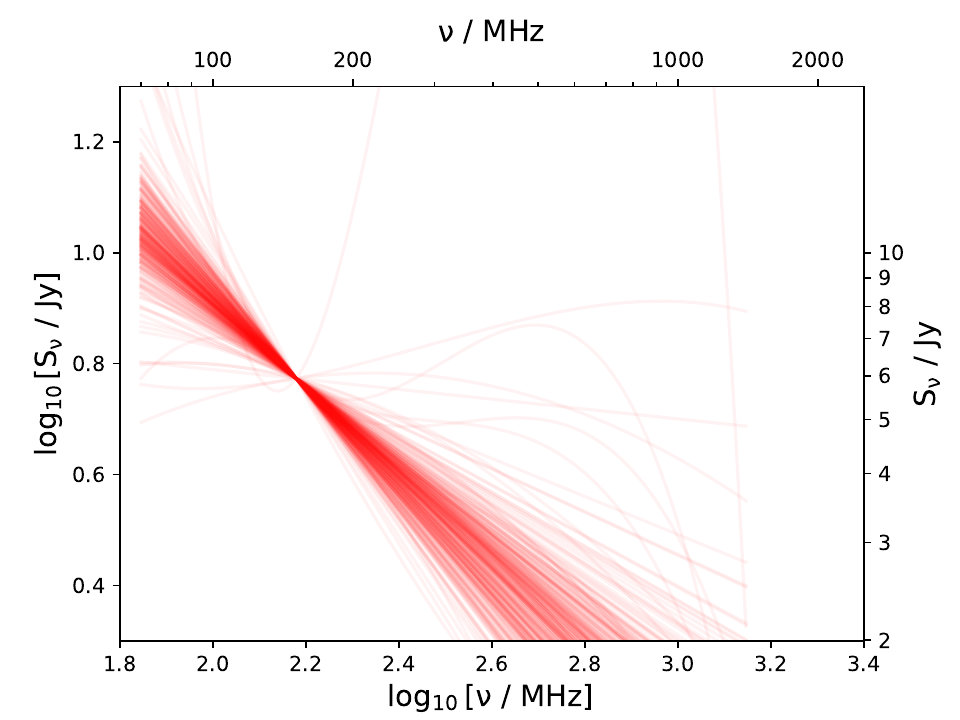} }
\subfigure[Sources with  SCI $> 0.15$]{
\includegraphics[scale=0.52]{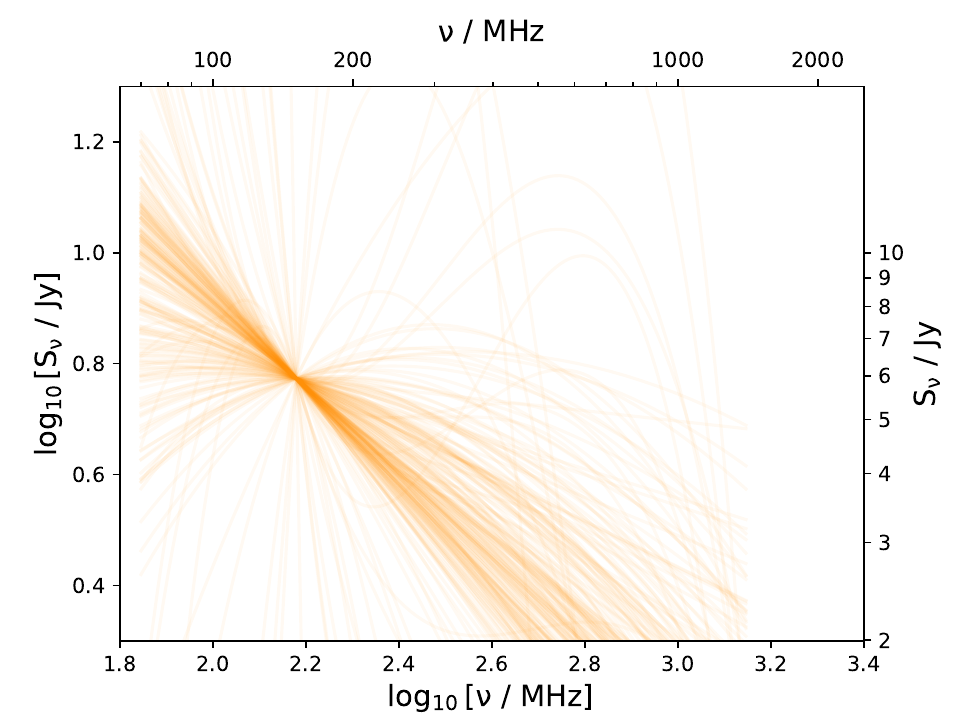} }
\caption{ (a) Showing the distribution in spectral curvature index (SCI) across the full G4Jy Sample, and so beyond that found by \citet{White2026b}. For each SCI interval, we also present radio SEDs (panels b to f). These have been normalised at 151\,MHz so that the variety of spectral shapes across the G4Jy Sample can be visualised.   }
\label{fig:SEDs_by_SCI}
\end{figure*}

Following \citet{White2026b}, we define the spectral-curvature index as SCI = $\alpha_{\mathrm{low}} - \alpha_{\mathrm{mid}}$ and present the two spectral indices against one another in Figure~\ref{fig:SEDs_by_SCI}a. As previously, sources with SCI $< -0.05$ have concave radio spectra (indicative of restarted AGN activity) whilst sources with SCI $> 0.05$ have convex radio spectra (which could be due to spectral ageing or absorption processes). However, within each SCI interval there are a number of sources where the radio emission is best-fit by a third-order polynomial (Figures~\ref{fig:SEDs_by_SCI}b--f).  

\subsection{Low- to mid-radio-frequency SEDs}
\label{sec:poly}

We fit the low-to-mid radio-frequency flux-densities in log-log space, with the highest polynomial considered being of third order: 
\begin{equation}
\log_{10}(S_{\nu}) = a \log_{10}(\nu)^3 + b \log_{10}(\nu)^2 + c \log_{10}(\nu) + d
    \label{eq:poly}
\end{equation}
where $a, b, c$ and $d$ are the polynomial coefficients. We use (reduced) $\chi^2$-minimisation to identify which sources are better-fit by a given polynomial, prioritising the fits where $\chi_{\mathrm{reduced}}^2 <2.0$ (Table~\ref{tab:poly}), and provide the fitted coefficients in the accompanying G4Jy catalogue. As part of the fitting, we consider the brightness-scale systematic errors added in quadrature to the flux-density errors. The former are Declination-dependent (8--13 per cent) for the GLEAM/G4Jy flux-densities \citep{HurleyWalker2017}, 16 per cent for the RACS-low flux-densities \citep{Hale2021}, 6 per cent for the RACS-mid flux-densities \citep{Duchesne2024}, and 10 per cent for the RACS-high flux-densities \citep{Duchesne2025}. Overall, we find that 89 per cent of the sample is well-fit by a first-order polynomial, i.e. by a power-law description of the radio emission. If the source is not well fit by any-order polynomial, we deem this as non-covergent behaviour and do not assign the source a $\chi^2_{\mathrm{reduced}}$ value.

\begin{table}
	\centering
	\caption{The distribution of polynomials (Equation~\ref{eq:poly}) fitted to the broadband (72-1700\,MHz) radio SEDs of the G4Jy Sample (Section~\ref{sec:poly}).}
	\label{tab:example_table}
	\begin{tabular}{lr} 
		\hline
		Characteristic & No. sources\\
		\hline
		Best-fit by a first-order polynomial  & 1,645\\
        Best-fit by a second-order polynomial & 159 \\
		Best-fit by a third-order polynomial & 42\\
        Not well-fitted by any polynomial considered & 17 \\

		\hline
	\end{tabular}
    \label{tab:poly}
\end{table}

\begin{figure*}
\centering
\subfigure[Sources at $0.0< z < 0.5$]{
\includegraphics[scale=0.52]{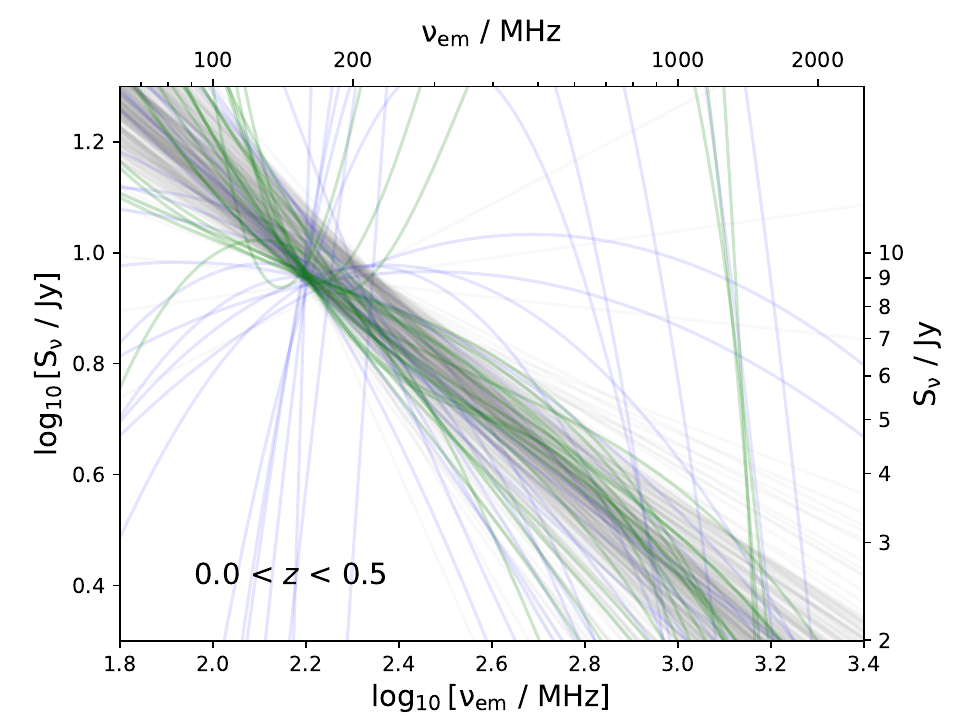} }
\subfigure[Sources at $0.5< z < 1.0$]{
\includegraphics[scale=0.52]{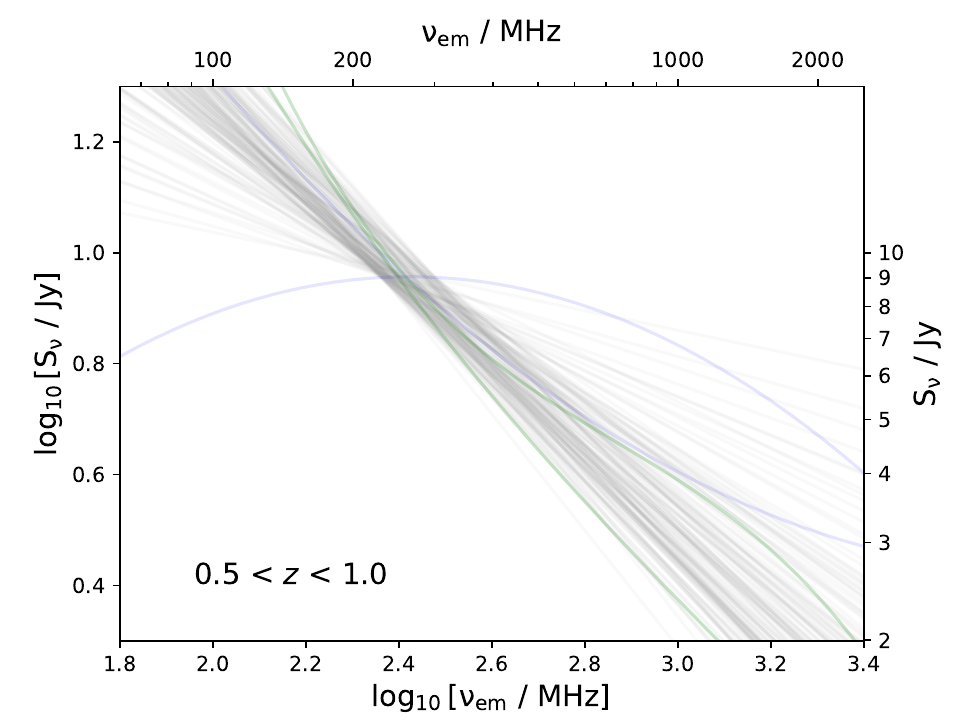} }
\subfigure[Sources at $1.0< z < 1.5$]{
\includegraphics[scale=0.52]{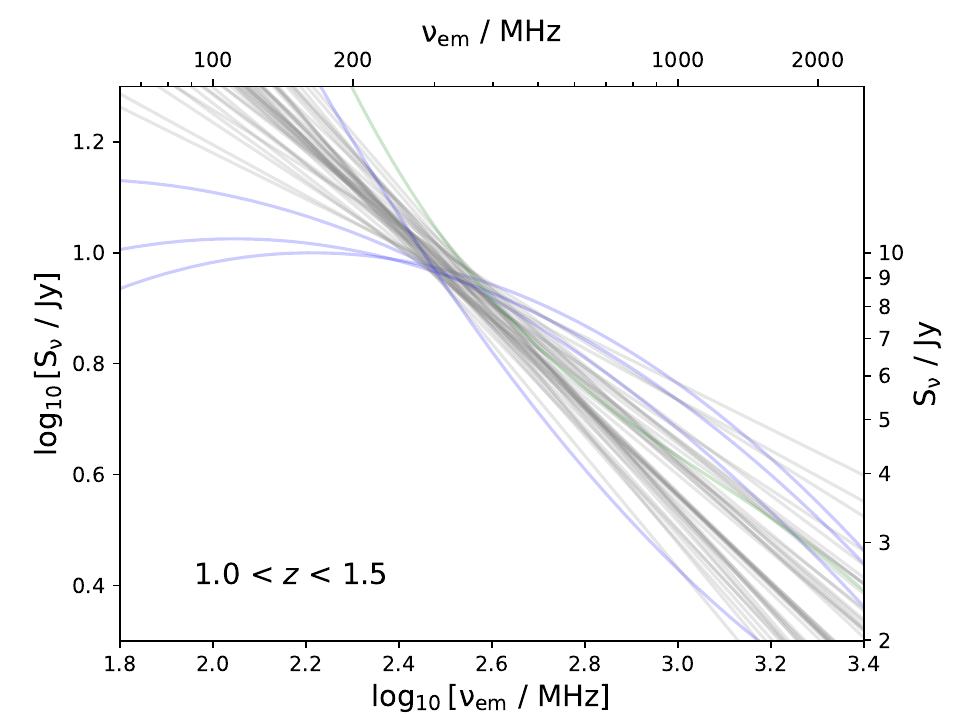} }
\subfigure[Sources at $1.5< z < 2.0$]{
\includegraphics[scale=0.52]{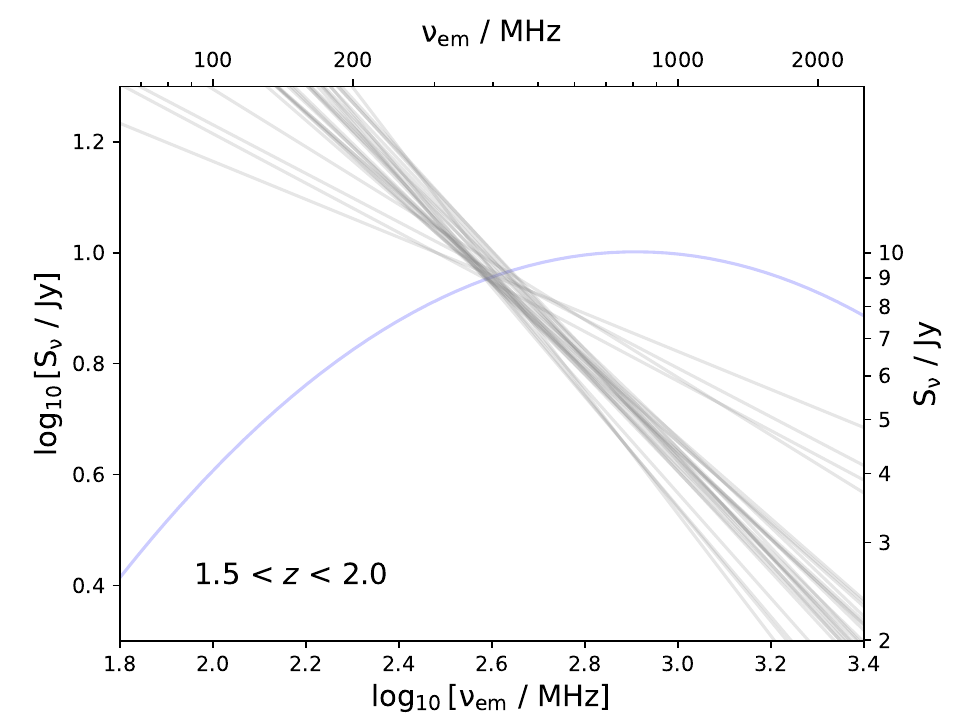} }
\subfigure[Sources at $2.0< z < 2.5$]{
\includegraphics[scale=0.52]{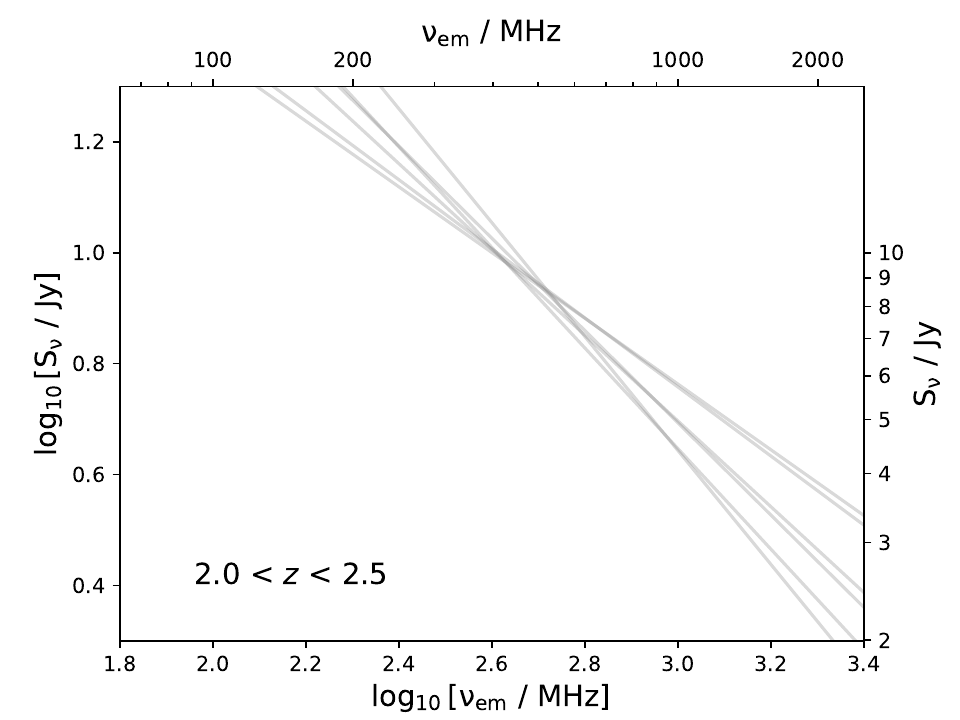} }
\subfigure[Sources at $2.5< z < 3.1$]{
\includegraphics[scale=0.52]{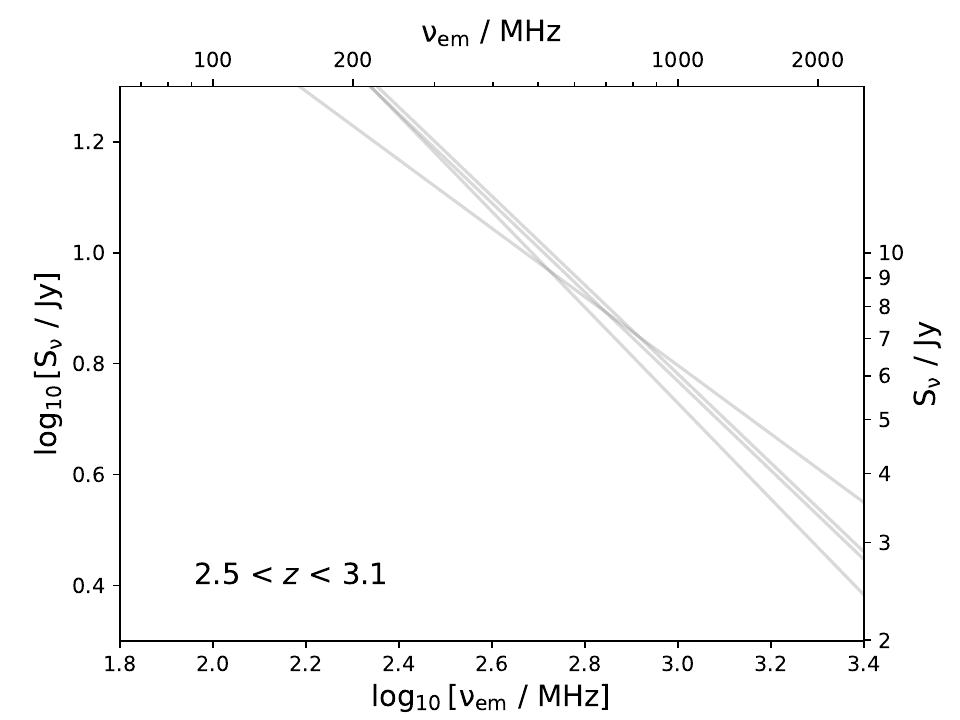} }
\caption{ Radio SEDs that have been normalised at 151\,MHz in the observed frame before being `deredshifted'. Grey lines represent power-law-fitted G4Jy sources, blue lines represent second-order polynomial-fits, and green lines represent third-order polynomial-fits (Section~\ref{sec:broadband_spectra}). }
\label{fig:deredshifted_SEDs}
\end{figure*}

For sources that have a spectroscopic redshift in the G4Jy catalogue \citep{White2026a,White2026b}, we also present the radio SEDs in the rest frame (Figure~\ref{fig:deredshifted_SEDs}) for interest. These have been evaluated at the median flux-density at 151\,MHz in the observed frame and then `deredshifted'. We note that only first-order polynomial fits are present for sources in the highest redshift-bins, which is likely a consequence of small-number statistics.

\begin{figure}
\centering
\includegraphics[scale=0.4]{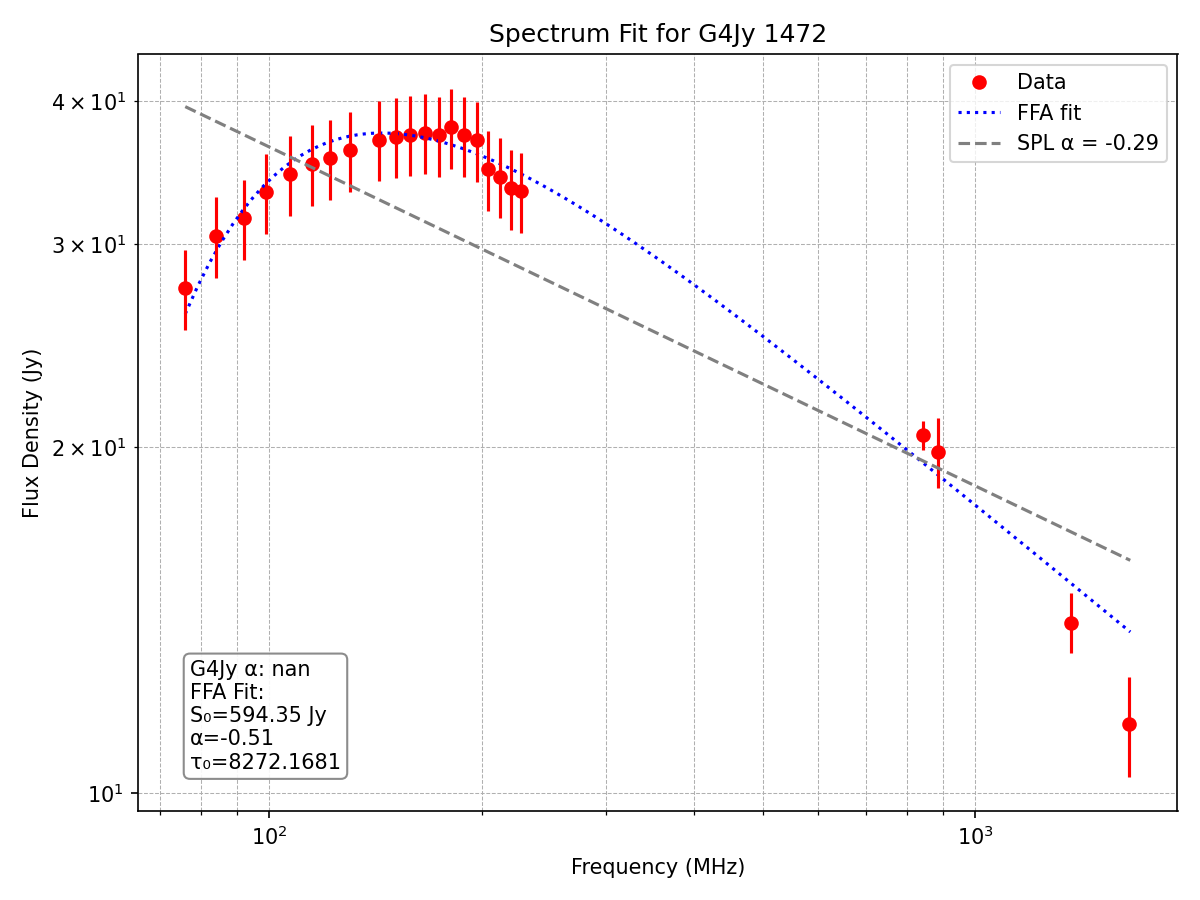} 
\caption{ An example of a free-free absorption (FFA) spectrum best-fitting the radio emission. This source's G4Jy spectral-index (G4Jy\_alpha) is `nan' due to the low-frequency spectrum showing a great degree of spectral curvature within the GLEAM band. In this particular case, SCI $> 0.15$ .}
\label{fig:ffa}
\end{figure}

\begin{figure}
\centering
\includegraphics[scale=0.4]{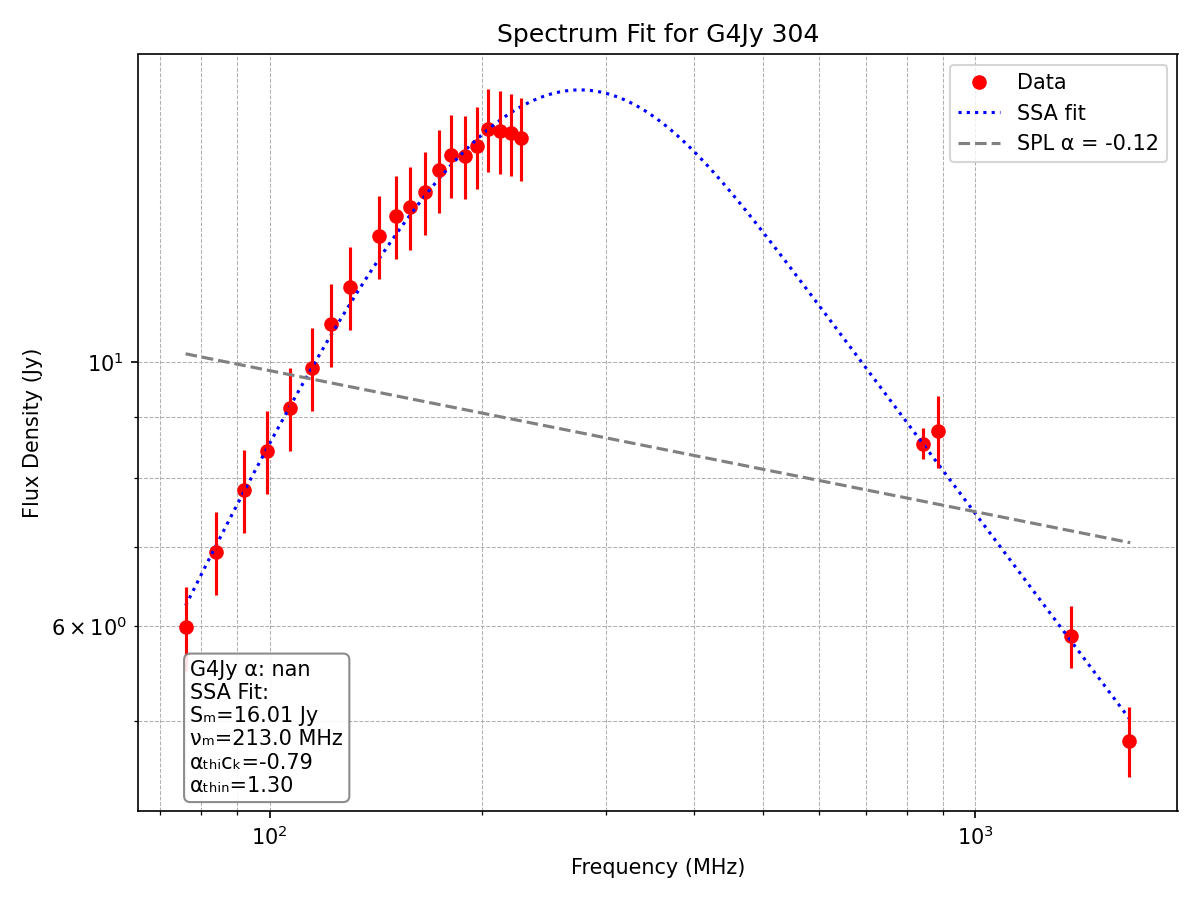} 
\caption{ An example of the synchrotron self-absorption-like (SSA-like) model best-fitting the radio emission. As for G4Jy~1472 in Figure~\ref{fig:ffa}, G4Jy~304 has a spectral curvature index (SCI) $> 0.15$. }
\label{fig:ssa}
\end{figure}

Next, we investigate whether free-free absorption (FFA; e.g. Figure~\ref{fig:ffa}) or synchrotron self-absorption (SSA; e.g. Figure~\ref{fig:ssa}) can explain the spectral behaviour of G4Jy sources that are not well-fitted by a power-law spectrum. In the case of FFA, we consider the scenario where the absorbing ionised medium is mixed with the synchrotron-emitting electrons. As such, the radio spectrum has the following functional form:
\begin{equation}
     S = S_0 \nu^{\alpha} \exp(-\tau_0 \nu) ^{-2.1}
\label{eqn:ffa}
\end{equation}
where $\tau_0$ governs the frequency at which the radio emission pivots. An alternative fitting is considered via the following parameterisation of a spectral turnover in the radio:
\begin{equation}
    S = S_m (x^{\alpha_{\mathrm{thick}}}) (1 - \exp(-x^ {\alpha_{\mathrm{thin}} - \alpha_{\mathrm{thick}}})) / (1 - e^{-1})
\label{eqn:ssa}
\end{equation}
where $x = \nu/\nu_m$, $S_m$ is the flux-density at frequency $\nu_m$, and $\alpha_{\mathrm{thin}}$ and $\alpha_{\mathrm{thick}}$ represent the spectral indices in the optically-thin and optically-thick regimes of the radio spectrum, respectively. This parameterisation reduces to the SSA equation in the event that $\alpha_{\mathrm{thick}}=+2.5$. Otherwise, it is assumed that this model will fit a convex radio spectrum (where $\alpha_{\mathrm{thick}}$ $>$ $\alpha_{\mathrm{thin}}$) and we find that this is the case where the fitting is robust (Table~\ref{tab:absorption}).

Table~\ref{tab:absorption} summarises the number of sources that are classed within either category (`FFA' or `SSA-like') for different SCI values. As before, the model with the lowest reduced $\chi^2$ value is interpreted as the best fit. We find that 3 per cent of (all) G4Jy sources exhibit FFA, whilst 8 per cent of G4Jy sources exhibit SSA-like spectra. Notably, the majority of such-fitted sources have SCI $> 0.15$.

\begin{table*}
	\centering
	\caption{The distribution of 206 sources across different categories when fitting the broadband (72-1700\,MHz) radio SEDs (Section~\ref{sec:broadband_spectra}). This is the subset that are not best-fit by a first-order polynomial (i.e. a power-law description) and are subsequently fitted by either a FFA or a SSA-like spectrum. We also provide the option of a simple power-law (SPL) function in the event of a poor fit by the FFA/SSA-like model. }
	\label{tab:example_table}
	\begin{tabular}{lccccc} 
		\hline
		Combination of the polynomial fitting & & & Number of sources & & \\
       and subsequent FFA/SSA-likw/SPL fitting & SCI $< -0.15$  & $-0.15< $ SCI $< -0.05$ & $-0.05<$ SCI $<0.05$ & $0.05<$ SCI $<0.15$ & SCI $>0.15$ \\
		\hline
		Best-fit by a second-order polynomial and FFA  & 7 & 3 & 0 & 2 & 16 \\
        Best-fit by a second-order polynomial and SSA-like & 0  & 3 & 10 & 25 & 72  \\
                Best-fit by a second-order polynomial and SPL & 0 & 0 & 2 & 0 & 0  \\
		Best-fit by a third-order polynomial and FFA & 5 & 6 & 3 & 1 & 9 \\
        Best-fit by a third-order polynomial and SSA-like & 2 & 4 & 14 & 1 & 18 \\
            Best-fit by a third-order polynomial and SPL & 0 & 0 & 1 & 0 & 2 \\
        \hline
Total & 14 & 16 & 30 & 29 & 117 \\
		\hline
	\end{tabular}
    \label{tab:absorption}
\end{table*}

\section{Broadband radio SEDs (72 MHz -- 20 GHz)}
\label{sec:atca_seds}

\begin{figure*}
\centering
\subfigure[G4Jy 216]{
\includegraphics[scale=0.52]{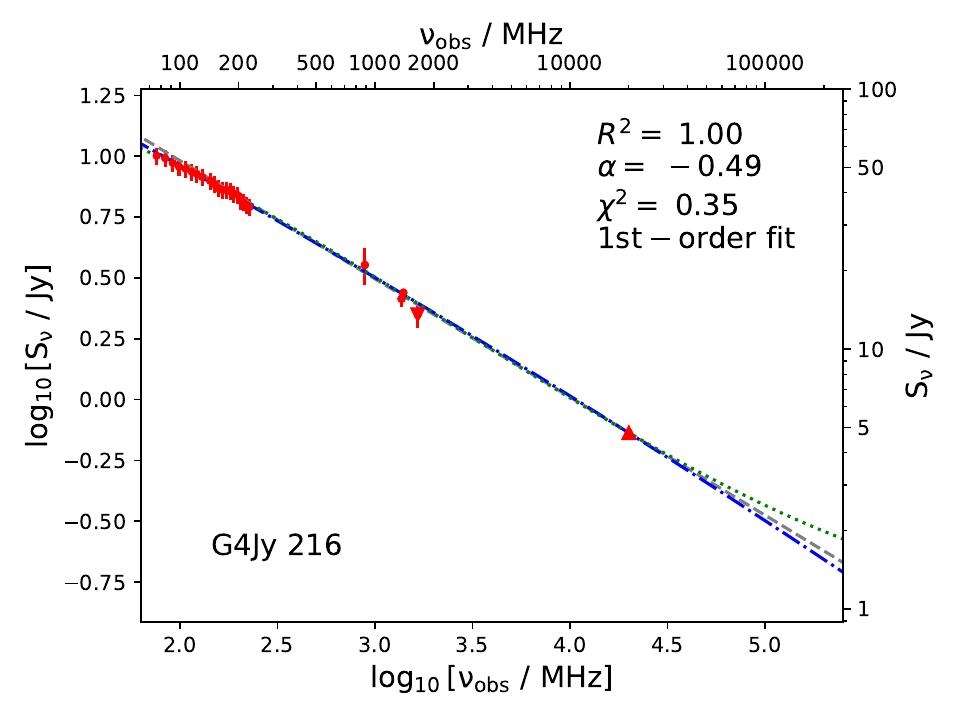} }
\subfigure[G4Jy 540]{
\includegraphics[scale=0.52]{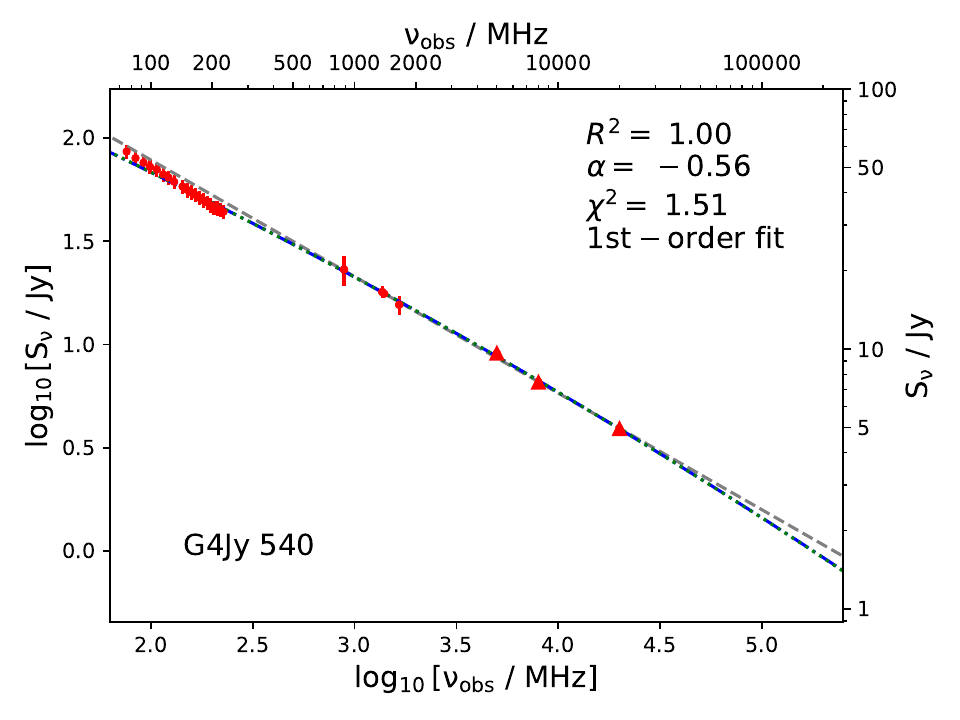} }
\subfigure[G4Jy 569]{
\includegraphics[scale=0.52]{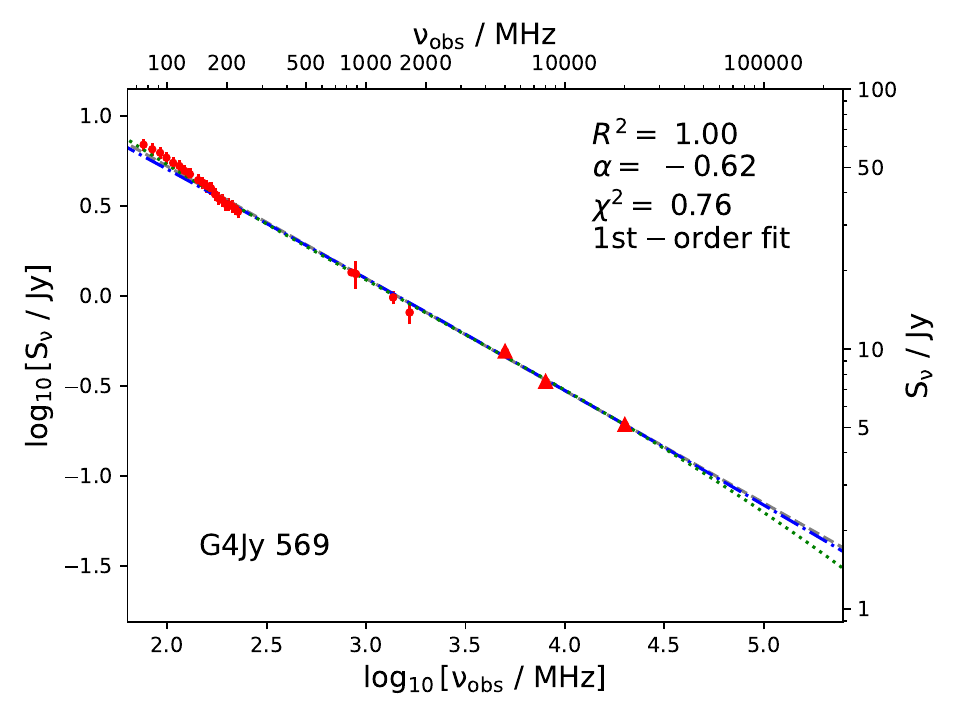} }
\subfigure[G4Jy 1225]{
\includegraphics[scale=0.52]{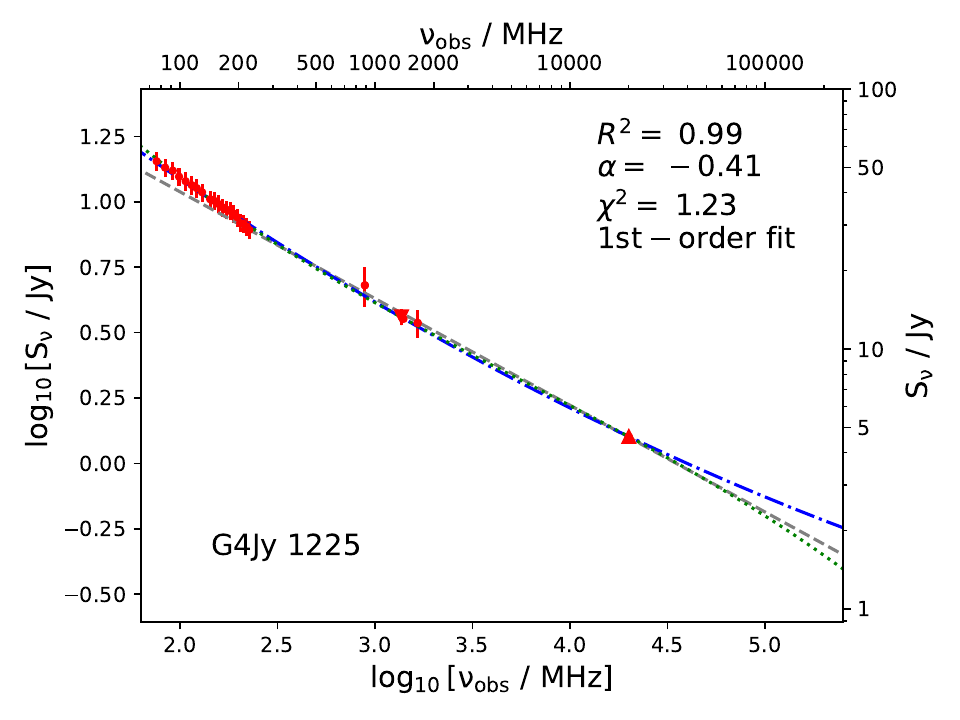} }
\subfigure[G4Jy 1788]{
\includegraphics[scale=0.52]{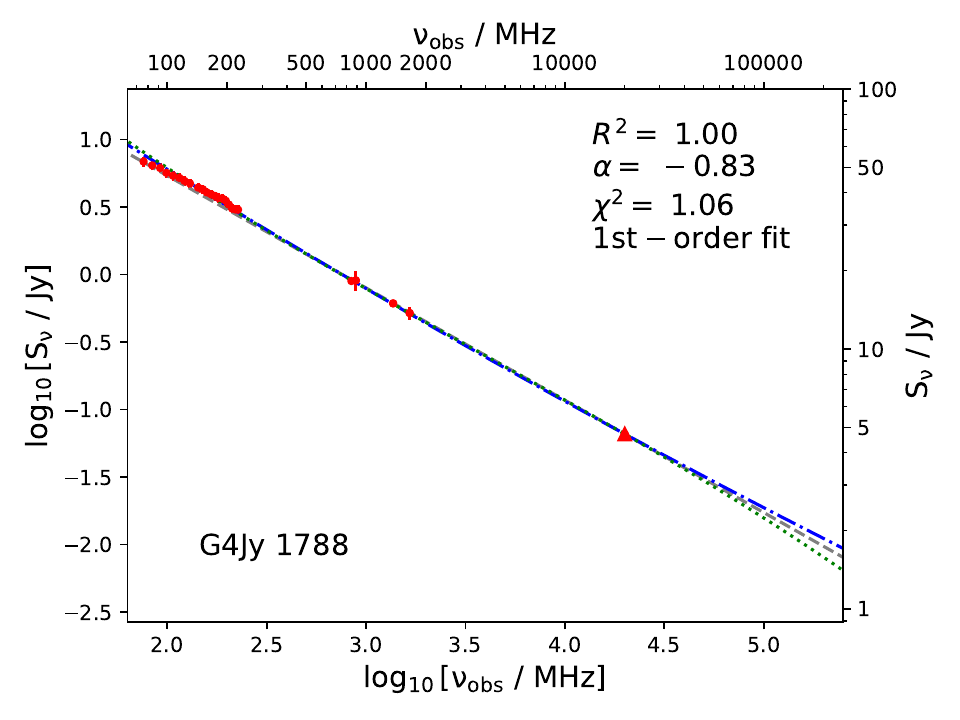} }
\subfigure[G4Jy 1800]{
\includegraphics[scale=0.52]{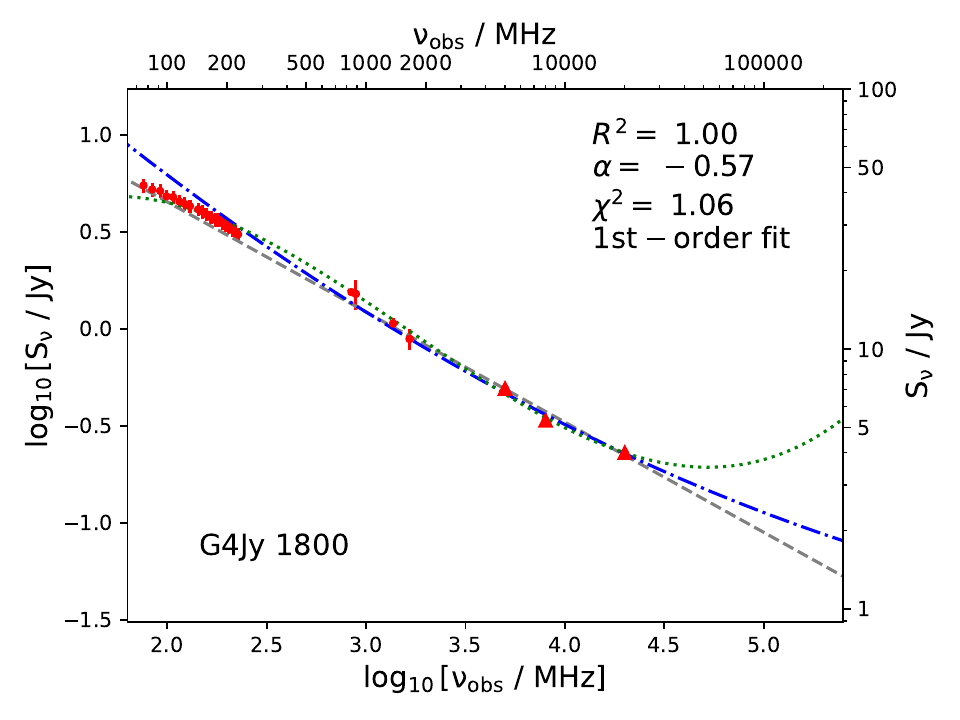} }
\caption{ Examples of G4Jy-ATCA sources that have their radio spectra best-fit by a first-order polynomial (i.e. a power-law function). This is represented by a grey dashed line, whilst a second-order polynomial is represented by a blue dash-dotted line and a third-order polynomial is represented by a green dotted line. Red dots represent G4Jy, SUMSS/NVSS and RACS flux-densities, whilst red triangles indicate that the ATCA flux-densities are lower limits. }
\label{fig:firstorder_fits}
\end{figure*}

\begin{figure*}
\centering
\subfigure[G4Jy 178]{
\includegraphics[scale=0.52]{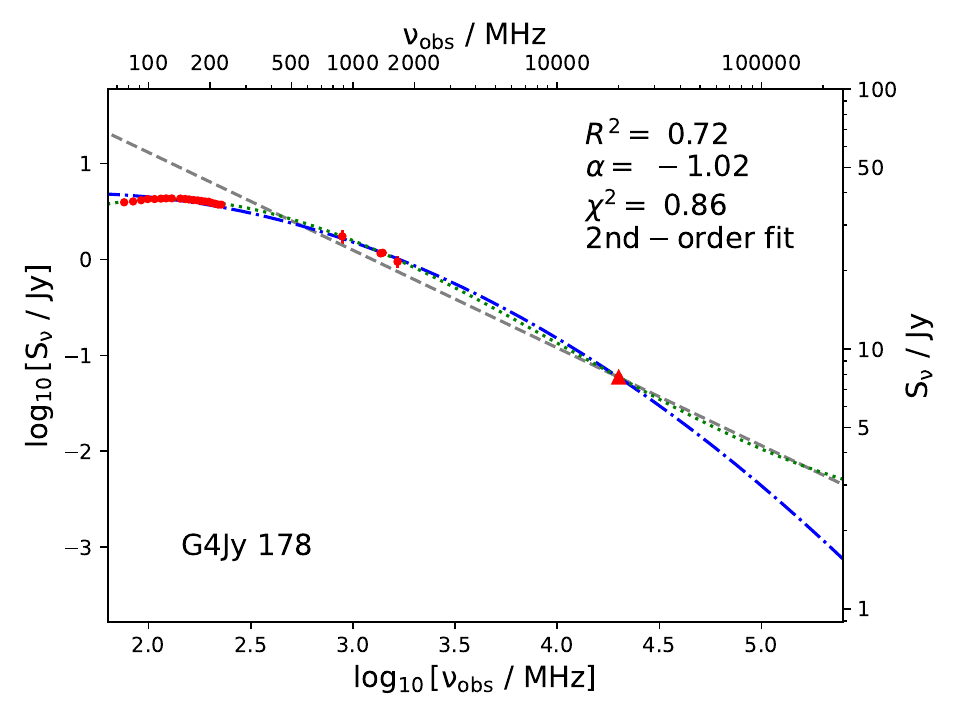} }
\subfigure[G4Jy 835]{
\includegraphics[scale=0.52]{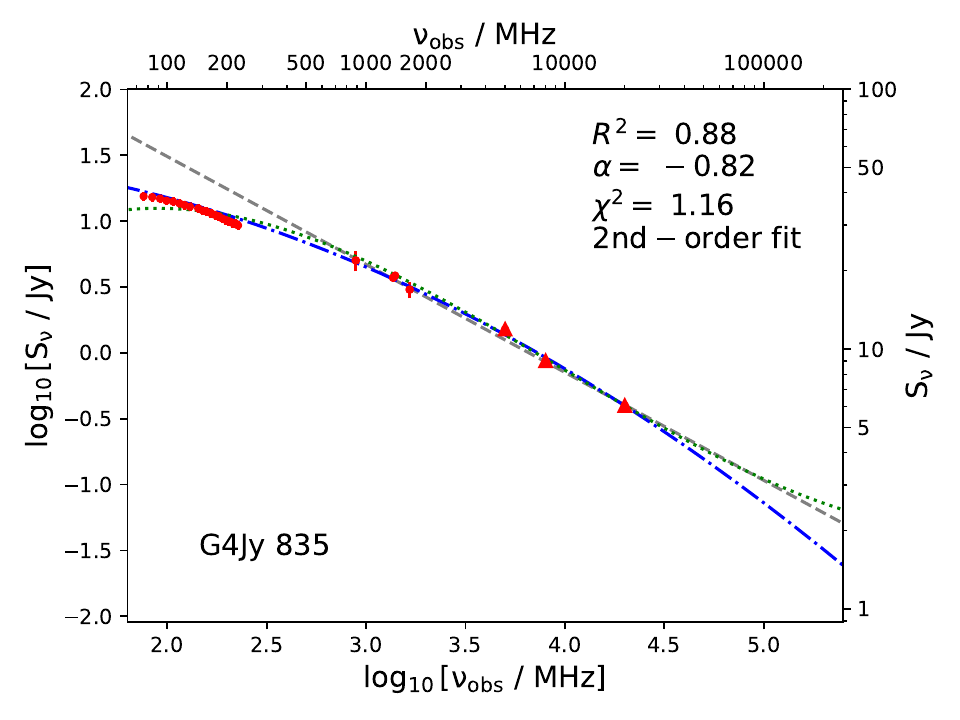} }
\subfigure[G4Jy 891]{
\includegraphics[scale=0.52]{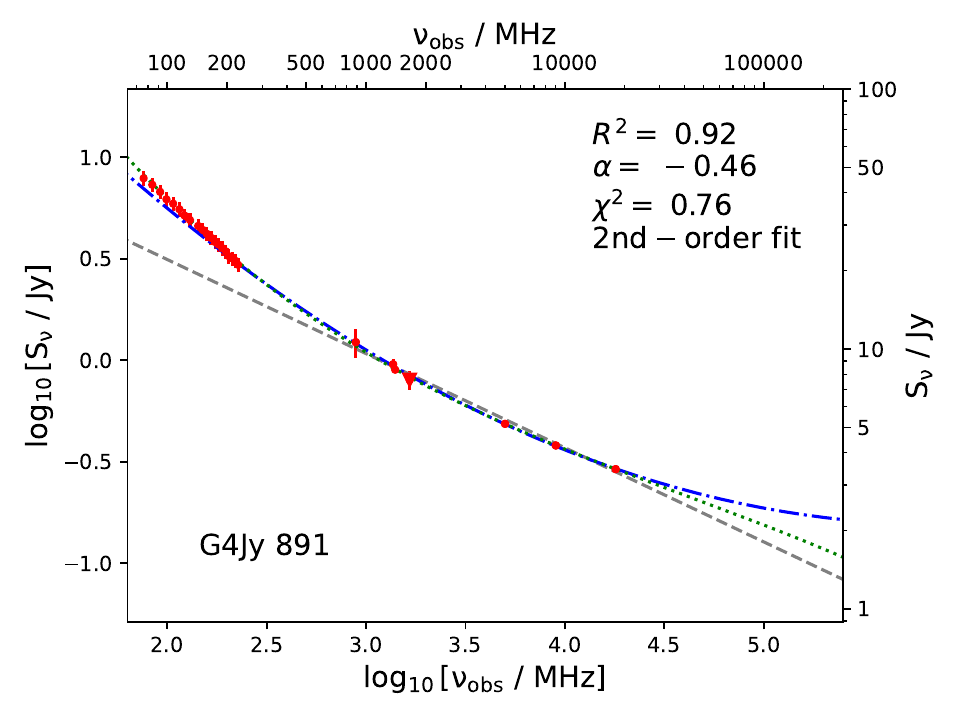} }
\subfigure[G4Jy 961]{
\includegraphics[scale=0.52]{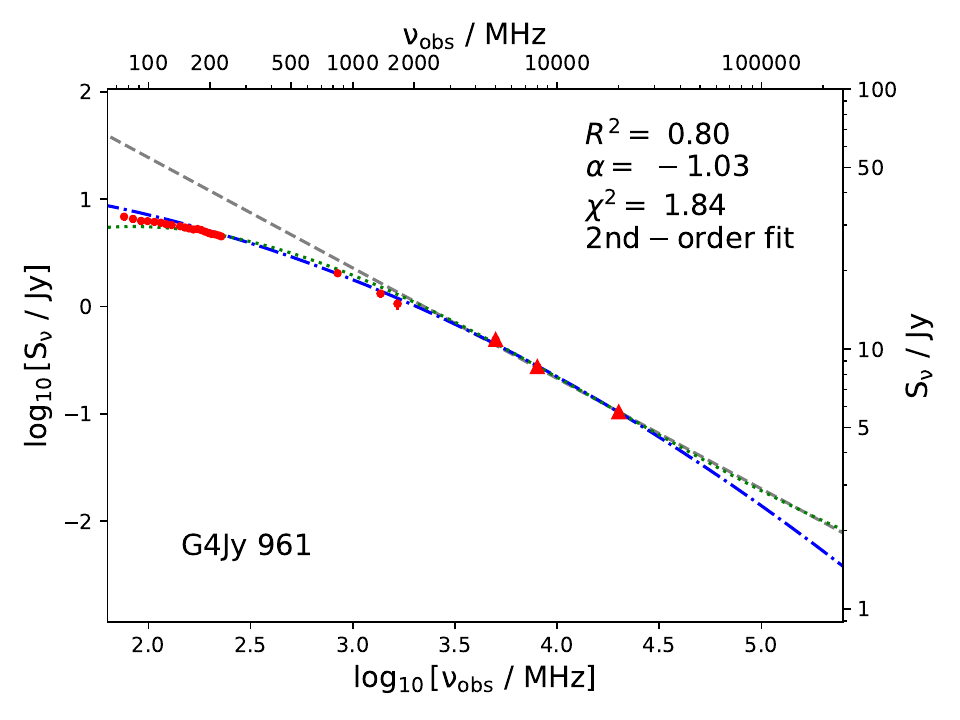} }
\subfigure[G4Jy 1769]{
\includegraphics[scale=0.52]{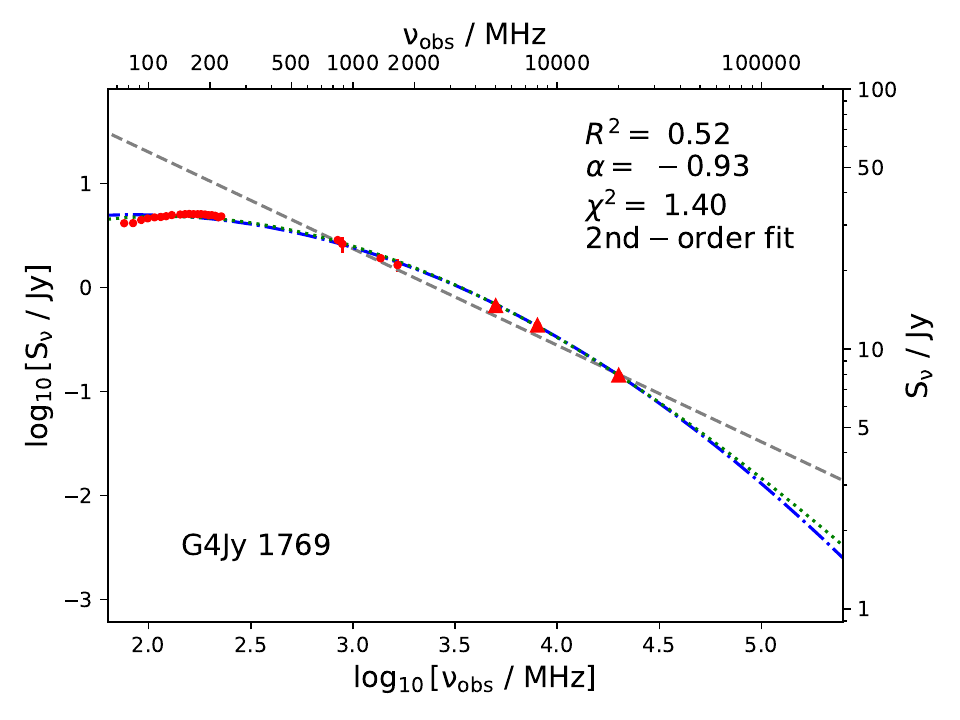} }
\subfigure[G4Jy 1777 (flat-spectrum radio quasar)]{
\includegraphics[scale=0.52]{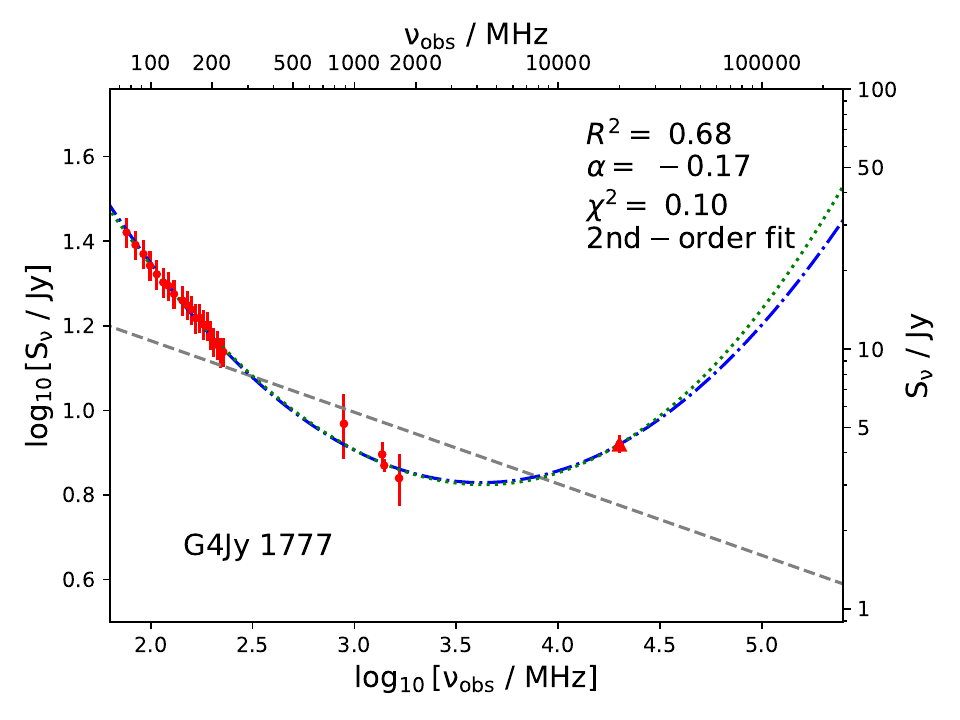} }
\caption{ Examples of G4Jy-ATCA sources that have their radio spectra best-fit by a second-order polynomial. This is represented by a blue dash-dotted line, whilst a first-order polynomial is represented by a grey dashed line and a third-order polynomial is represented by a green dotted line. Red dots represent G4Jy, SUMSS/NVSS and RACS flux-densities, whilst red triangles indicate that the ATCA flux-densities are lower limits. }
\label{fig:secondorder_fits}
\end{figure*}

\begin{figure*}
\centering
\subfigure[G4Jy 104]{
\includegraphics[scale=0.52]{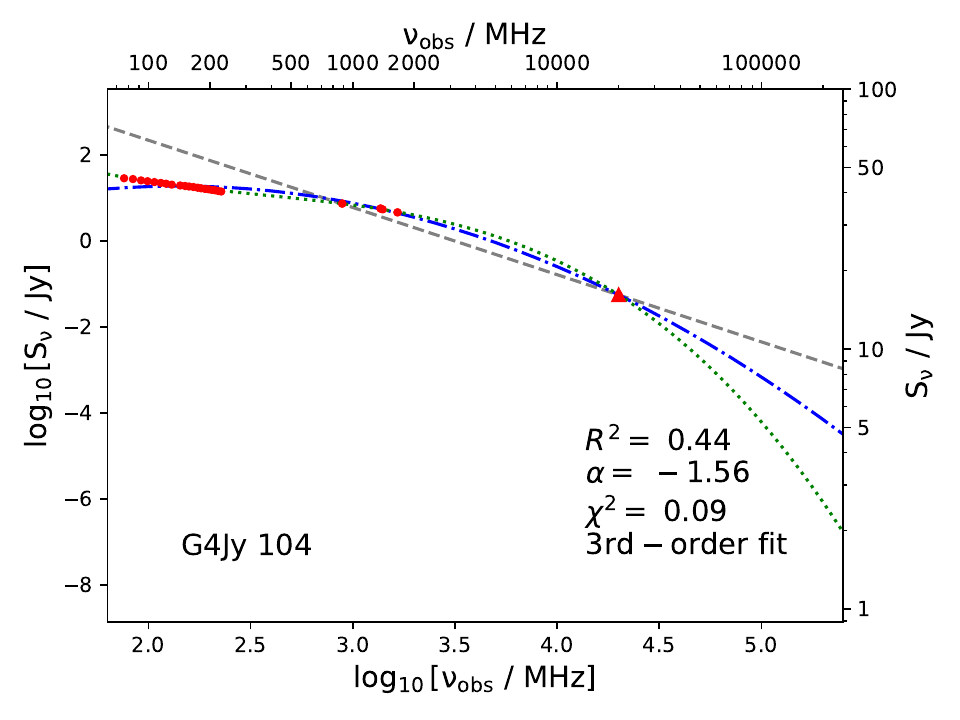} }
\subfigure[G4Jy 212]{
\includegraphics[scale=0.52]{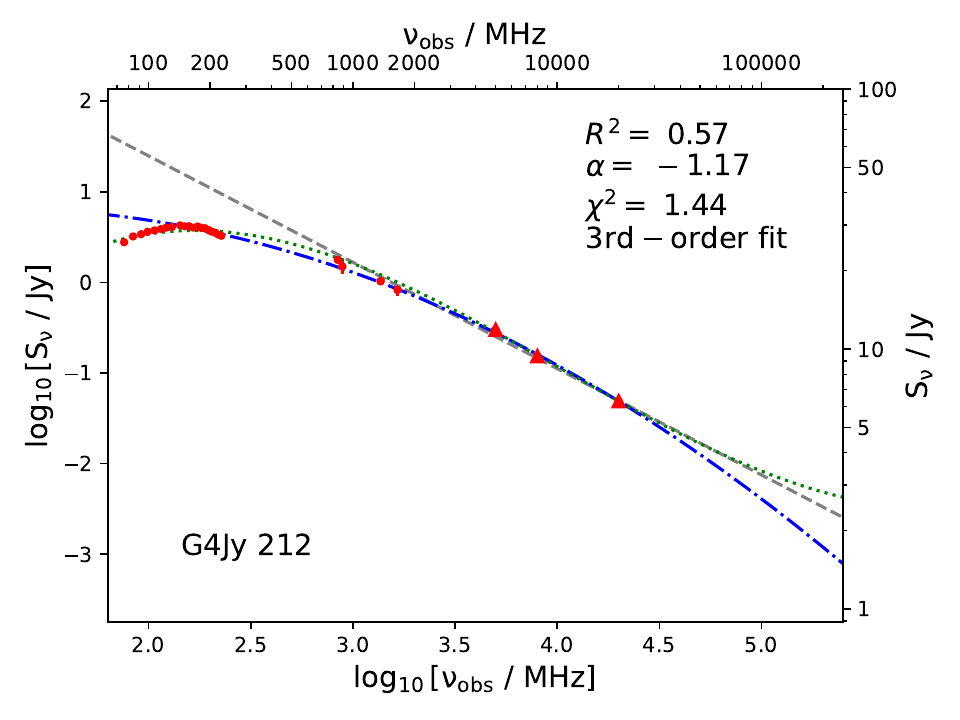} }
\subfigure[G4Jy 819]{
\includegraphics[scale=0.52]{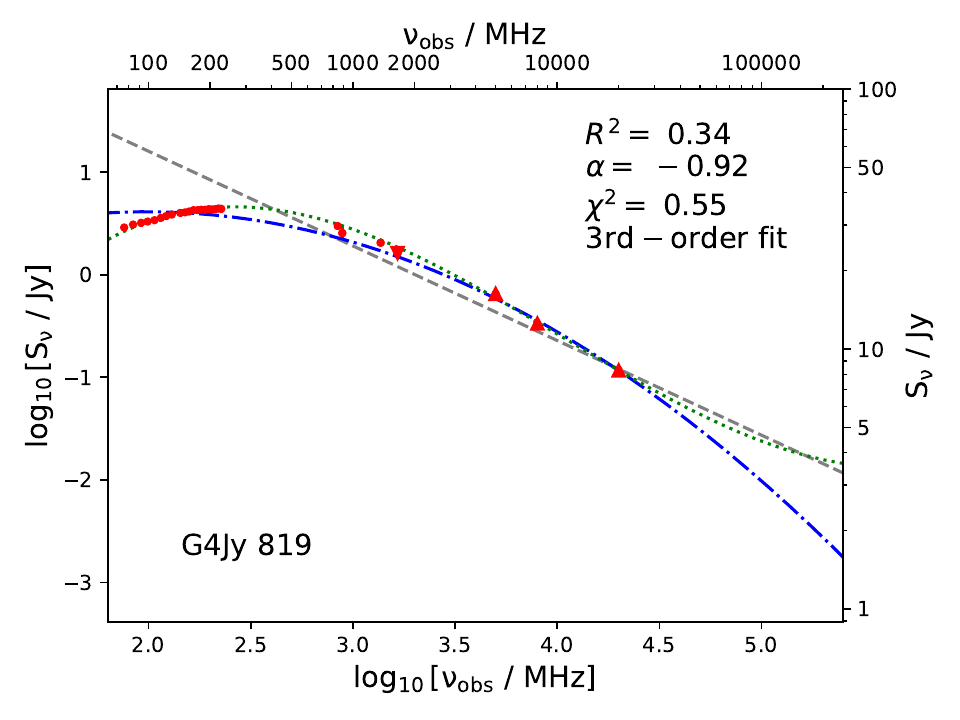} }
\subfigure[G4Jy 1168 (flat-spectrum radio quasar)]{
\includegraphics[scale=0.52]{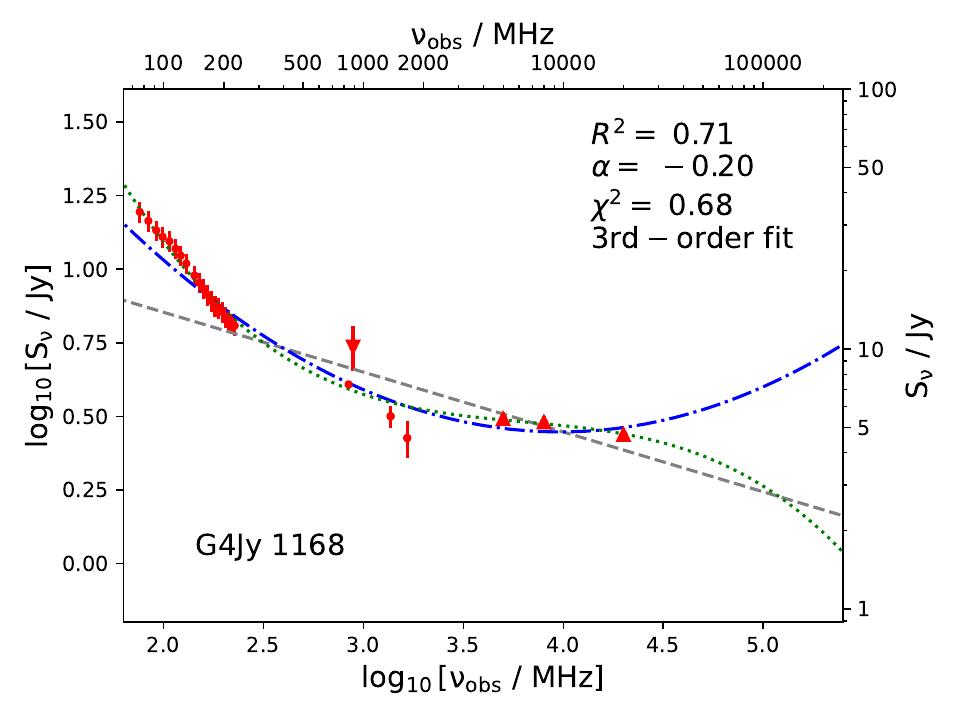} }
\subfigure[G4Jy 1487]{
\includegraphics[scale=0.52]{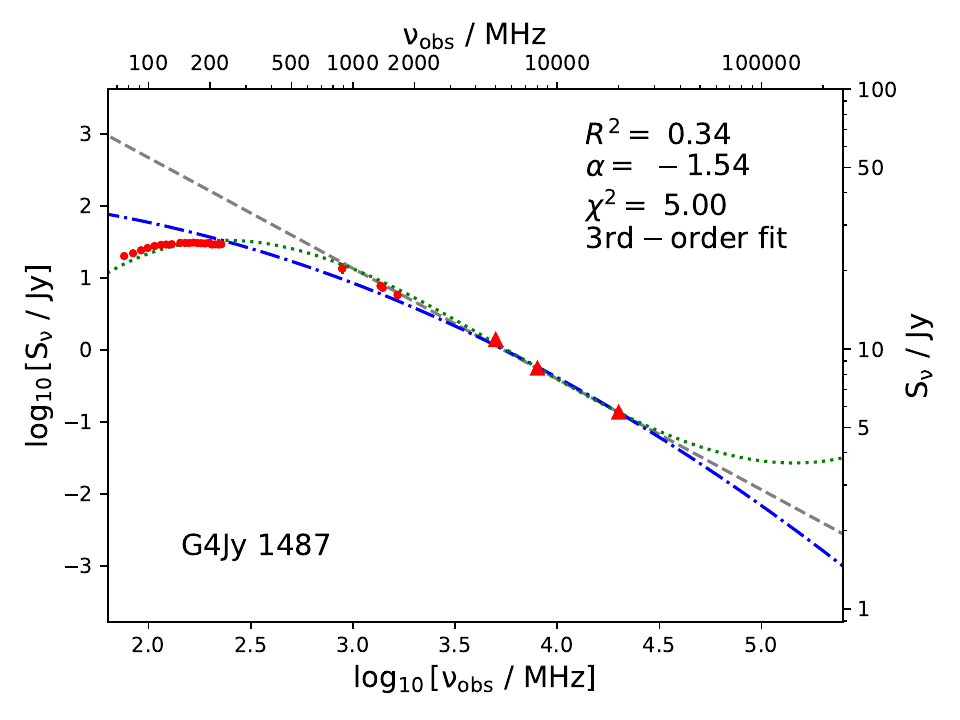} }
\subfigure[G4Jy 1850]{
\includegraphics[scale=0.52]{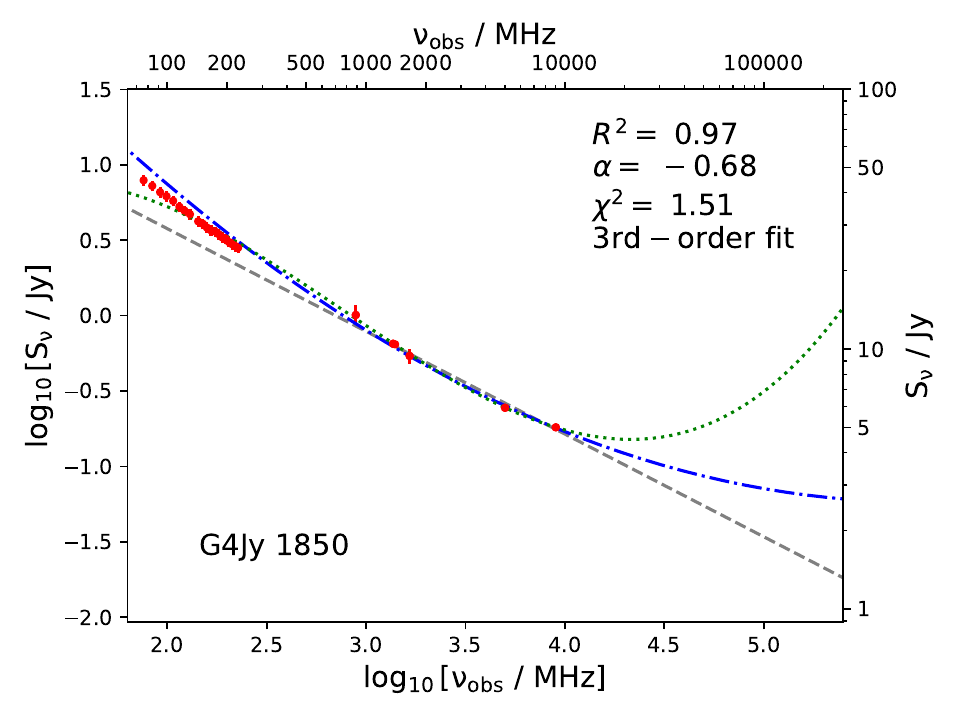} }
\caption{ Examples of G4Jy-ATCA sources that have their radio spectra best-fit by a third-order polynomial. This is represented by a green dotted line, whilst a first-order polynomial is represented by a grey dashed line and a second-order polynomial is represented by a blue dash-dotted line. Red dots represent G4Jy, SUMSS/NVSS and RACS flux-densities, whilst red triangles indicate that the ATCA flux-densities are lower limits. }
\label{fig:thirdorder_fits}
\end{figure*}

\begin{figure}
\centering
\includegraphics[scale=0.52]{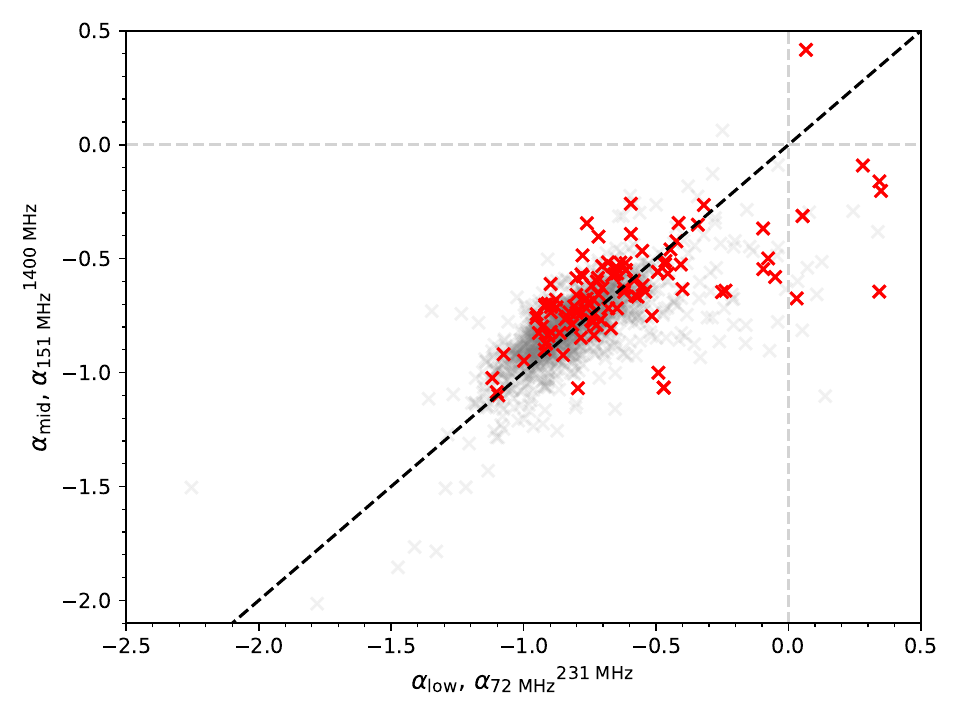} 
\caption{ The distribution of `NVSS-ATCA' sources in the $\alpha_{\mathrm{low}}-\alpha_{\mathrm{mid}}$ plane, which suggests that this subset is unbiased with respect to the degree (and sign) of spectral curvature. }
\label{fig:alphalow_alphamid_ATCA}
\end{figure}

For a subset of 177 G4Jy sources, we have ATCA flux-density measurements that allow us to study the radio spectrum from 72\,MHz to 20\,GHz (e.g. see Figs.~\ref{fig:firstorder_fits} to \ref{fig:thirdorder_fits}). We therefore re-fit polynomials (Equation~\ref{eq:poly}) to the data and provide updated coefficients in the accompanying G4Jy catalogue. Noting that the majority of the ATCA flux-densities are via the AT20G catalogue \citep{Murphy2010}, which is a blind-scan survey, we expect this subset to be biased towards sources that have their radio-jet axis close to the line-of-sight. As such, we may also expect the radio spectrum to exhibit a concave spectral shape, but this is not borne out (as illustrated by Fig.~\ref{fig:alphalow_alphamid_ATCA}). This is likely because hotspots, which are also traced by 20-GHz emission, do not have `directionality' connected to their detection. (Note that we have also checked the literature to determine whether the source in question is a blazar, with cross-identifications presented in Appendix~\ref{app:blazars}.)

\begin{figure*}
\centering
\subfigure[G4Jy 487]{
\includegraphics[scale=0.52]{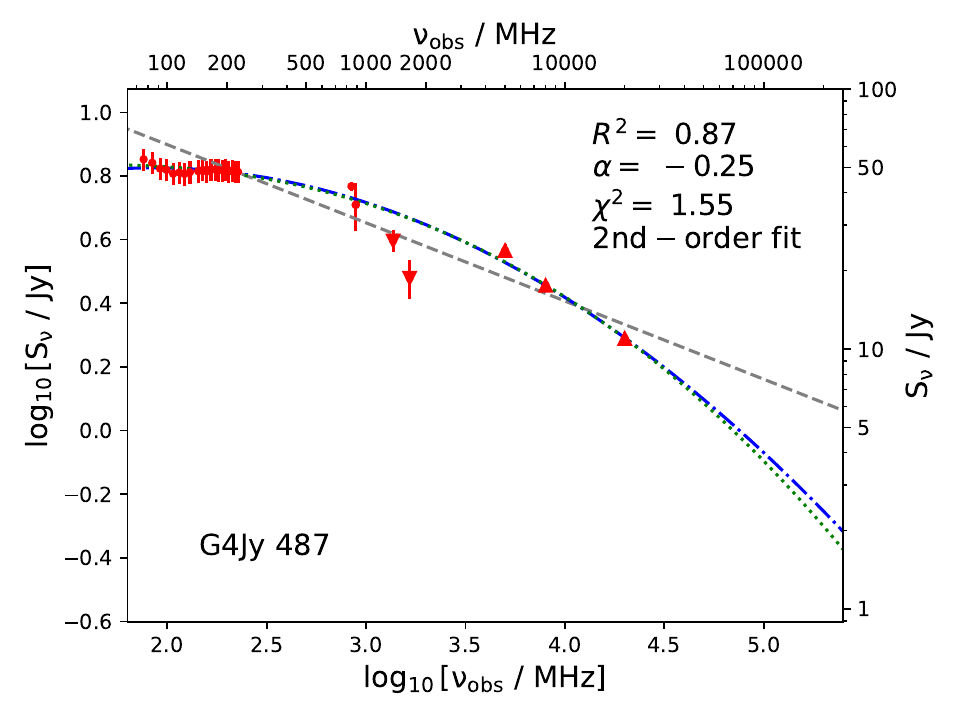} }
\subfigure[G4Jy 1260]{
\includegraphics[scale=0.52]{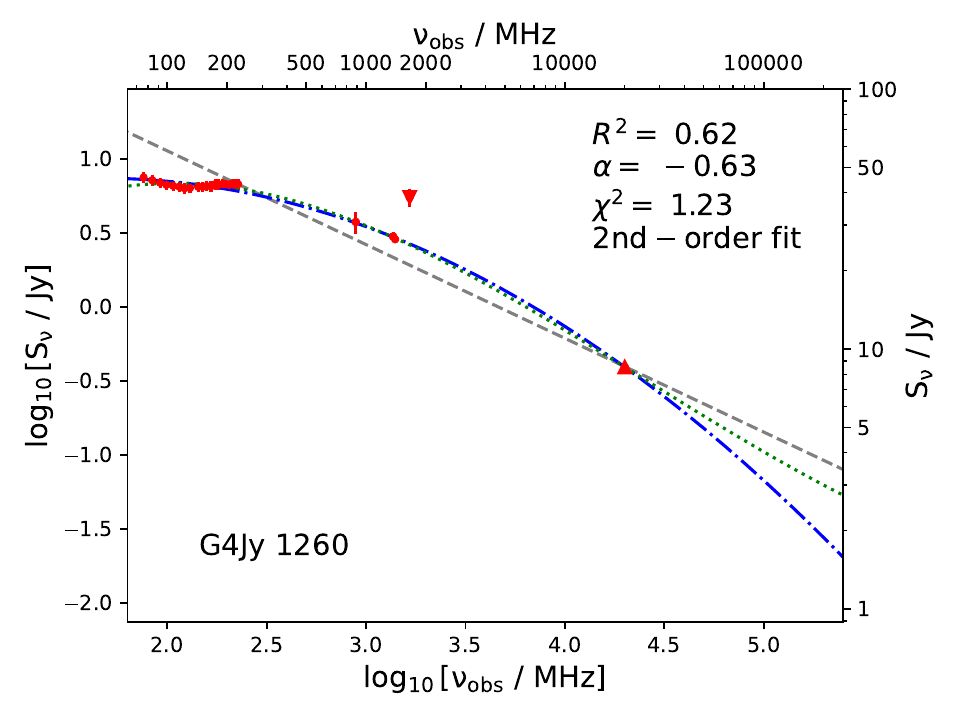} }
\subfigure[G4Jy 1264]{
\includegraphics[scale=0.52]{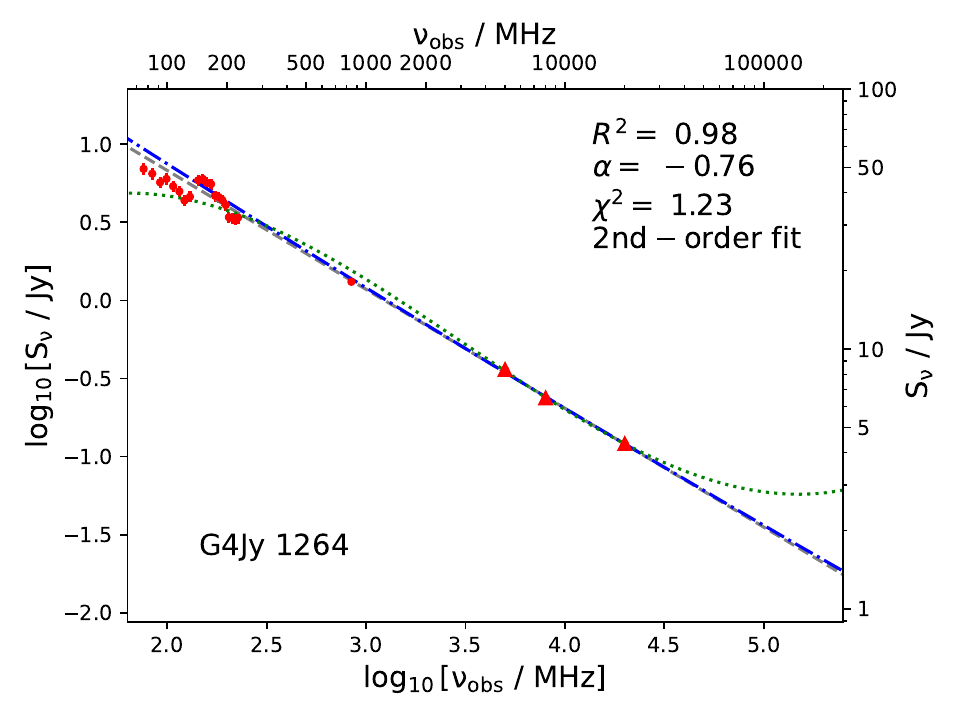} }
\subfigure[G4Jy 1286]{
\includegraphics[scale=0.52]{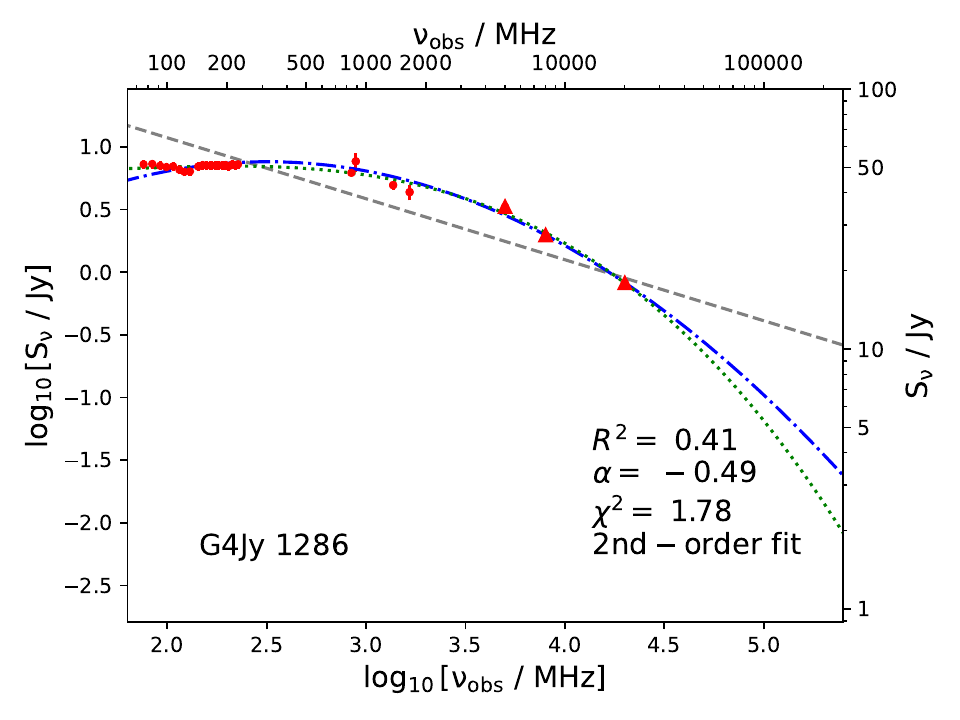} }
\caption{ Four G4Jy-ATCA sources that display a `kink' within the GLEAM region of the broadband radio spectrum. The grey dashed line represents a power-law fit, the blue dash-dotted line represents a second-order polynomial fit, and the green dotted line represents a third-order polynomial fit. Red dots represent G4Jy, SUMSS/NVSS and RACS flux-densities, whilst red triangles indicate that the ATCA flux-densities are lower limits. }
\label{fig:kinked_spectra}
\end{figure*}

As part of visually inspecting the broadband radio SEDs, we identify sources where the contiguous coverage provided by GLEAM reveals a `kink' in the radio spectrum. We show a few examples of these in Fig.~\ref{fig:kinked_spectra}. Whilst multiple epochs of AGN activity could result in these distinctive shapes, we caution that this could also be the result of unrelated radio sources having their radio emission blended by the 2-arcmin MWA beam. 

\section{Modelling the radio emission}
\label{sec:raise_models}

With broadband radio SEDs in place, we are now in prime position for their physical interpretation. For this we make use of RAiSE (Radio AGN in Semi-analytic Environments) models \citep[][]{Turner2015, Turner2018, Turner2023}, which provide estimates of the source age and jet power across the sample (Section~\ref{sec:raise_results}). Firstly, we introduce RAiSE (Section~\ref{sec:raise_overview}) and note the considerations that were made for the present work (Section~\ref{sec:raise_details}). 

\subsection{The RAiSE framework}
\label{sec:raise_overview}

The RAiSE model captures the dynamics of the jet, lobe formation, and active-to-relic expansion phases of the resulting radio lobe \citep{Turner2023}. The conservation equations for a relativistic fluid are used to derive an analytic description for the expansion of the jet, while the lobe evolution is similarly modelled with the inclusion of a non-trivial adiabatic equation of state. The RAiSE framework solves a system of ordinary differential equations to self-consistently transition the dynamics from the analytic solutions describing the jet and lobe expansion.
The properties of the host group/cluster environment, into which the jet expands, are described by a density profile informed by X-ray observations of nearby relaxed clusters \citep{Vikhlinin2006}. The core density and virial radius (overdensity of 200) of these profiles are scaled for a given halo mass to provide a range of plausible host environments; we assume the mean density profile and temperature for each halo mass following \citet{Turner2015}. The semi-analytic approach of the RAiSE model predicts very comparable evolution in the size, volume and pressure of jets/lobes to hydrodynamic simulations \citep[][their fig. 3]{Turner2023}.

The radio emissivity of the RAiSE lobes at a given frequency, due to synchrotron-emission from shock-accelerated particles, is calculated based on their volume and the time-evolution of their magnetic field; related to the thermal pressure by some factor $q$ of the equipartition value \citep{Turner2015, KDA1997}. This analytic calculation includes both adiabatic expansion and radiative loss process (i.e., inverse-Compton and synchrotron). \citet{Turner2018} considered the spatial distribution of the time of last shock-acceleration for `packets' of synchrotron-emitting electrons, with \citet{Turner2023} including local and large-scale inhomogeneities in the magnetic field. This yields surface brightness images consistent with those of hydrodynamic simulations \citep[e.g.][]{Stewart2025}, and consequently, very comparable estimates for the integrated flux density \citet[][their fig. 11]{Turner2023}.  

The analytic underpinning of the RAiSE framework, and comparable analytic models \citep[for review, see][]{Turner2023rev}, is critical to perform a Bayesian inversion from observations to intrinsic parameters. \citet{Turner2015} employ a grid-based maximum likelihood algorithm to transform the $P$--$D$ diagram to the jet power--source age parameter space for each host environment and redshift. \citet{Turner2018b} constrain the lobe magnetic field, in addition the jet power and source age, using spectral curvature and propose prior probability density functions for sources with more limited spectral coverage; \citet{Quici2022} introduce further constraints based on the spatial surface brightness distribution to measure the `on' and `off' times of relic radio galaxies. The \citet{Turner2026a} version of RAiSE is used in this work, taking advantage of its greatly improved computational efficiency when applied to simpler parameter inversions. This permits a fine sampling of the jet power and source age in the hyper-dimensional parameter space.

\subsection{Considerations for the present work}
\label{sec:raise_details}

We perform a Bayesian inversion following the approach of \citet{Turner2018b} based on the RACS-mid flux density, angular size of the source, and the best redshift estimate for the identified host galaxy. The 1367.5~MHz observer-frame frequency of the RACS-mid images largely excludes the low-frequency spectral turnover of compact sources \citep[e.g.,][]{Young2025}, while simultaneously probing the `young' synchrotron population that has experienced minimal radiative losses; although we note that such radiative losses are explicitly modelled in RAiSE. We include the $\alpha_{\mathrm{mid}}$ spectral-index and SCI as additional constraints in the inversion for a subset of 405 objects with $\alpha_{\mathrm{mid}}< 0.0$ and SCI $>$ 0.0; this subset excludes peaked-spectrum sources not captured by the \citet{Turner2026a} version of RAiSE. We consider an additional systematic uncertainty in the spectral index due to the unknown value of the freshly shock-accelerated population; we assume $\Delta \alpha_{\mathrm{mid}} = 0.117$ based on the spectral fits by \citet{Turner2018b}. The properties of the host group/cluster environments are modelled for three plausible halo masses: $10^{12.5}$, $10^{13.5}$ and $10^{14.5}$~M$_\odot$. These values are in agreement with halo masses derived in previous work \citep[e.g.][]{Wake2008,Mandelbaum2009}. The high central density ($\sim\rho = 10^{21}$\,kg\,m$^{-3}$) of the $10^{14.5}$~M$_\odot$ halo provides a rough approximation for the interaction of compact sources with a denser galaxy environment.

We generate a dense grid of synthetic RAiSE sources for the following set of parameters: jet power ($Q = 10^{36}$ to $10^{42}$~W with a 0.1~dex step size; both jets), source age ($t = 0.1$ to $1000$~Myr with a 0.1~dex step size), and the redshift dependence of the radiative losses and cosmological scaling of the halo virial radius ($z = 0.1$ to 1 with a 0.1 step size; the exact redshift is used elsewhere). The remaining model parameters adopt the default values of \citet{Turner2023}, including an electron–proton plasma in the lobe, an equipartition factor of $q = 10^{-1.5\pm0.5}$, an injection spectral index of $\alpha = -0.7$, and a lobe axis ratio of $A = 2.83$. The Bayesian inversion is performed for each source (converting the physical size and luminosity from RAiSE to angular size and flux density based on the host redshift) with the median and 16/84th percentiles of the resulting marginal posterior probability density functions recorded. This yields three estimates for the jet power and source age of each object in our sample; one corresponding to each plausible halo mass.

\subsection{Results of RAiSE modelling}
\label{sec:raise_results}

\begin{figure*}
\centering
\subfigure[]{
\includegraphics[scale=0.85]{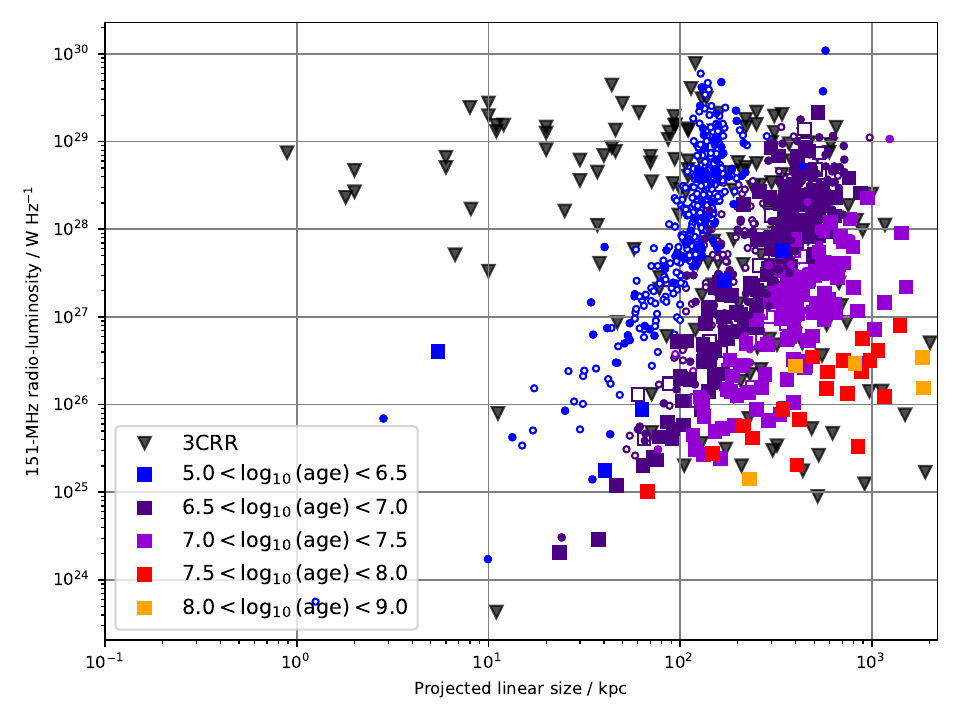} }
\subfigure[]{
\includegraphics[scale=0.85]{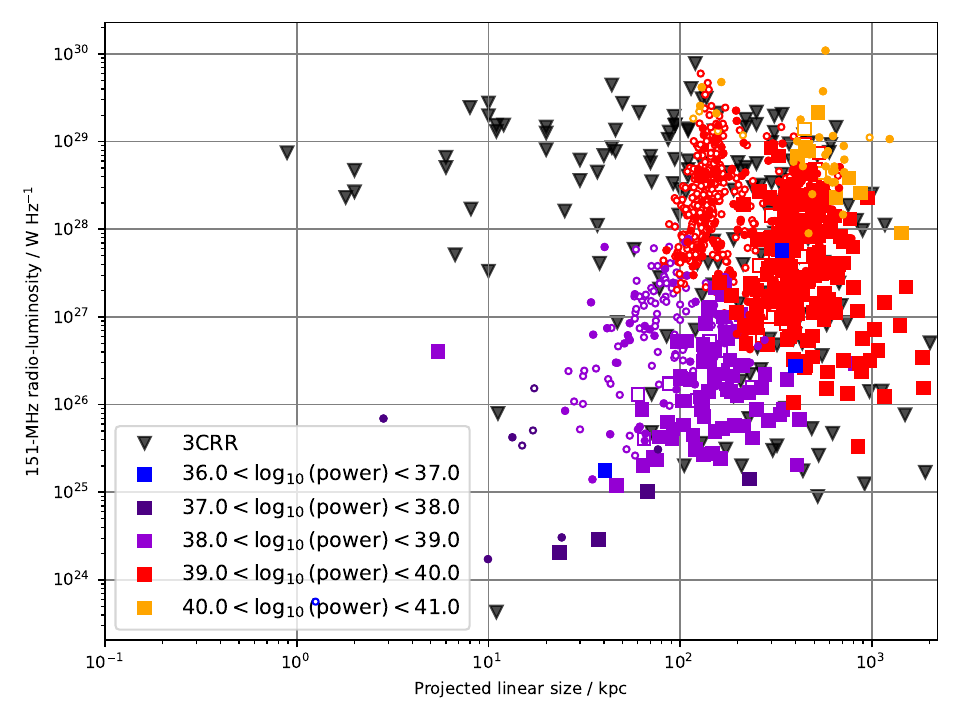} } 
\caption{ The $P$--$D$ diagram plotted as a function of (a) source age and (b) jet power, assuming a halo mass of $10^{12.5}$\,$\Msol$ (Section~\ref{sec:raise_results}). Unfilled symbols represent sources with upper limits in their linear size (as a consequence of their angular sizes being upper limits). Datapoints for the 3CRR sample \citep{Laing1983} are plotted for comparison, whilst `single' (circular datapoints) and `double' (square datapoints) refer to the radio-morphology classification completed by \citet{White2020a,White2020b}. }
\label{fig:PD_diagram_1}
\end{figure*}

As discussed in the literature, the $P$--$D$ diagram \citep{Baldwin1982} allows the evolutionary stage of radio galaxies to be appreciated as a function of radio-power ($P$) and linear size ($D$). Theory dictates that sources traverse this parameter space from left to right, corresponding to a radio galaxy showing (at first) increasing radio luminosity with source size. After time, the radio galaxy then fades due to synchrotron radiation and adiabatic expansion, resulting in these sources appearing in the lower right-hand region of the $P$--$D$ diagram. This is exactly what we see for the $P$--$D$ diagram plotted as a function of source age (Fig.~\ref{fig:PD_diagram_1}a) -- the first time that this has been presented for such a large sample of AGN. 

Accompanying this is the $P$--$D$ diagram as a function of intrinsic jet-power (Fig.~\ref{fig:PD_diagram_1}b), noting the formal difference between radio luminosity and radio power\footnote{To avoid ambiguity, it may be better to refer to the $P$--$D$ diagram as the {\it luminosity}--size, $L$--$D$, diagram.}. The former is an intrinsic property of the radio galaxies calculated directly from the measured flux-density (in this case, at 151\,MHz), whilst the latter is inferred via modelling of the radio galaxy's interaction with the ambient medium. To simplify the analysis, our modelling assumes a single halo mass ($10^{12.5}$, $10^{13.5}$, or $10^{14.5}$\,$\Msol$) is applicable for the whole sample (Section~\ref{sec:raise_details}). In addition, we only consider the modelling of sources that have `single' or `double' radio morphologies, for which the RAiSE results are most applicable \citep[as opposed to `triple' and `complex' radio morphologies;][]{White2020a,White2020b}.

\begin{figure*}
\centering
\subfigure[]{
\includegraphics[scale=0.85]{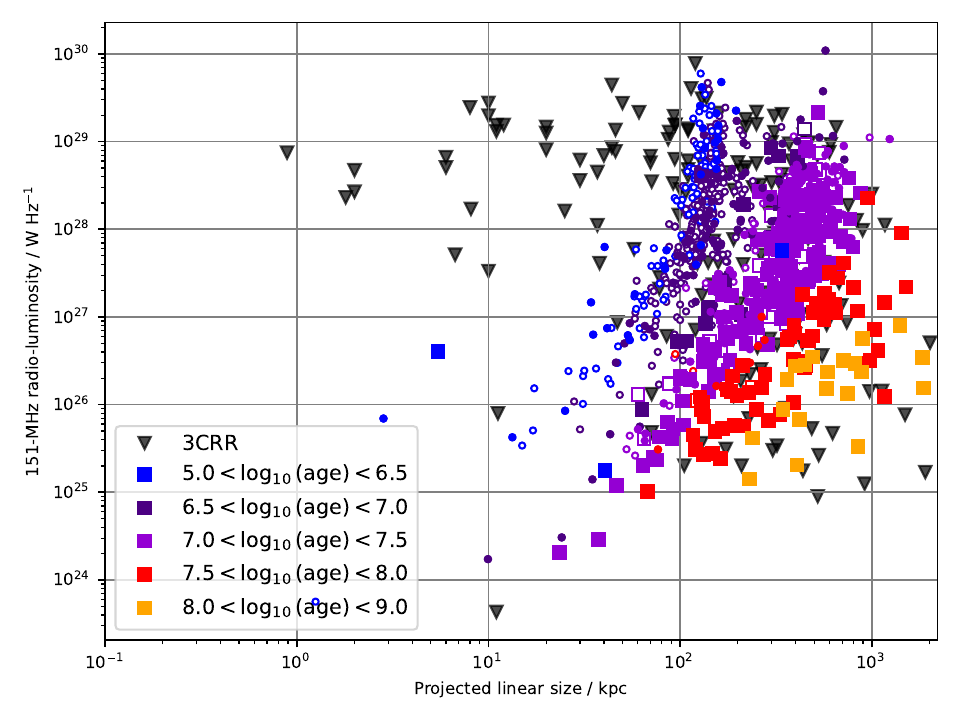} }
\subfigure[]{
\includegraphics[scale=0.85]{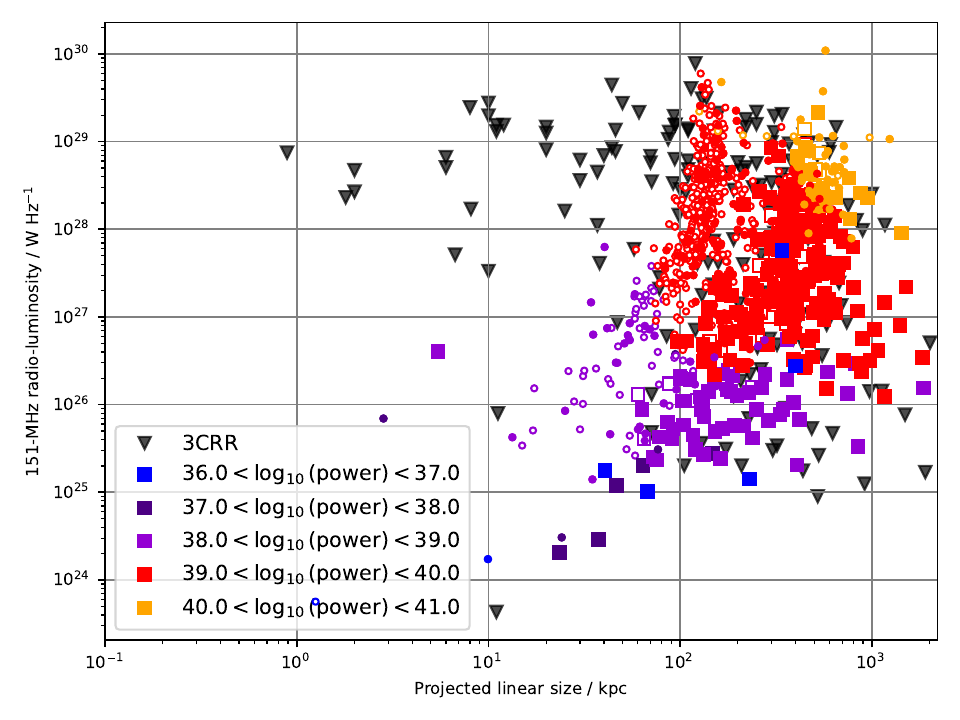} } 
\caption{ The $P$--$D$ diagram plotted as a function of (a) source age and (b) jet power, assuming a halo mass of $10^{13.5}$\,$\Msol$ (Section~\ref{sec:raise_results}). Unfilled symbols represent sources with upper limits in their linear size (as a consequence of their angular sizes being upper limits). Datapoints for the 3CRR sample \citep{Laing1983} are plotted for comparison, whilst `single' (circular datapoints) and `double' (square datapoints) refer to the radio-morphology classification completed by \citet{White2020a,White2020b}. }
\label{fig:PD_diagram_2}
\end{figure*}

\begin{figure*}
\centering
\subfigure[]{
\includegraphics[scale=0.85]{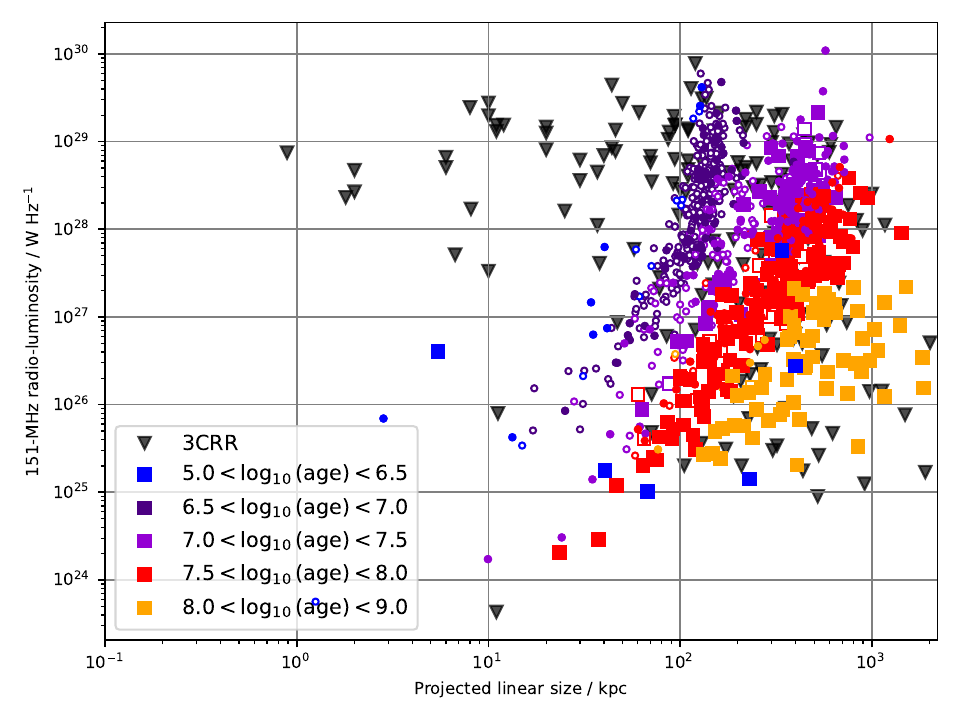} }
\subfigure[]{
\includegraphics[scale=0.85]{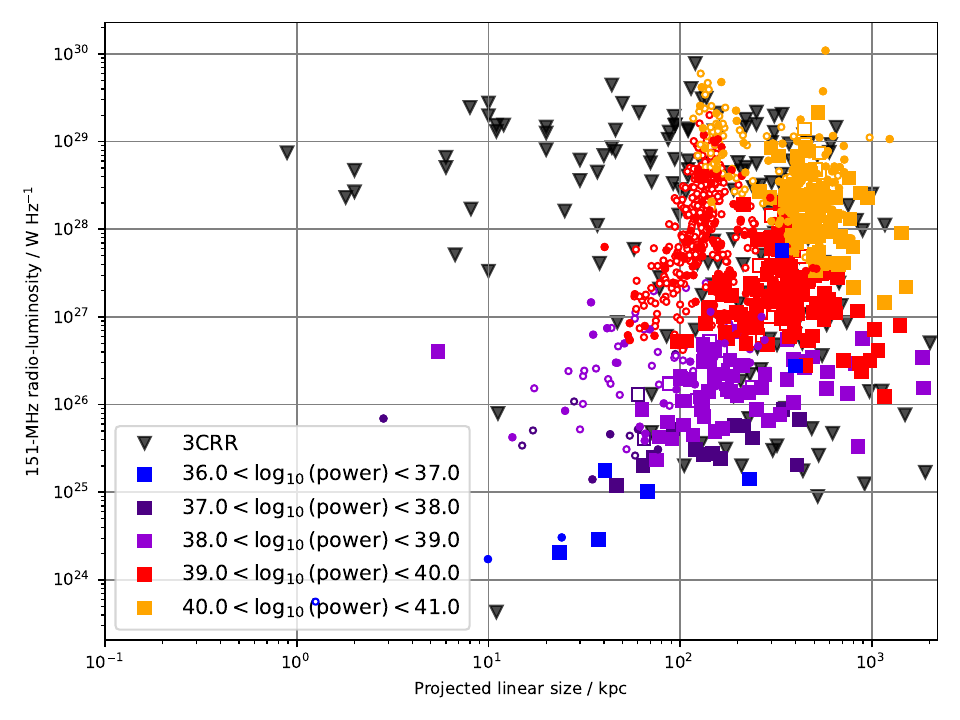} } 
\caption{ The $P$--$D$ diagram plotted as a function of (a) source age and (b) jet power, assuming a halo mass of $10^{14.5}$\,$\Msol$ (Section~\ref{sec:raise_results}). Unfilled symbols represent sources with upper limits in their linear size (as a consequence of their angular sizes being upper limits). Datapoints for the 3CRR sample \citep{Laing1983} are plotted for comparison, whilst `single' (circular datapoints) and `double' (square datapoints) refer to the radio-morphology classification completed by \citet{White2020a,White2020b}. }
\label{fig:PD_diagram_3}
\end{figure*}

Through comparing the $P$--$D$ diagrams for different halo masses (Figs.~\ref{fig:PD_diagram_1}--\ref{fig:PD_diagram_3}), we see a general trend of increasing source age and jet power with increasing halo mass. This can be explained as follows: with higher halo mass, the virial mass of the galaxy is larger \citep[e.g.][]{Croton2006}, resulting in higher densities of gas/material in the core. The impact that this has is to slow the expansion of the radio jets/lobes, necessitating higher jet-powers and longer source-ages in order to reach the same size. However, the relationship is complicated by a competing effect: higher densities also lead to higher pressures and magnetic fields in the lobes, and consequently higher radio luminosities, thus restricting the increase in jet power with halo mass. 

\begin{figure}
\centering
\includegraphics[scale=0.52]{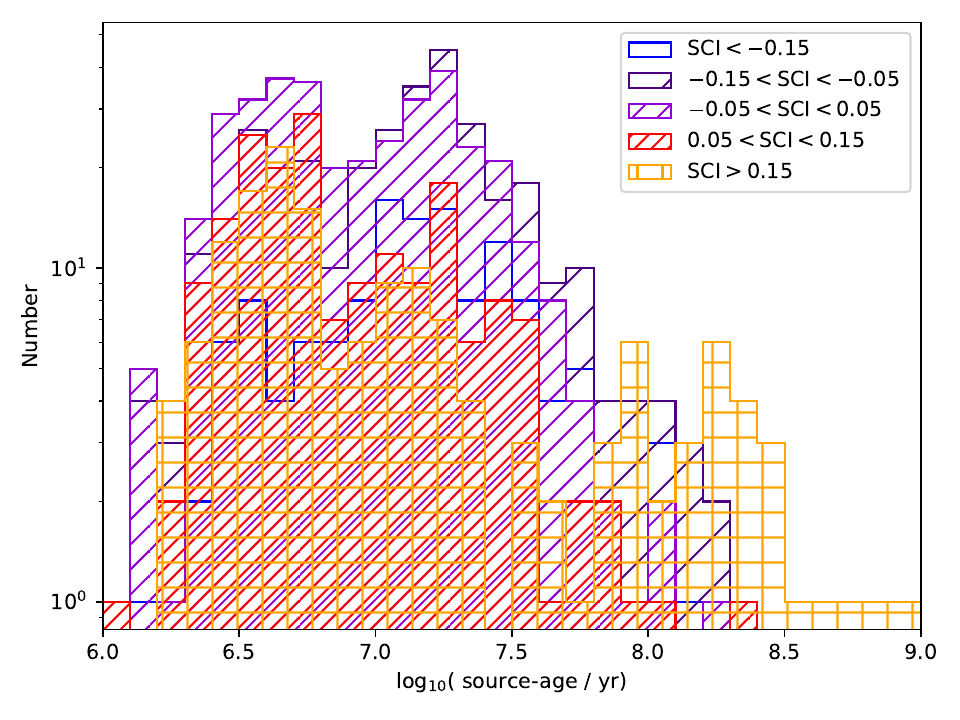} 
\caption{ The distribution in spectral source age (Section~\ref{sec:raise_results}) as a function of spectral curvature, SCI = $\alpha_{\mathrm{low}} - \alpha_{\mathrm{mid}}$. For this figure we have assumed that the typical halo mass is $10^{13.5}\,\Msol$. }
\label{fig:source_age_by_SCI}
\end{figure}

To tie this work in with the earlier results regarding spectral curvature across the sample (Section~\ref{sec:alpha_mid}), we plot the distribution in source age as a function of SCI in Fig.~\ref{fig:source_age_by_SCI}. As expected, the oldest sources have the highest degree of spectral curvature, likely due to radiative losses through inverse-Compton and synchrotron processes. However, the SCI\,$>0.15$ sources do also span a wide range in source ages, indicating that we cannot assume that high spectral curvature corresponds to ageing radio-sources \citep[cf.][]{White2026b}. Nevertheless, this category gives us a firm basis for exploring the remnant radio-galaxy fraction in future work. Related to this, we note that \citet{Dutta2026} find that remnant radio galaxies (SCI\,$ \gtrsim $\,0.5) have spectral ages of $\sim$8--42\,Myr, signalling that the remnant phase may be more short-lived than originally assumed. 

\begin{figure*}
\centering
\subfigure[]{
\includegraphics[scale=0.52]{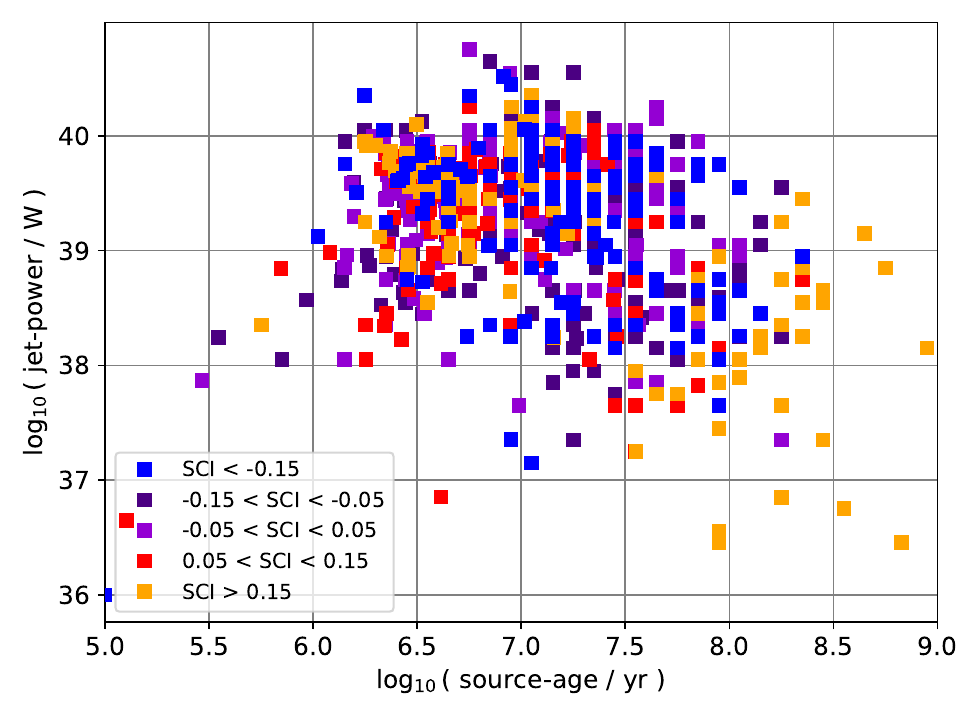} }
\subfigure[]{
\includegraphics[scale=0.52]{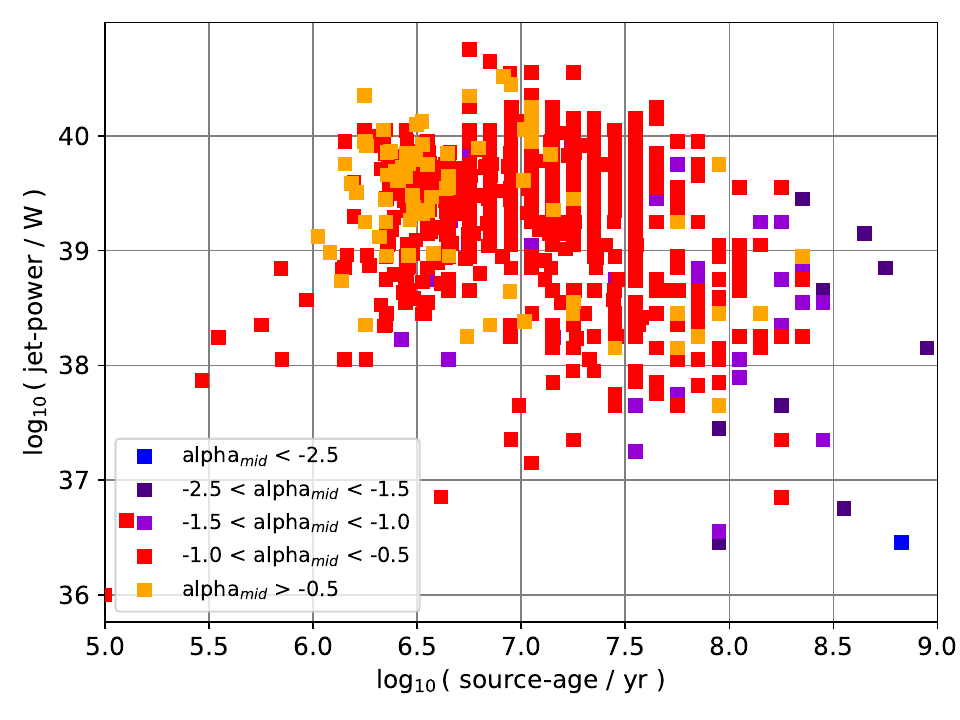} }
\subfigure[]{
\includegraphics[scale=0.52]{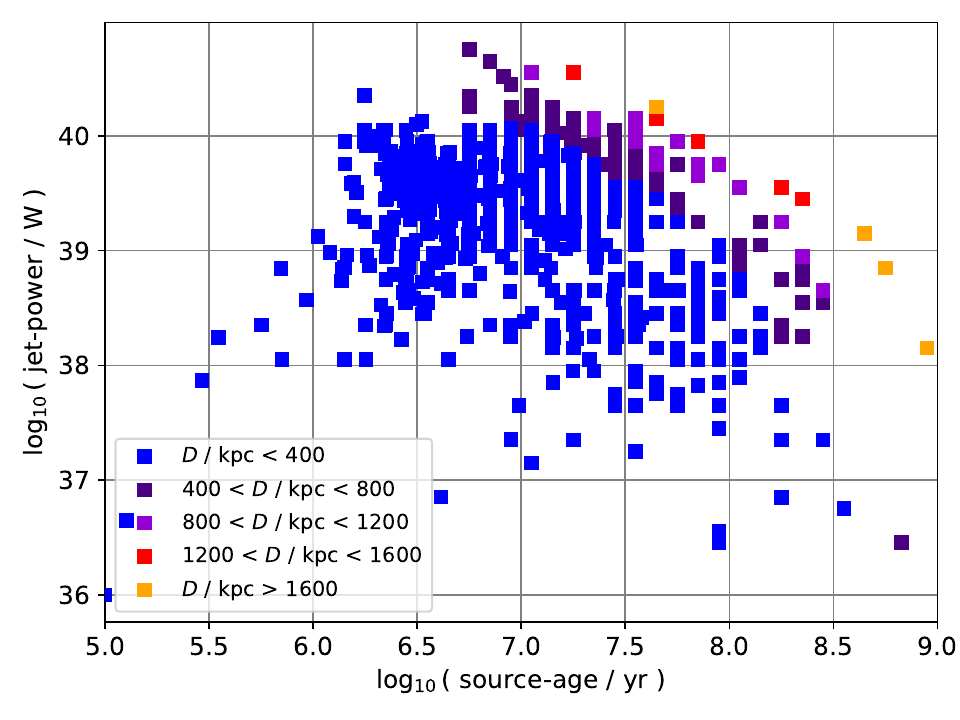} }
\subfigure[]{
\includegraphics[scale=0.52]{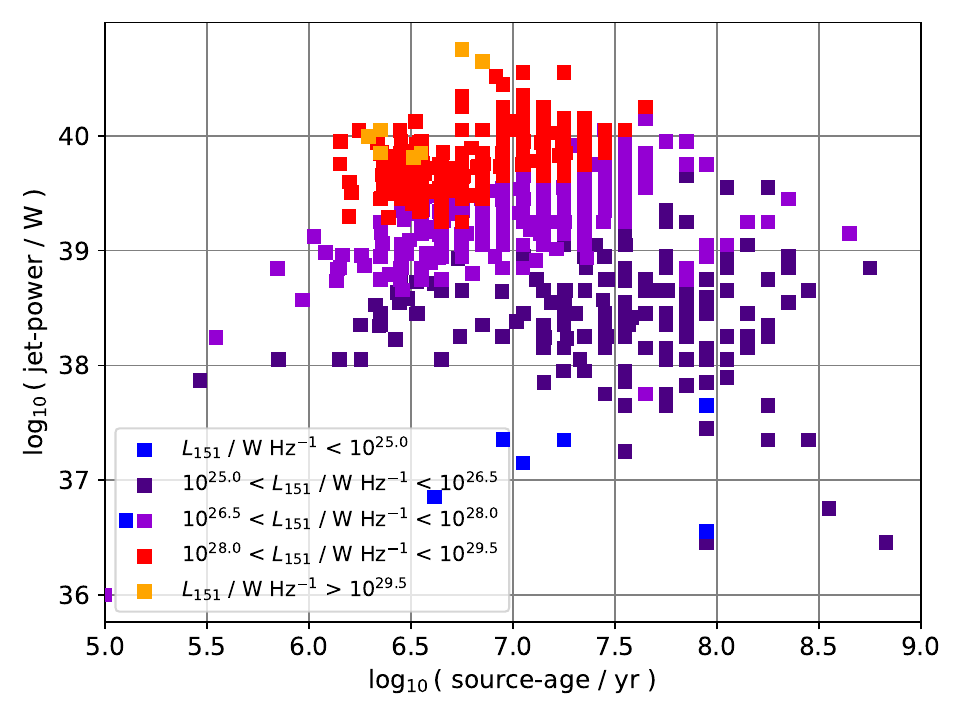} }
\caption{ The distribution in jet power versus source age (Section~\ref{sec:raise_results}) as a function of (a) spectral curvature index (SCI), (b) spectral index ($\alpha_{\mathrm{mid}}$), (c) linear size and (d) radio luminosity ($L_{\mathrm{151\,MHz}}$). These results are from assuming a halo mass of $10^{13.5}\Msol$. }
\label{fig:Qt_diagrams}
\end{figure*}

In Fig.~\ref{fig:Qt_diagrams} we explore the results of the RAiSE modelling in further detail. We start by considering the SCI distribution in radio-power($Q$)--source-age($t$) space, and note that there appear to be two sub-populations with SCI\,$>$\,0.15 (Fig.~\ref{fig:Qt_diagrams}a). Those with longer source ages are coupled with lower jet powers, which suggests that the power is not large enough to replenish the electrons that result in synchrotron radiation (leading to steepening of the spectrum). This is supported by the corresponding spectral index (Fig.~\ref{fig:Qt_diagrams}b) being steeper than the general population for this subset. The second population of SCI$>0.15$ sources have lower source ages coupled with higher jet powers, but these do not dominate the general population's higher jet powers. This could be a consequence of assuming a single halo mass (and therefore magnetic-field structure) across the sample, when in fact radio sources inhabit a wide range of environments. 

As expected, the largest sizes are achieved through a combination of high radio-jet power and long source age, with Fig.~\ref{fig:Qt_diagrams}c also demonstrating the relative rarity of giant radio-galaxies across the G4Jy Sample. We also present the $Q-t$ distribution as a function of radio luminosity (Fig.~\ref{fig:Qt_diagrams}d), and note the diagonal trend in the colouration of the datapoints. This can be explained by our earlier comment on the density of the material in the halo core: that is, the more confined the radio galaxy is, the higher the radio-power and/or source age that is needed in order to reach a given luminosity.

\section{Conclusions}
\label{sec:conclusions}

To better understand the interplay of radio galaxies and their environments, it is important to use a complete sample that is unbiased with respect to the orientation of the radio-jet axis. Hence our focus on the G4Jy Sample, which is a compilation of the radio-brightest sources in the southern sky, made in preparation for studies with the Square Kilometre Array (SKA). In the meantime, the sample is being used for commissioning of SKA-Low, and a subset of sources have been observed through MeerKAT Open Time \citep[][Clews et al., 2026; Mitha et al., in prep.]{Sejake2023}.   

In this work we have enhanced previous studies of the G4Jy Sample by considering the extended radio emission that is exhibited at each of the RACS observing frequencies (887.5, 1367.5 and 1655.5\,MHz). Correctly determining the associated mid-frequency radio-emission involves a significant amount of manual work, and we find that in 1.6 per cent of cases {\sc PyBDSF} fails to accurately characterise diffuse radio emission. We update the G4Jy catalogue to include the newly-determined RACS radio flux-densities and use these to construct broadband radio SEDs for the G4Jy Sample. These SEDs are extended to 20\,GHz for a subset of 177 G4Jy sources, thanks to further measurements from ATCA, and they demonstrate a wide variety of spectral behaviour. Convex spectra (SCI > 0) are indicative of spectral ageing or low-frequency absorption, whilst concave spectra (SCI < 0) are indicative of restarted AGN activity or blazars. The full set of flux densities and redshifts (where available) are used by RAiSE modelling to determine the source age and jet power for a given, assumed density profile of the surrounding halo. For a halo mass of 13.5\,\Msol, the source age ranges from 0.1 to 891\,Myr, whilst the jet power ranges from 10$^{36}$ to 10$^{41}$\,W. The radio-power--size ($P$--$D$) diagram shows the expected distribution when plotted as a function of source age -- the first time that this has been presented for such a large sample. Meanwhile, non-linearities in the jet-power distribution suggest that the environment is constricting the expansion of the radio jets/lobes for some sources. Future work will include investigating the impact of the environment on a source-by-source basis.      

\section*{Acknowledgements}

This work is based on the research supported in part by the National Research Foundation of South Africa (Grant Number 151060). We thank Stefan Duchesne for helpful discussion of the RACS data, and Alessandro Peca for advice regarding ingestion of the G4Jy catalogue into the Active Galactic Nuclei DataBase \citep[AGN-DB;][]{Peca2026}.

This scientific work uses data obtained from Inyarrimanha Ilgari Bundara, the CSIRO Murchison Radio-astronomy Observatory. We acknowledge the Wajarri Yamaji People as the Traditional Owners and native title holders of the Observatory site. CSIRO's ASKAP radio telescope is part of the Australia Telescope National Facility (\url{https://ror.org/05qajvd42}). Operation of ASKAP is funded by the Australian Government with support from the National Collaborative Research Infrastructure Strategy. ASKAP uses the resources of the Pawsey Supercomputing Research Centre. Establishment of ASKAP, Inyarrimanha Ilgari Bundara, the CSIRO Murchison Radio-astronomy Observatory and the Pawsey Supercomputing Research Centre are initiatives of the Australian Government, with support from the Government of Western Australia and the Science and Industry Endowment Fund.

The Australia Telescope Compact Array is part of the Australia Telescope National Facility (\url{https://ror.org/05qajvd42}) which is funded by the Australian Government for operation as a National Facility managed by CSIRO. We acknowledge the Gomeroi people as the Traditional Owners of the Observatory site.

\section*{Data Availability}

RACS contour plots, individual RACS component lists, broadband SEDs, and an updated G4Jy catalogue are all made available via the G4Jy repository on Zenodo: \url{https://zenodo.org/communities/g4jy/records}.



\bibliographystyle{mnras}
\bibliography{G4Jy_Paper_V} 




\appendix

\section{The most-extended radio galaxies}
\label{app:extended}

\begin{landscape}
\begin{figure}
\centering
\includegraphics[scale=0.16]{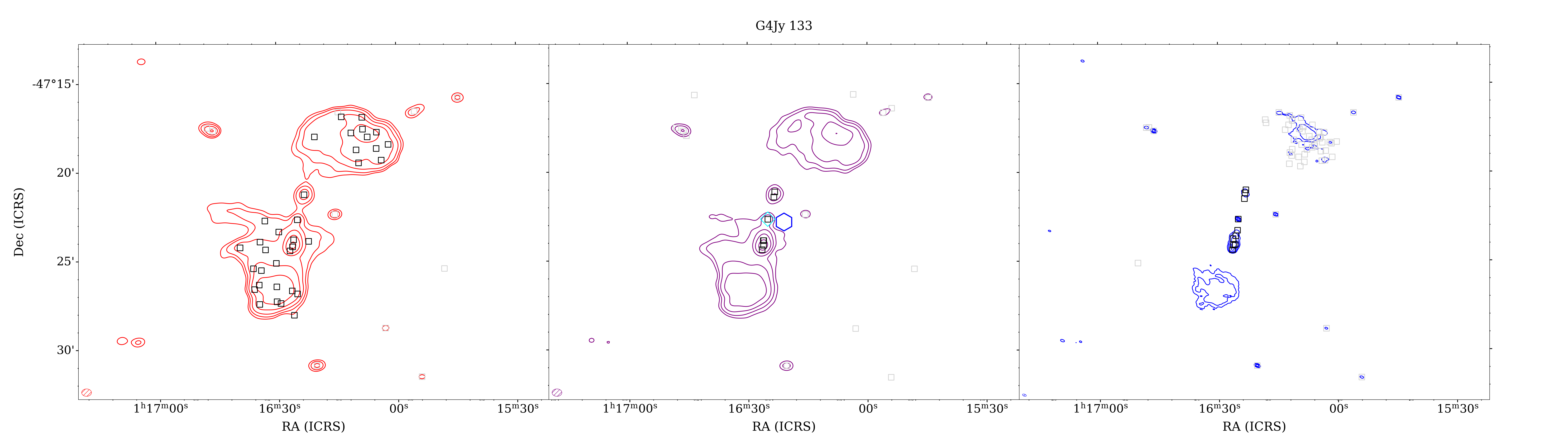} 
\includegraphics[scale=0.16]{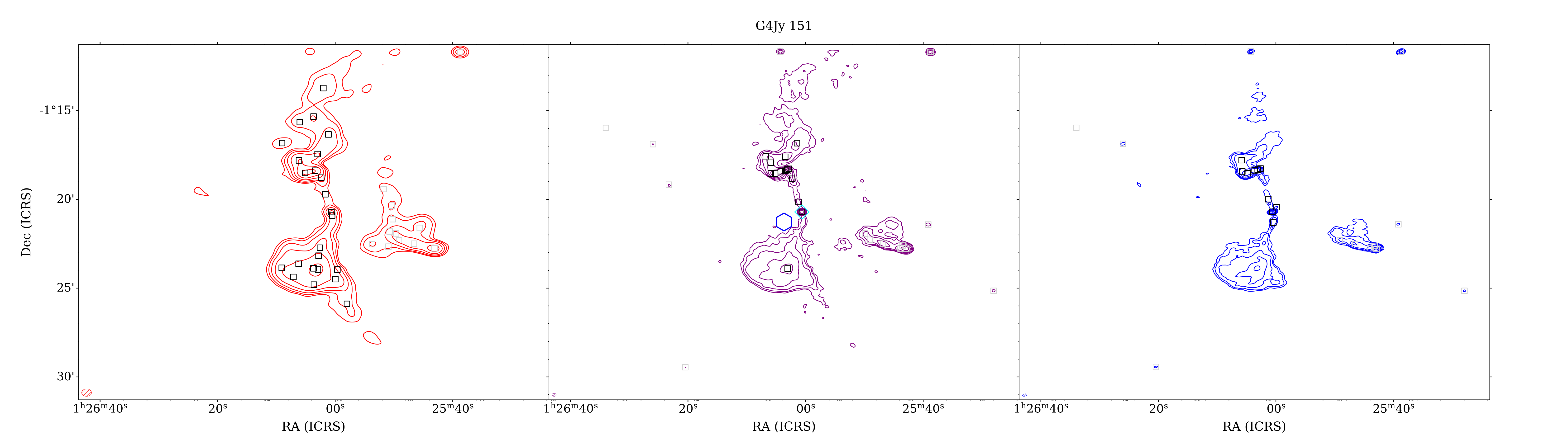} 
\caption{ Two examples of the RACS-low (red), RACS-mid (purple), and RACS-high (blue) contour plots that are inspected for the most-extended radio-galaxies (Appendix~\ref{app:extended}). The lowest contour is at 6 times the local rms ($\sigma$), and doubles with each subsequent contour. Meanwhile, the RACS beam is indicated in the lower left-hand corner of each sub-plot. Light-grey squares represent all RACS components that are within 30\,arcmin of the radio centroid position (blue hexagon), whilst black squares represent RACS components that are deemed to be associated with the G4Jy source. The host galaxy position is marked with a cyan diamond and determines the starting point for the flood-fill algorithm (Section~\ref{sec:floodfill}). }
\label{fig:extended_1}
\end{figure}
\end{landscape}

\begin{landscape}
\begin{figure}
\centering
\includegraphics[scale=0.16]{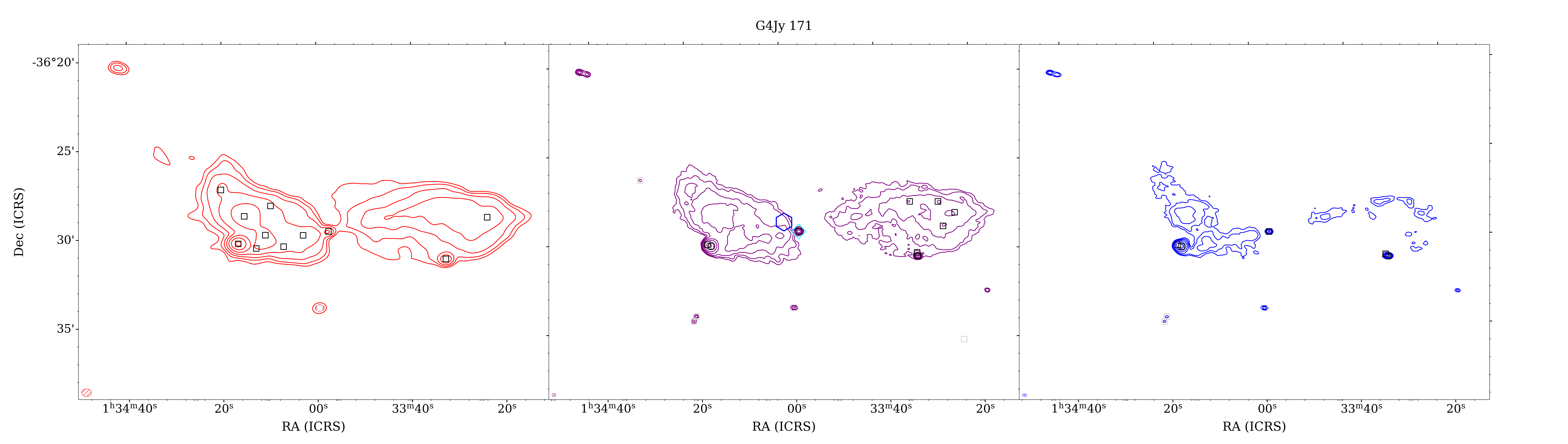} 
\includegraphics[scale=0.16]{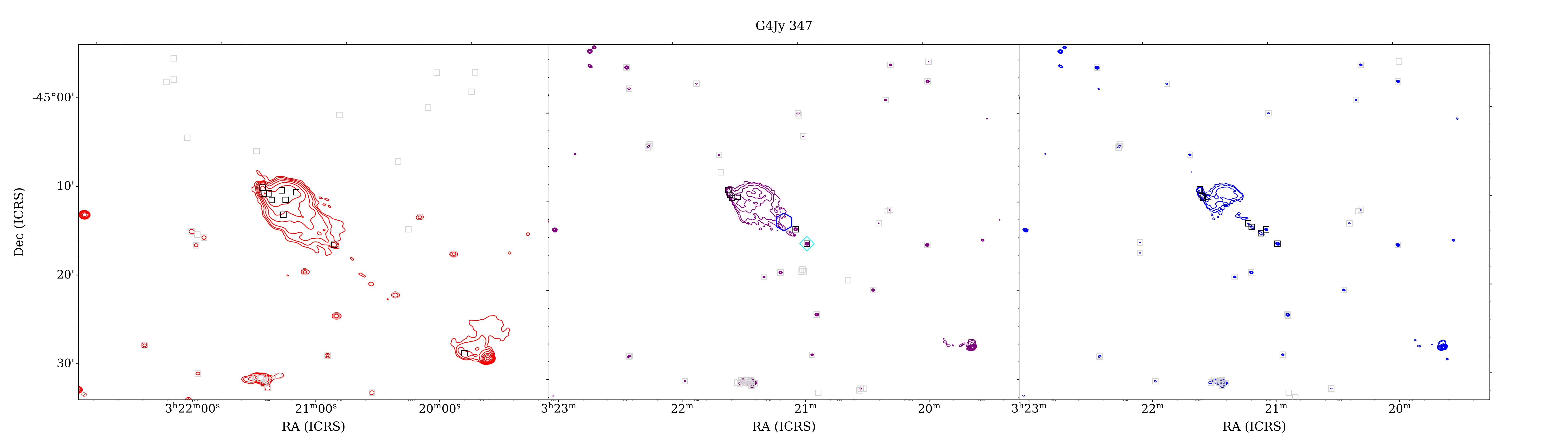} 
\caption{ Two examples of the RACS-low (red), RACS-mid (purple), and RACS-high (blue) contour plots that are inspected for the most-extended radio-galaxies (Appendix~\ref{app:extended}). The lowest contour is at 6 times the local rms ($\sigma$), and doubles with each subsequent contour. Meanwhile, the RACS beam is indicated in the lower left-hand corner of each sub-plot. Light-grey squares represent all RACS components that are within 30\,arcmin of the radio centroid position (blue hexagon), whilst black squares represent RACS components that are deemed to be associated with the G4Jy source. The host galaxy position is marked with a cyan diamond and determines the starting point for the flood-fill algorithm (Section~\ref{sec:floodfill}). }
\label{fig:extended_2}
\end{figure}
\end{landscape}

\begin{landscape}
\begin{figure}
\centering
\includegraphics[scale=0.16]{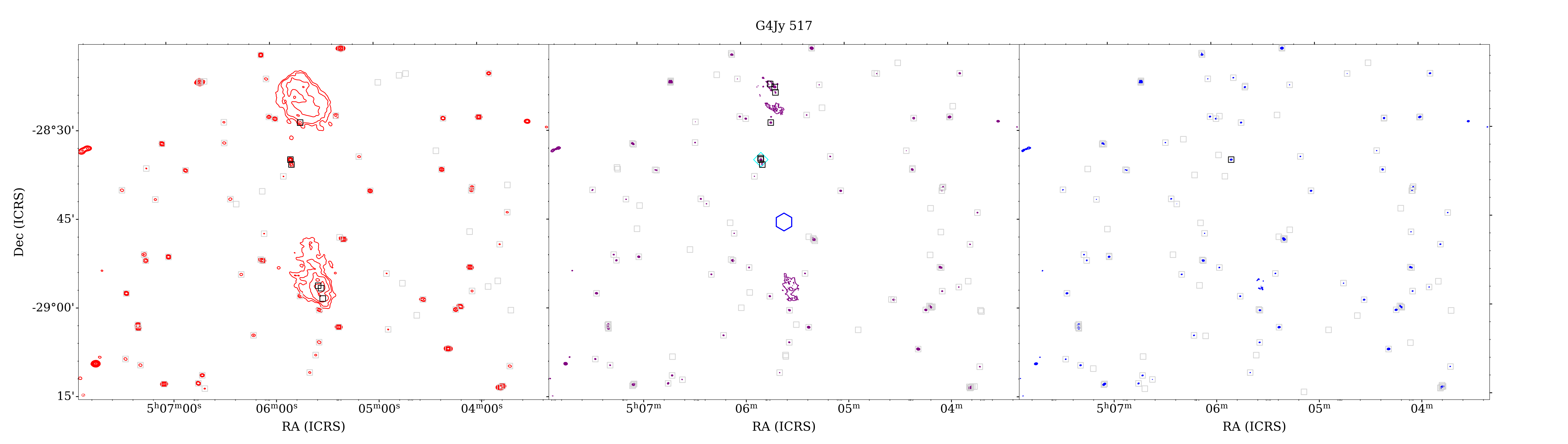} 
\includegraphics[scale=0.16]{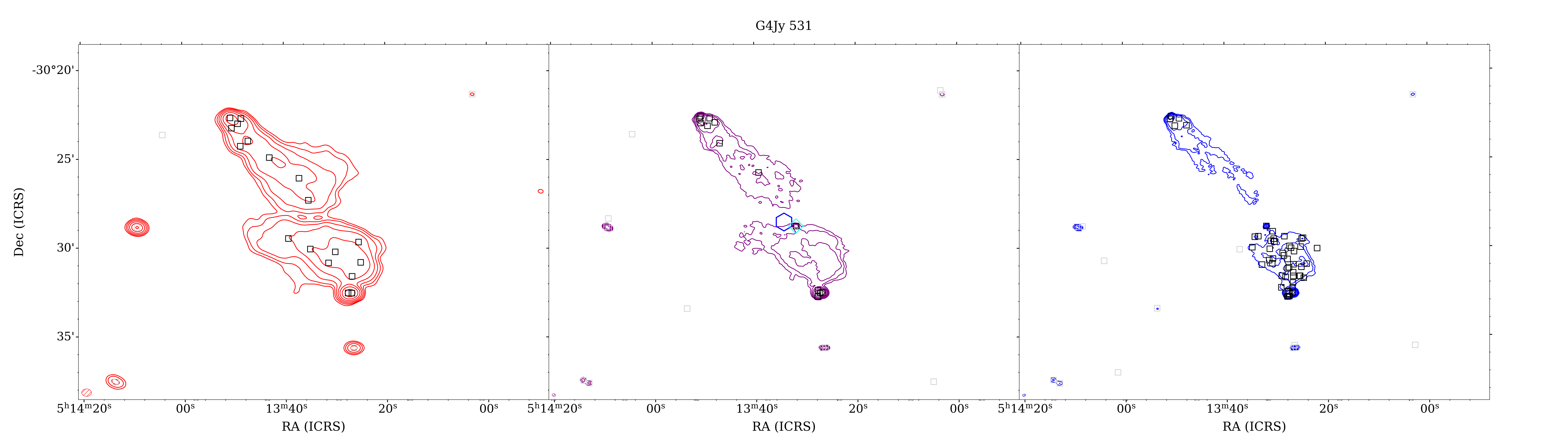} 
\caption{ Two examples of the RACS-low (red), RACS-mid (purple), and RACS-high (blue) contour plots that are inspected for the most-extended radio-galaxies (Appendix~\ref{app:extended}). The lowest contour is at 6 times the local rms ($\sigma$), and doubles with each subsequent contour. Meanwhile, the RACS beam is indicated in the lower left-hand corner of each sub-plot. Light-grey squares represent all RACS components that are within 30\,arcmin of the radio centroid position (blue hexagon), whilst black squares represent RACS components that are deemed to be associated with the G4Jy source. The host galaxy position is marked with a cyan diamond and determines the starting point for the flood-fill algorithm (Section~\ref{sec:floodfill}). }
\label{fig:extended_3}
\end{figure}
\end{landscape}

\begin{landscape}
\begin{figure}
\centering
\includegraphics[scale=0.16]{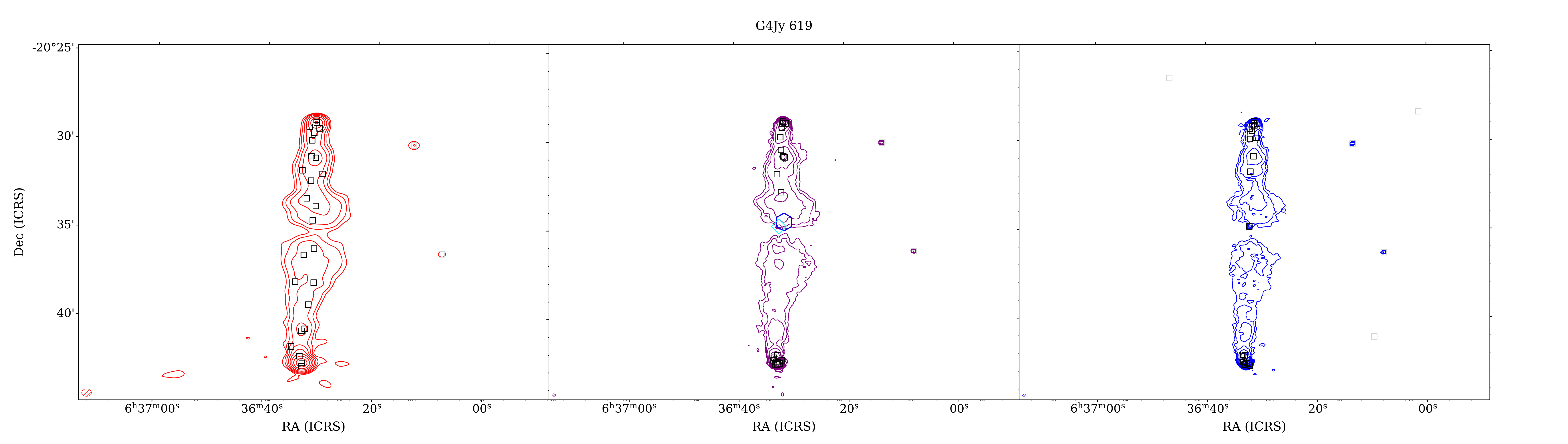} 
\includegraphics[scale=0.16]{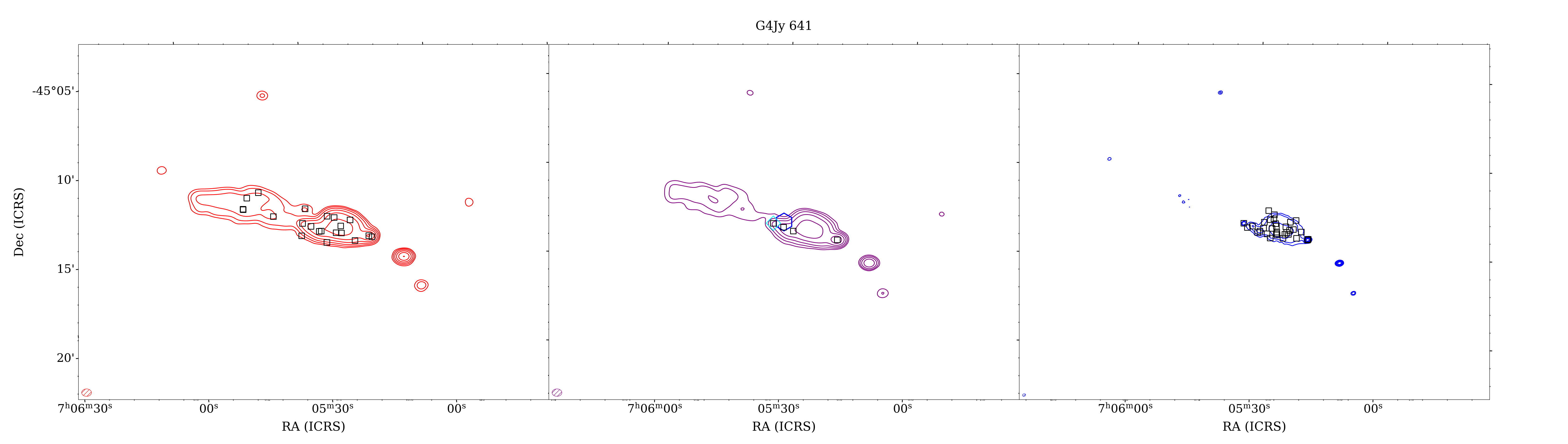} 
\caption{ Two examples of the RACS-low (red), RACS-mid (purple), and RACS-high (blue) contour plots that are inspected for the most-extended radio-galaxies (Appendix~\ref{app:extended}). The lowest contour is at 6 times the local rms ($\sigma$), and doubles with each subsequent contour. Meanwhile, the RACS beam is indicated in the lower left-hand corner of each sub-plot. Light-grey squares represent all RACS components that are within 30\,arcmin of the radio centroid position (blue hexagon), whilst black squares represent RACS components that are deemed to be associated with the G4Jy source. The host galaxy position is marked with a cyan diamond and determines the starting point for the flood-fill algorithm (Section~\ref{sec:floodfill}). }
\label{fig:extended_4}
\end{figure}
\end{landscape}

\begin{landscape}
\begin{figure}
\centering
\includegraphics[scale=0.16]{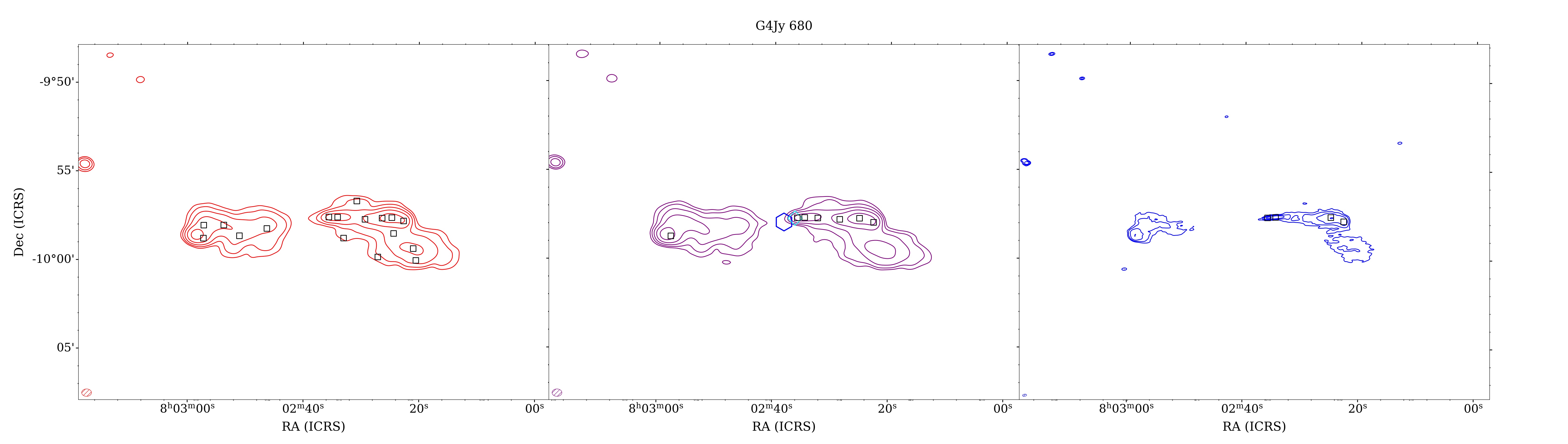} 
\includegraphics[scale=0.16]{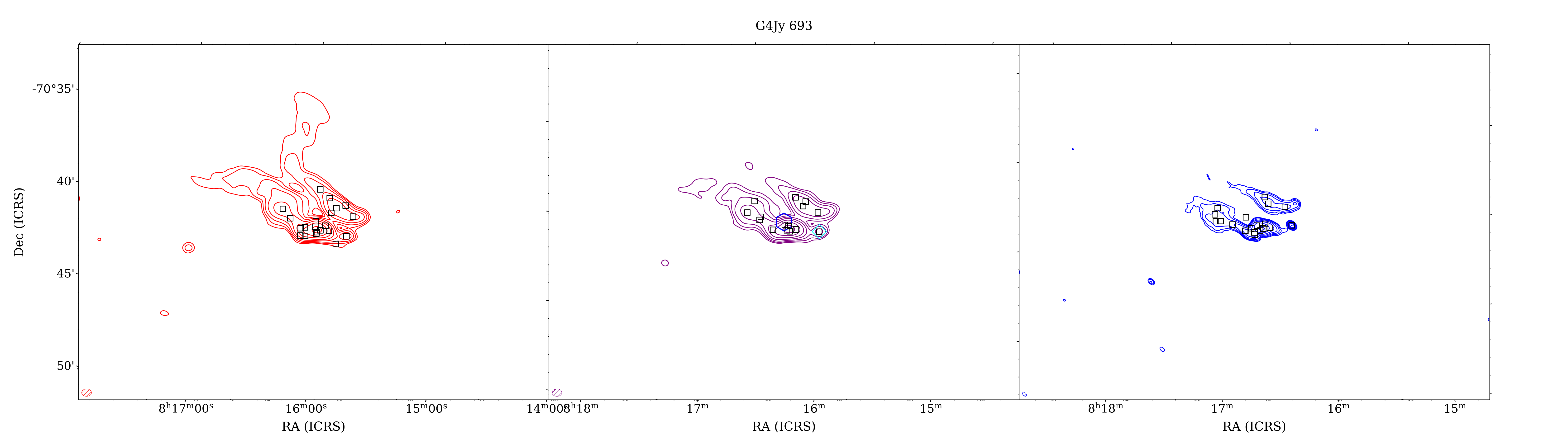} 
\caption{ Two examples of the RACS-low (red), RACS-mid (purple), and RACS-high (blue) contour plots that are inspected for the most-extended radio-galaxies (Appendix~\ref{app:extended}). The lowest contour is at 6 times the local rms ($\sigma$), and doubles with each subsequent contour. Meanwhile, the RACS beam is indicated in the lower left-hand corner of each sub-plot. Light-grey squares represent all RACS components that are within 30\,arcmin of the radio centroid position (blue hexagon), whilst black squares represent RACS components that are deemed to be associated with the G4Jy source. The host galaxy position is marked with a cyan diamond and determines the starting point for the flood-fill algorithm (Section~\ref{sec:floodfill}). }
\label{fig:extended_5}
\end{figure}
\end{landscape}

\begin{landscape}
\begin{figure}
\centering
\includegraphics[scale=0.16]{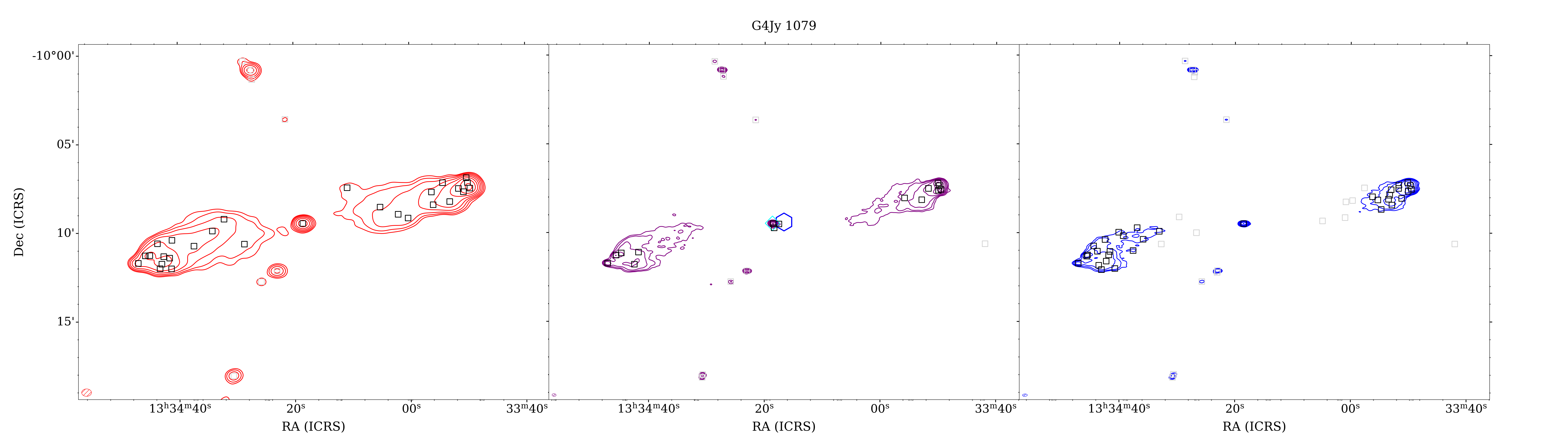} 
\includegraphics[scale=0.16]{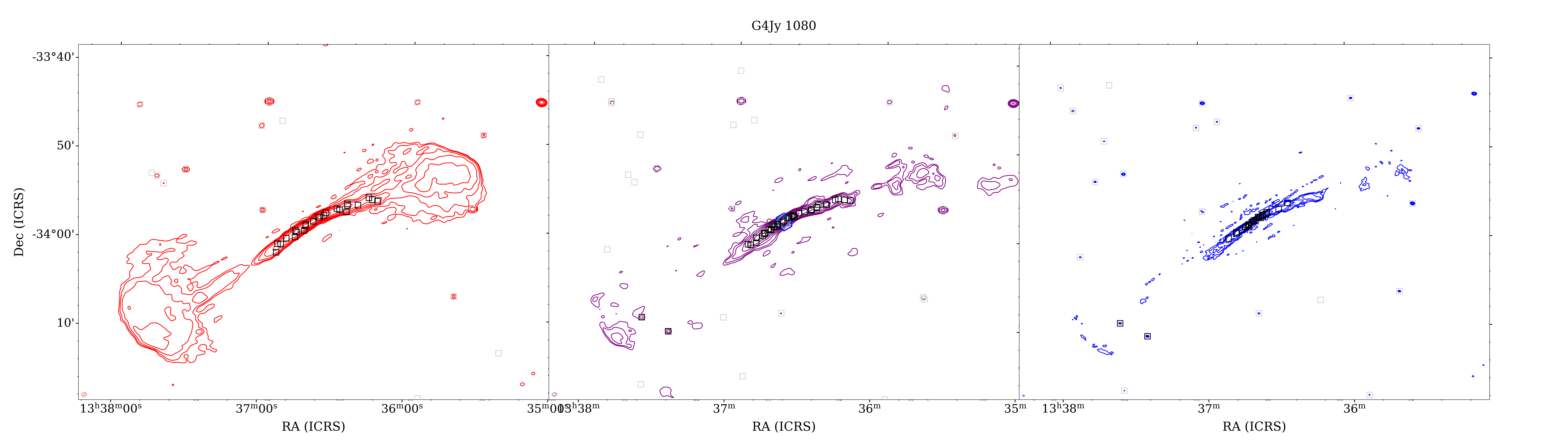} 
\caption{ Two examples of the RACS-low (red), RACS-mid (purple), and RACS-high (blue) contour plots that are inspected for the most-extended radio-galaxies (Appendix~\ref{app:extended}). The lowest contour is at 6 times the local rms ($\sigma$), and doubles with each subsequent contour. Meanwhile, the RACS beam is indicated in the lower left-hand corner of each sub-plot. Light-grey squares represent all RACS components that are within 30\,arcmin of the radio centroid position (blue hexagon), whilst black squares represent RACS components that are deemed to be associated with the G4Jy source. The host galaxy position is marked with a cyan diamond and determines the starting point for the flood-fill algorithm (Section~\ref{sec:floodfill}). }
\label{fig:extended_6}
\end{figure}
\end{landscape}

\begin{landscape}
\begin{figure}
\centering
\includegraphics[scale=0.16]{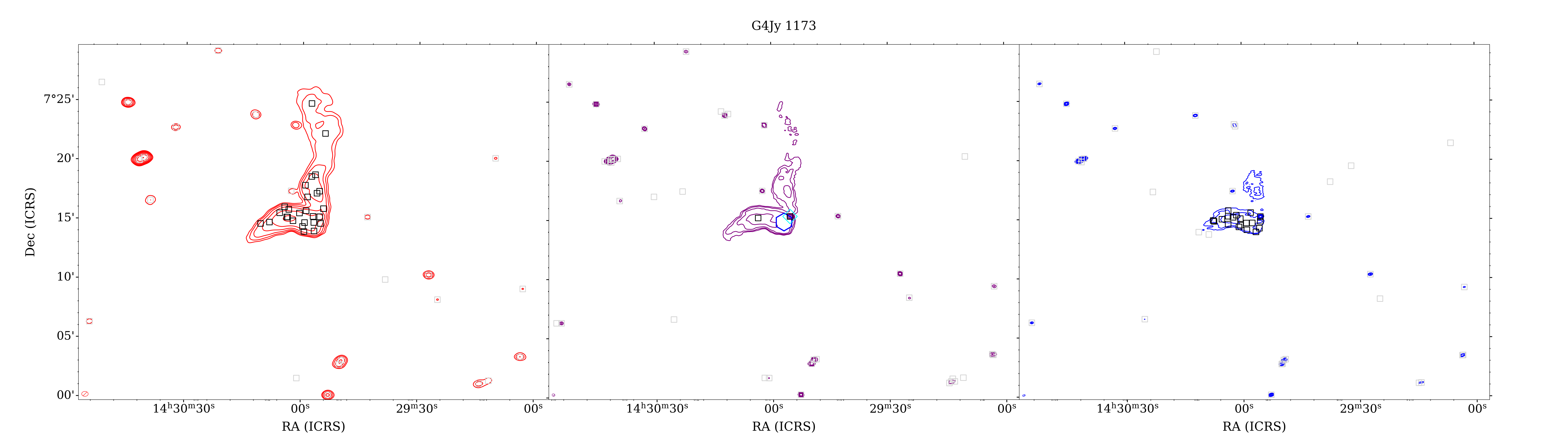} 
\includegraphics[scale=0.16]{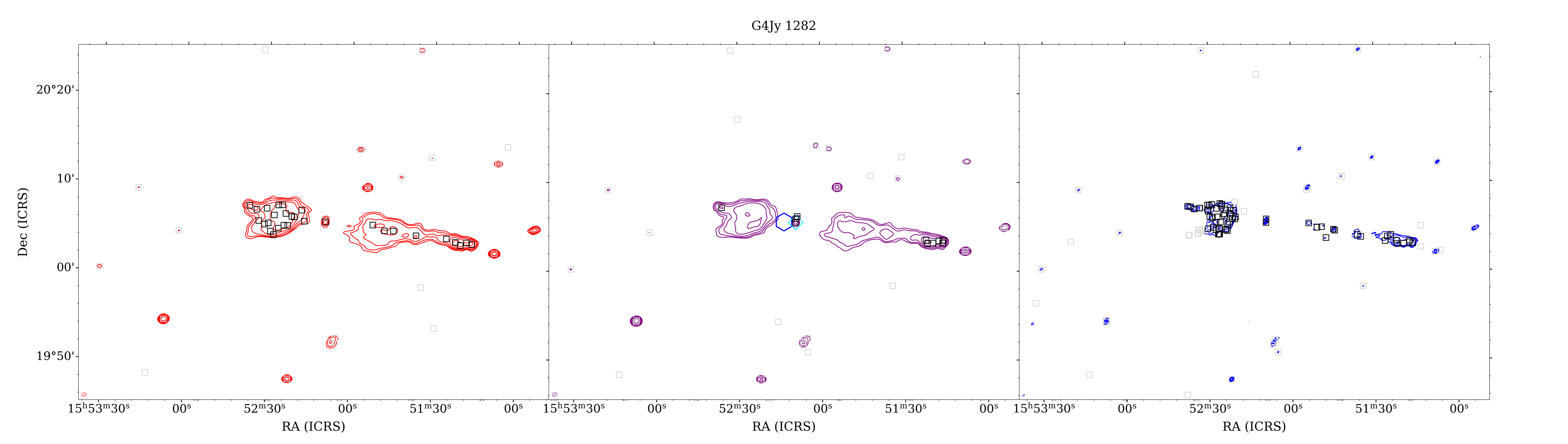} 
\caption{ Two examples of the RACS-low (red), RACS-mid (purple), and RACS-high (blue) contour plots that are inspected for the most-extended radio-galaxies (Appendix~\ref{app:extended}). The lowest contour is at 6 times the local rms ($\sigma$), and doubles with each subsequent contour. Meanwhile, the RACS beam is indicated in the lower left-hand corner of each sub-plot. Light-grey squares represent all RACS components that are within 30\,arcmin of the radio centroid position (blue hexagon), whilst black squares represent RACS components that are deemed to be associated with the G4Jy source. The host galaxy position is marked with a cyan diamond and determines the starting point for the flood-fill algorithm (Section~\ref{sec:floodfill}). }
\label{fig:extended_7}
\end{figure}
\end{landscape}

\begin{landscape}
\begin{figure}
\centering
\includegraphics[scale=0.16]{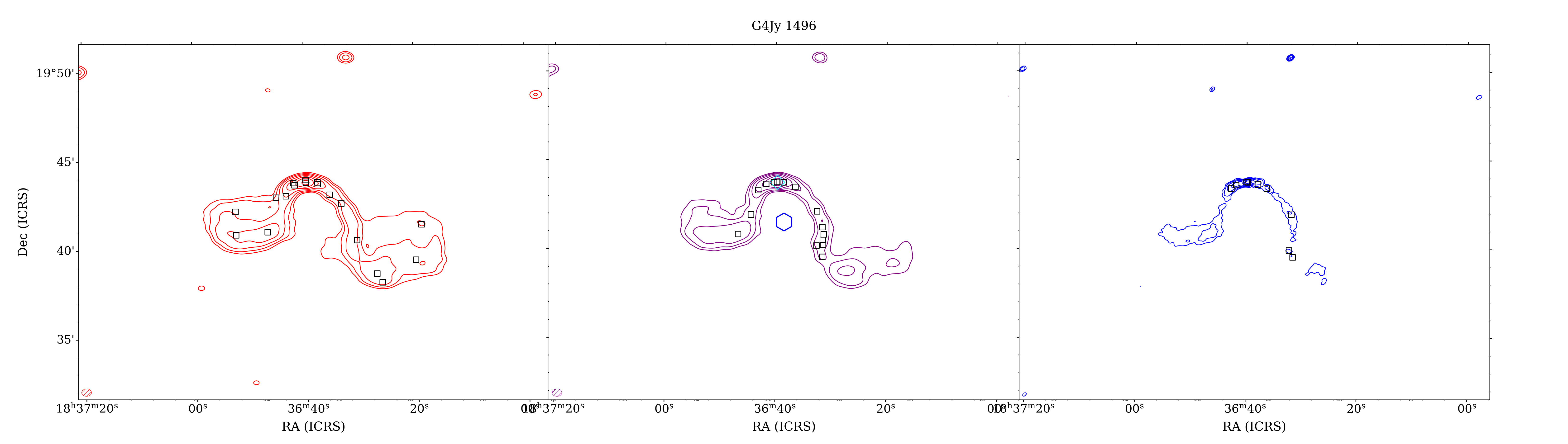} 
\includegraphics[scale=0.16]{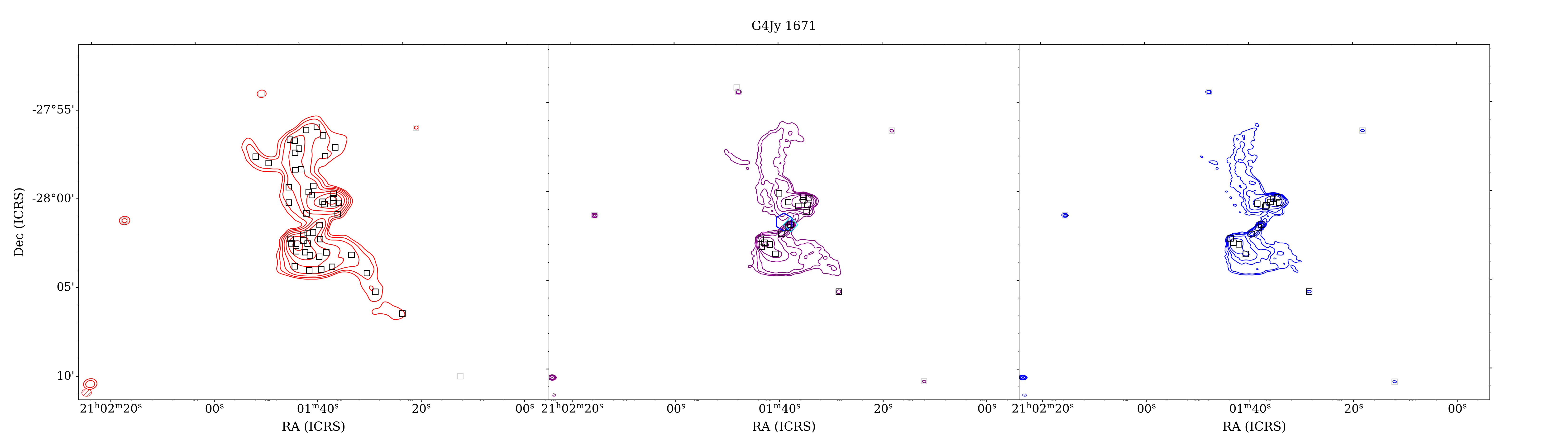} 
\caption{ Two examples of the RACS-low (red), RACS-mid (purple), and RACS-high (blue) contour plots that are inspected for the most-extended radio-galaxies (Appendix~\ref{app:extended}). The lowest contour is at 6 times the local rms ($\sigma$), and doubles with each subsequent contour. Meanwhile, the RACS beam is indicated in the lower left-hand corner of each sub-plot. Light-grey squares represent all RACS components that are within 30\,arcmin of the radio centroid position (blue hexagon), whilst black squares represent RACS components that are deemed to be associated with the G4Jy source. The host galaxy position is marked with a cyan diamond and determines the starting point for the flood-fill algorithm (Section~\ref{sec:floodfill}). }
\label{fig:extended_8}
\end{figure}
\end{landscape}

\begin{landscape}
\begin{figure}
\centering
\includegraphics[scale=0.16]{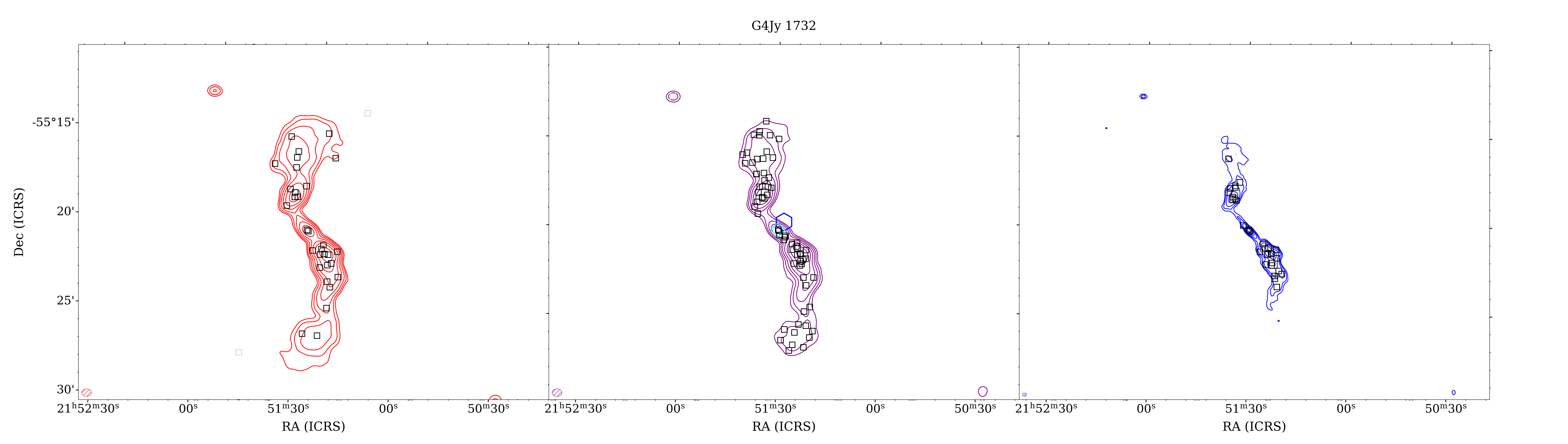} 
\includegraphics[scale=0.16]{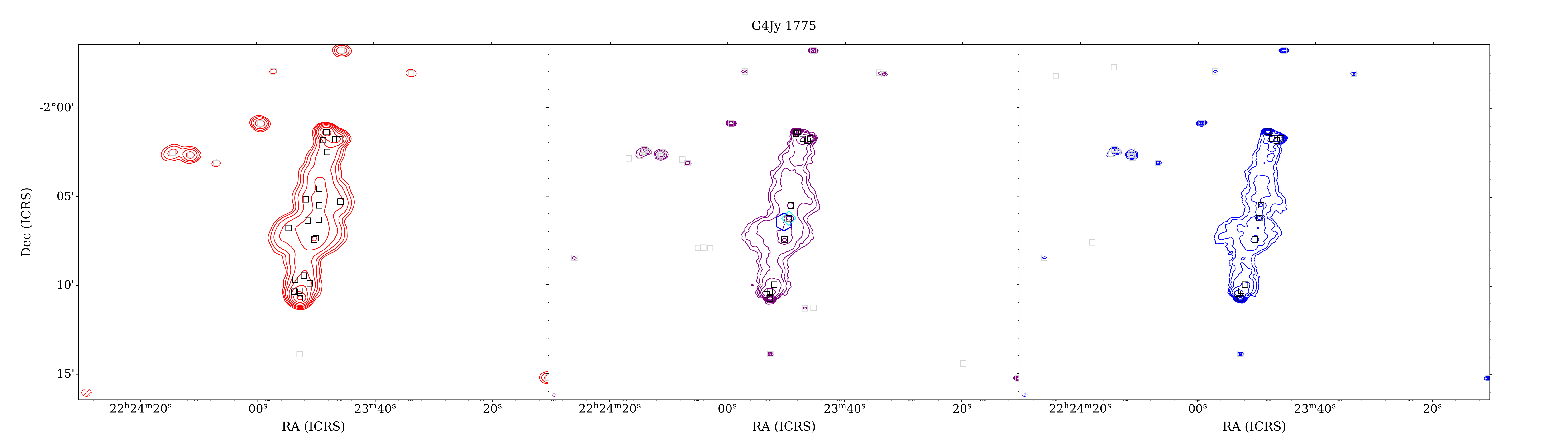} 
\caption{ Two examples of the RACS-low (red), RACS-mid (purple), and RACS-high (blue) contour plots that are inspected for the most-extended radio-galaxies (Appendix~\ref{app:extended}). The lowest contour is at 6 times the local rms ($\sigma$), and doubles with each subsequent contour. Meanwhile, the RACS beam is indicated in the lower left-hand corner of each sub-plot. Light-grey squares represent all RACS components that are within 30\,arcmin of the radio centroid position (blue hexagon), whilst black squares represent RACS components that are deemed to be associated with the G4Jy source. The host galaxy position is marked with a cyan diamond and determines the starting point for the flood-fill algorithm (Section~\ref{sec:floodfill}). }
\label{fig:extended_9}
\end{figure}
\end{landscape}

In this Appendix we present the G4Jy sources that have the largest angular size (Figures~\ref{fig:extended_1}-\ref{fig:extended_9}). As such, the default search radius for identifying RACS components (Section~\ref{sec:floodfill}) is increased accordingly from 5\,arcmin to a maximum of 30\,arcmin. 

\section{Radio spectra for known blazars}
\label{app:blazars}

\begin{figure}
\centering
\vspace{-0.5cm}
\includegraphics[scale=0.3]{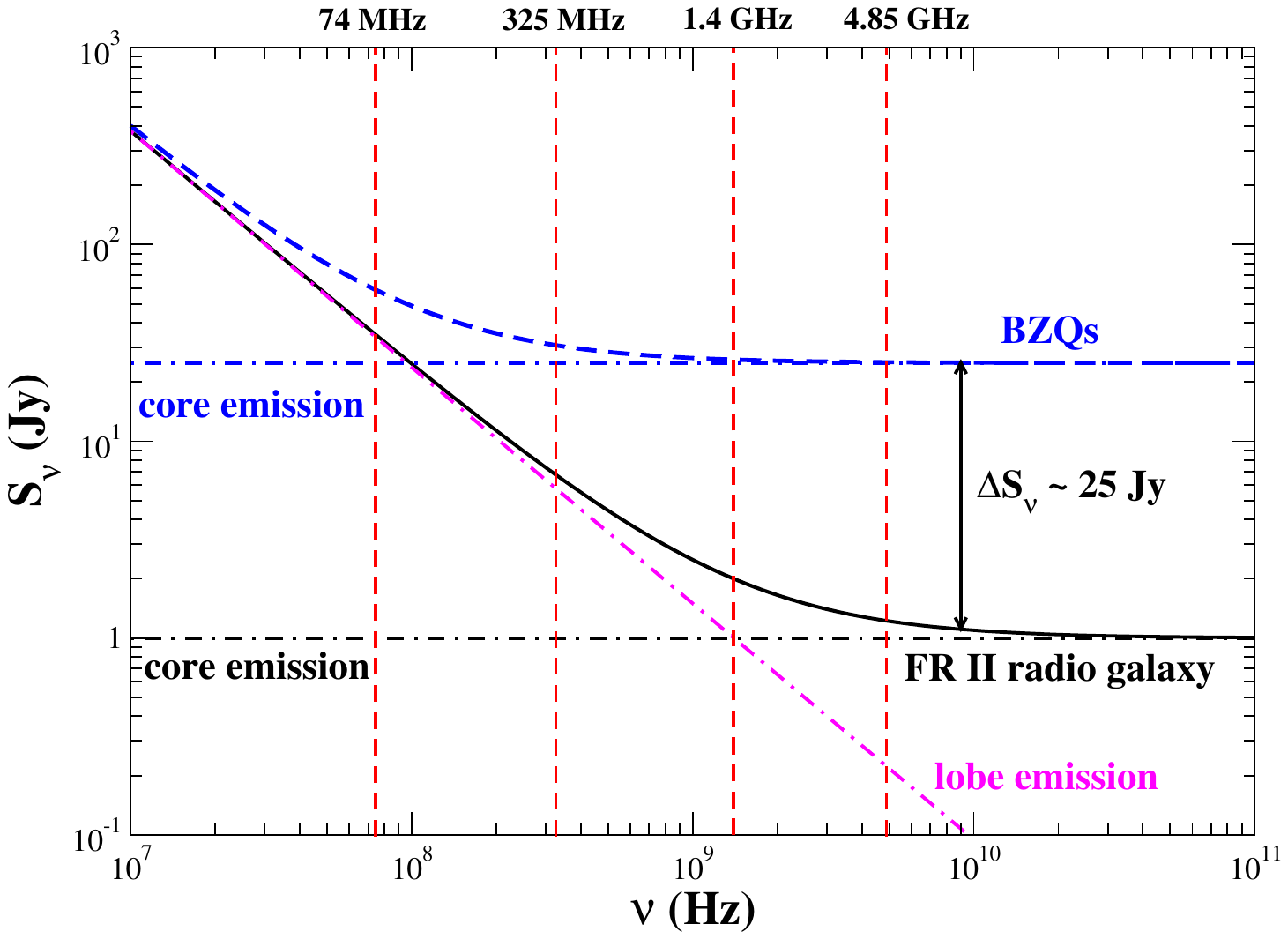} 
\caption{ Figure courtesy of \citet{Massaro2013}, shown here to aid interpretation of concave radio spectra (where SCI $< 0$; Appendix~\ref{app:blazars}). The lobe emission leads to a steep spectrum at low frequencies, whilst emission from the core dominates at high frequencies. `BZQ' is the name given to blazars with a quasar-like optical spectrum, whilst `FR II' refers to the \citet{Fanaroff1974} morphology classification of an edge-brightened radio galaxy.  }
\label{fig:concave}
\end{figure}

\begin{figure*}
\centering
\subfigure[G4Jy 134 (flat-spectrum radio quasar)]{
\includegraphics[scale=0.52]{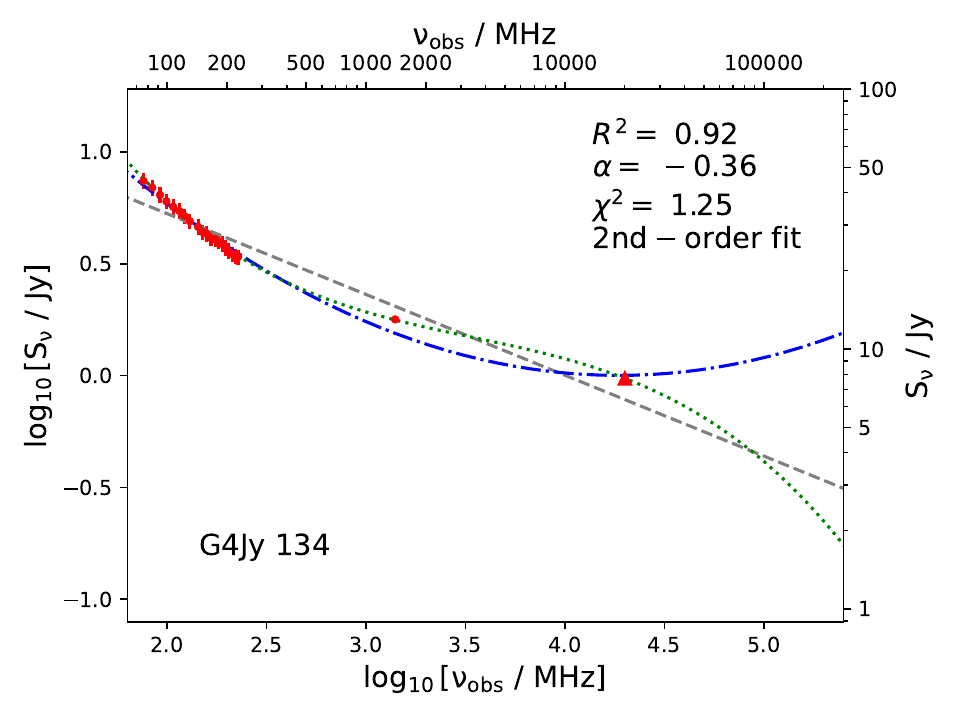} }
\subfigure[G4Jy 229 (flat-spectrum radio quasar)]{
\includegraphics[scale=0.52]{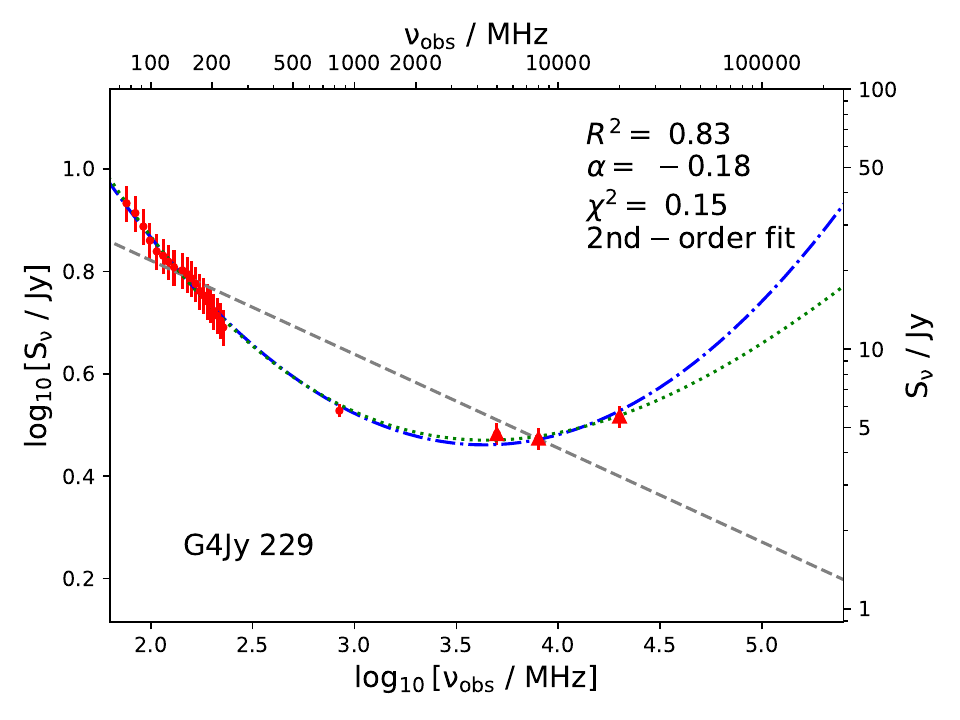} }
\subfigure[G4Jy 292 (flat-spectrum radio quasar)]{
\includegraphics[scale=0.52]{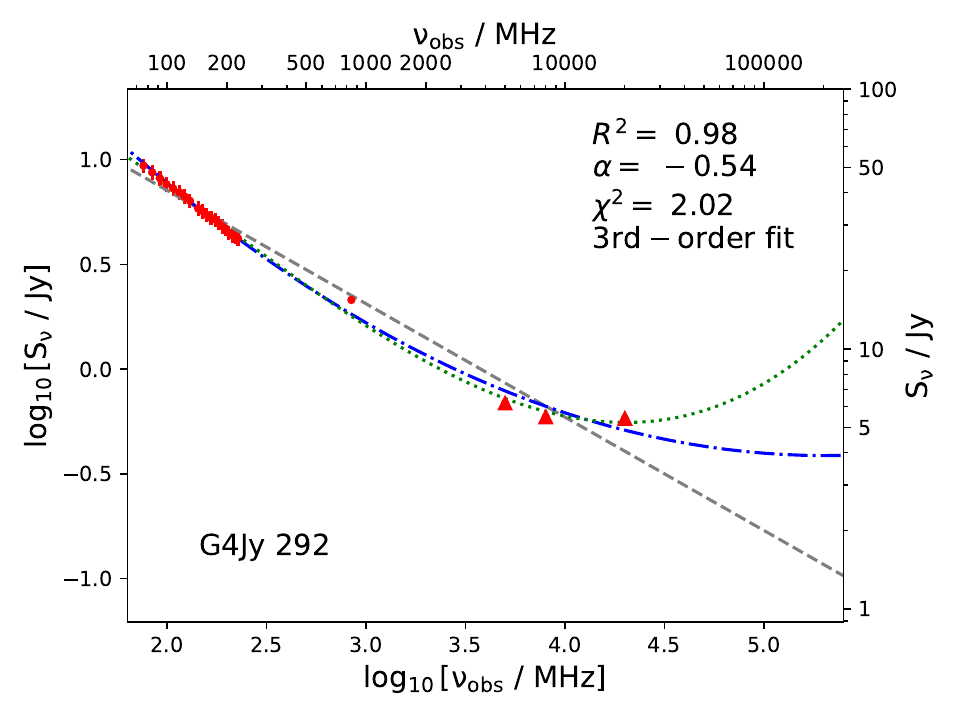} }
\subfigure[G4Jy 344 (flat-spectrum radio quasar)]{
\includegraphics[scale=0.52]{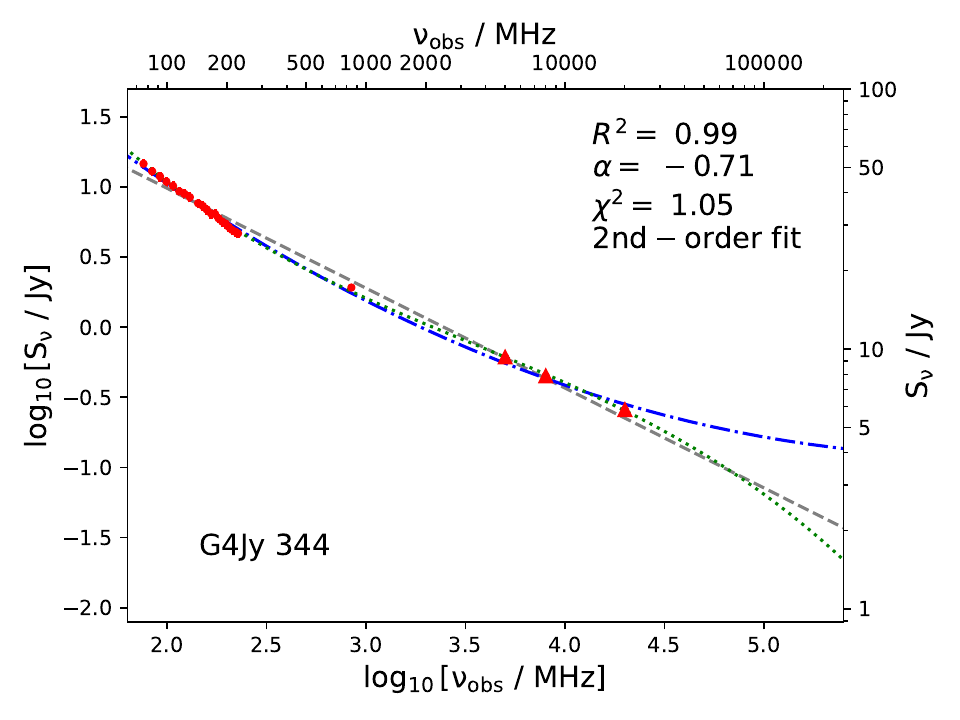} }
\subfigure[G4Jy 415 (blazar of unknown type)]{
\includegraphics[scale=0.52]{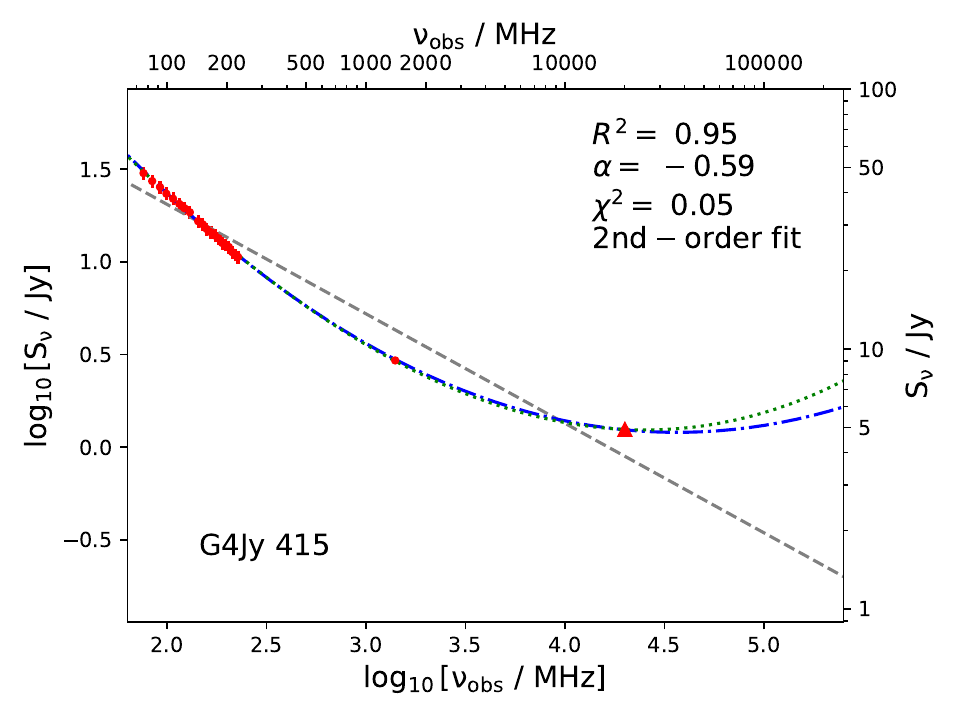} }
\subfigure[G4Jy 483 (flat-spectrum radio-quasar)]{
\includegraphics[scale=0.52]{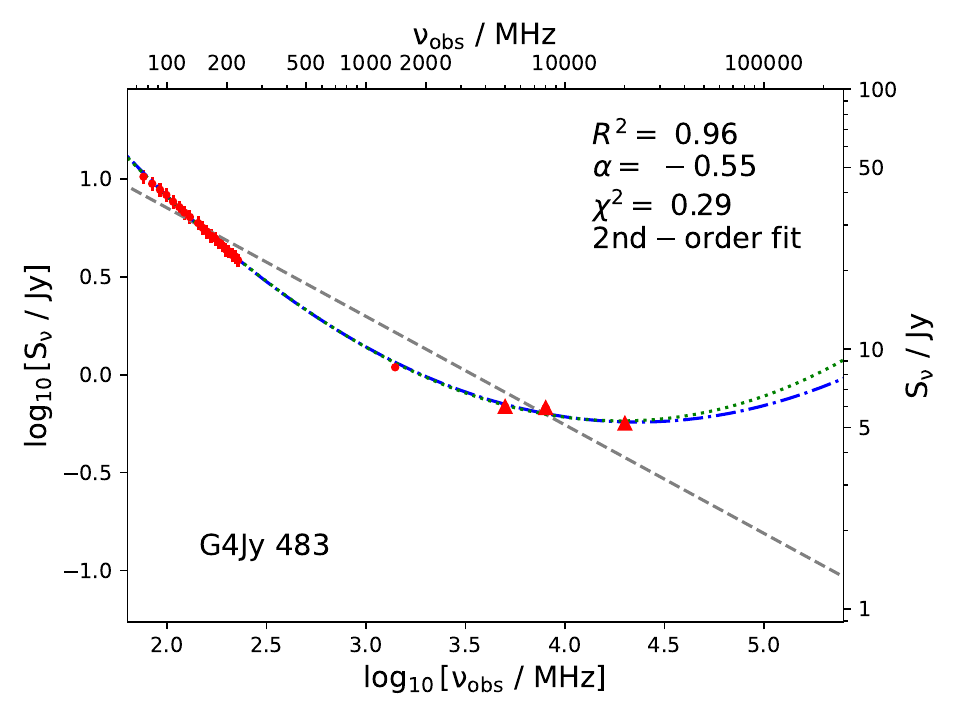} }
\caption{ The radio spectra of G4Jy-ATCA sources that appear in the BLAZE blazar catalogue \citep{Haemmerich2026}. A first-order polynomial fit to the data is represented by a grey dashed line, a second-order polynomial fit is represented by a blue dash-dotted line, and a third-order polynomial fit is represented by a green dotted line. Red dots represent G4Jy and SUMSS/NVSS flux-densities, whilst red triangles indicate that the ATCA flux-densities are lower limits. }
\label{fig:blazars1}
\end{figure*}

\begin{figure*}
\centering
\subfigure[G4Jy 509 (flat-spectrum radio quasar)]{
\includegraphics[scale=0.52]{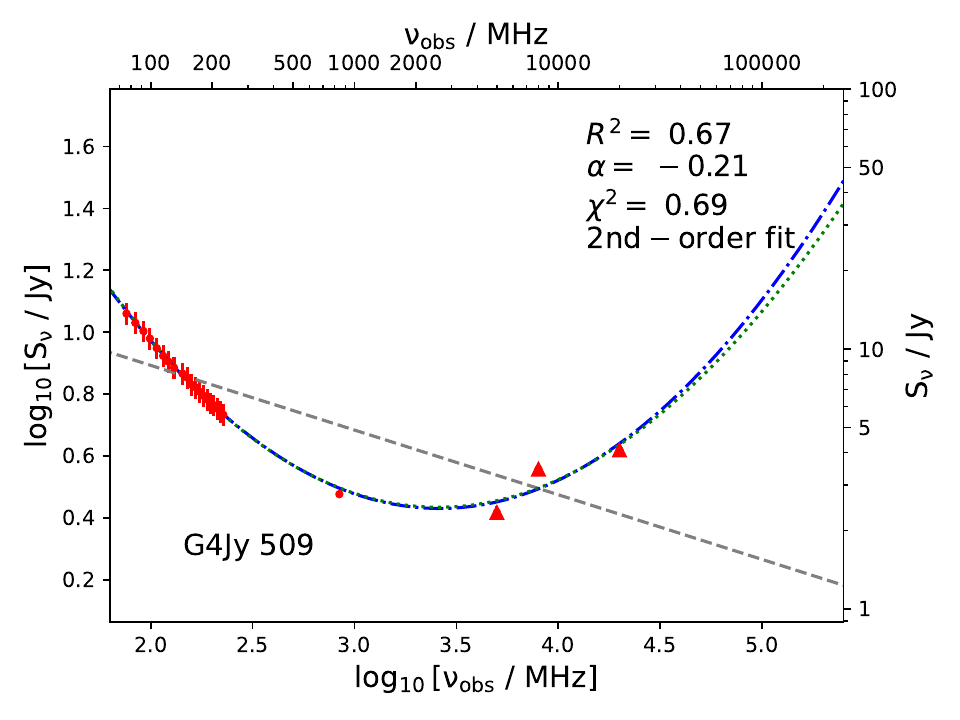} }
\subfigure[G4Jy 518 (flat-spectrum radio quasar)]{
\includegraphics[scale=0.52]{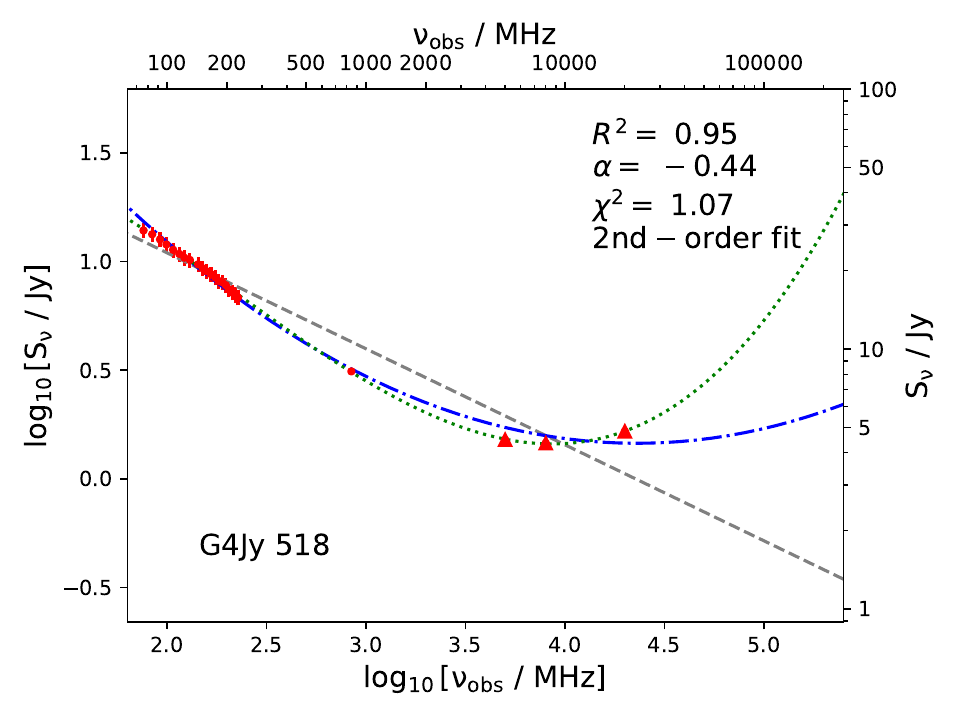} }
\subfigure[G4Jy 871 (flat-spectrum radio quasar)]{
\includegraphics[scale=0.52]{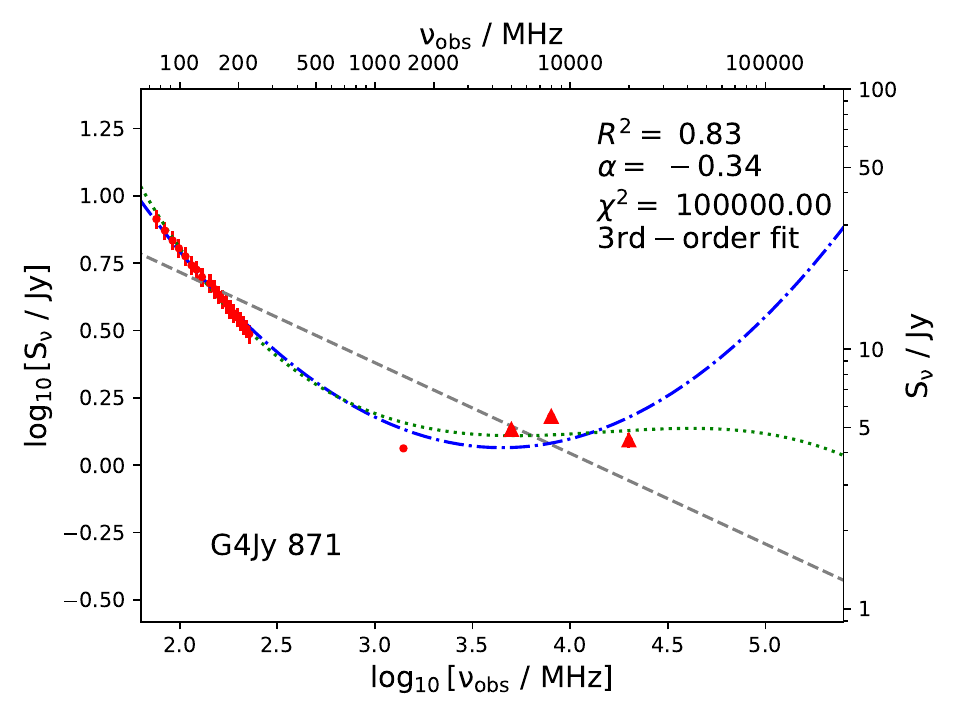} }
\subfigure[G4Jy 1202 (flat-spectrum radio quasar)]{
\includegraphics[scale=0.52]{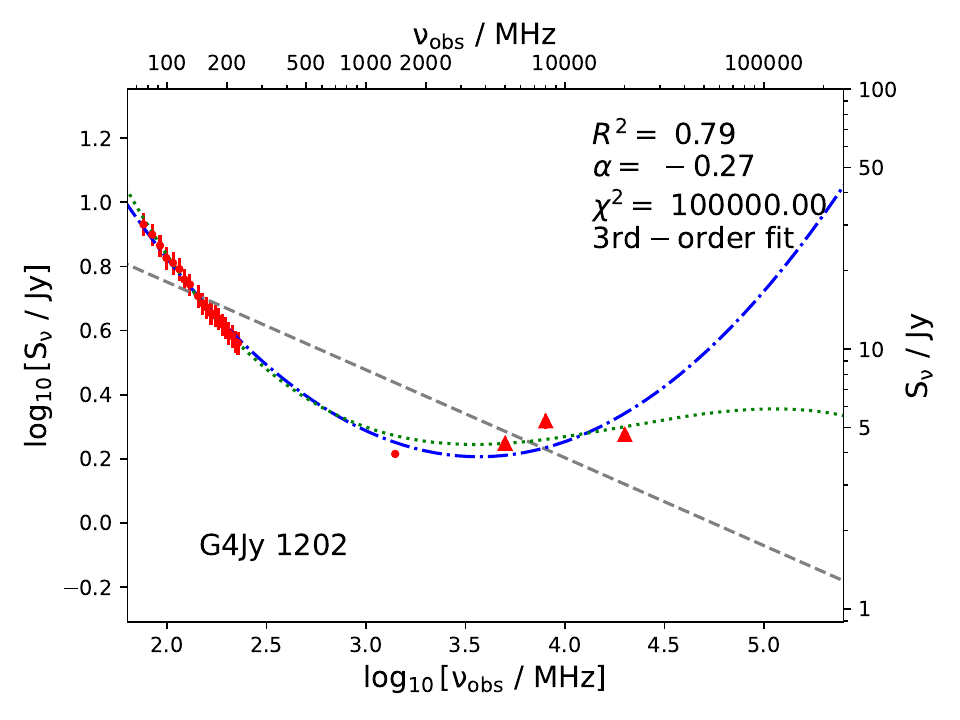} }
\subfigure[G4Jy 1360 (BL Lac)]{
\includegraphics[scale=0.52]{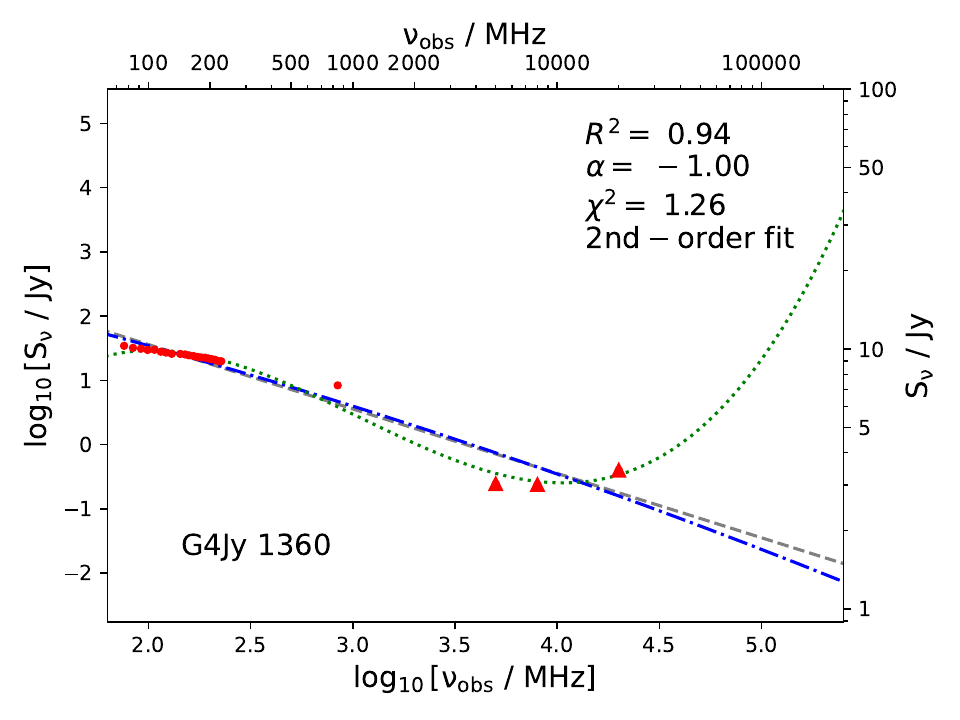} }
\subfigure[G4Jy 1521 (candidate BL Lac)]{
\includegraphics[scale=0.52]{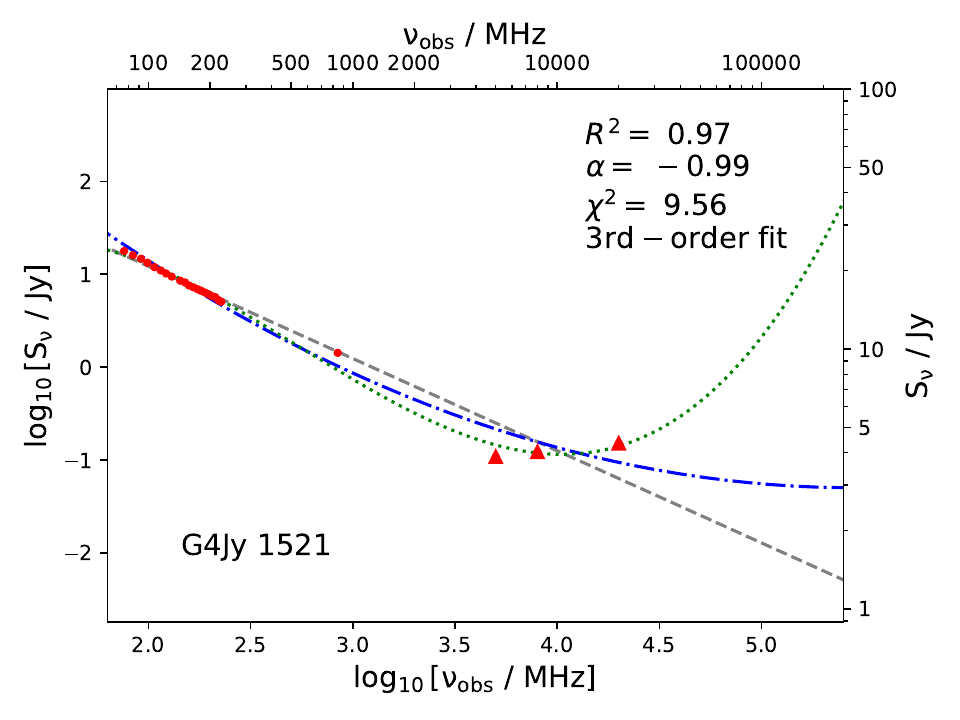} }
\caption{ $Continued$ -- The radio spectra of G4Jy-ATCA sources that appear in the BLAZE blazar catalogue \citep{Haemmerich2026}. A first-order polynomial fit to the data is represented by a grey dashed line, a second-order polynomial fit is represented by a blue dash-dotted line, and a third-order polynomial fit is represented by a green dotted line. Red dots represent G4Jy and SUMSS/NVSS flux-densities, whilst red triangles indicate that the ATCA flux-densities are lower limits. }
\label{fig:blazars2}
\end{figure*}

\begin{figure*}
\centering
\subfigure[G4Jy 1533 (flat-spectrum radio quasar)]{
\includegraphics[scale=0.52]{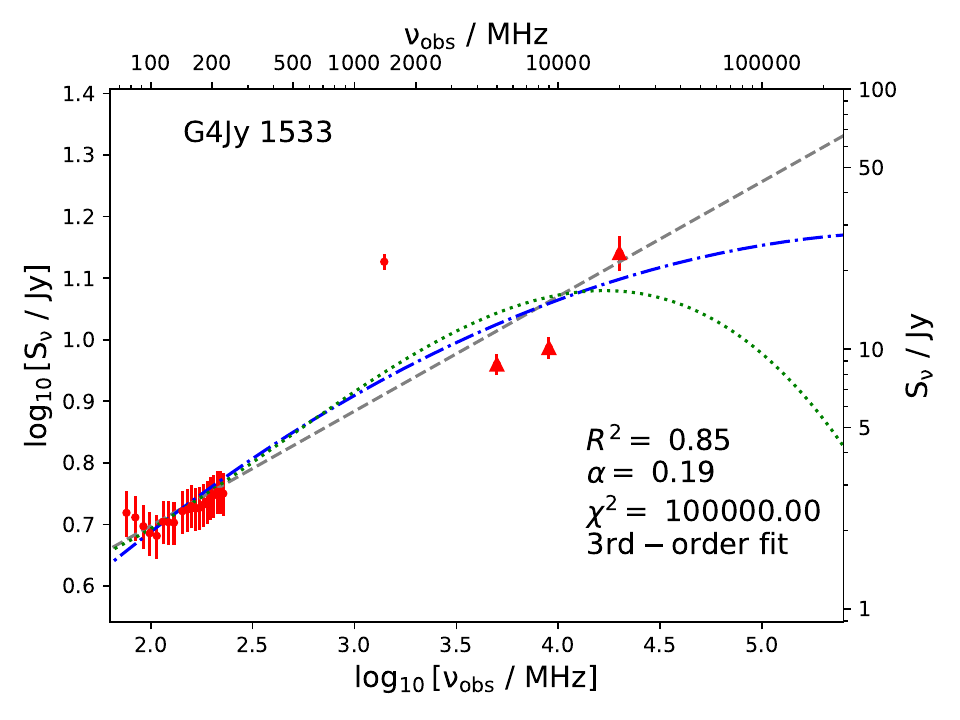} }
\subfigure[G4Jy 1621 (flat-spectrum radio quasar)]{
\includegraphics[scale=0.52]{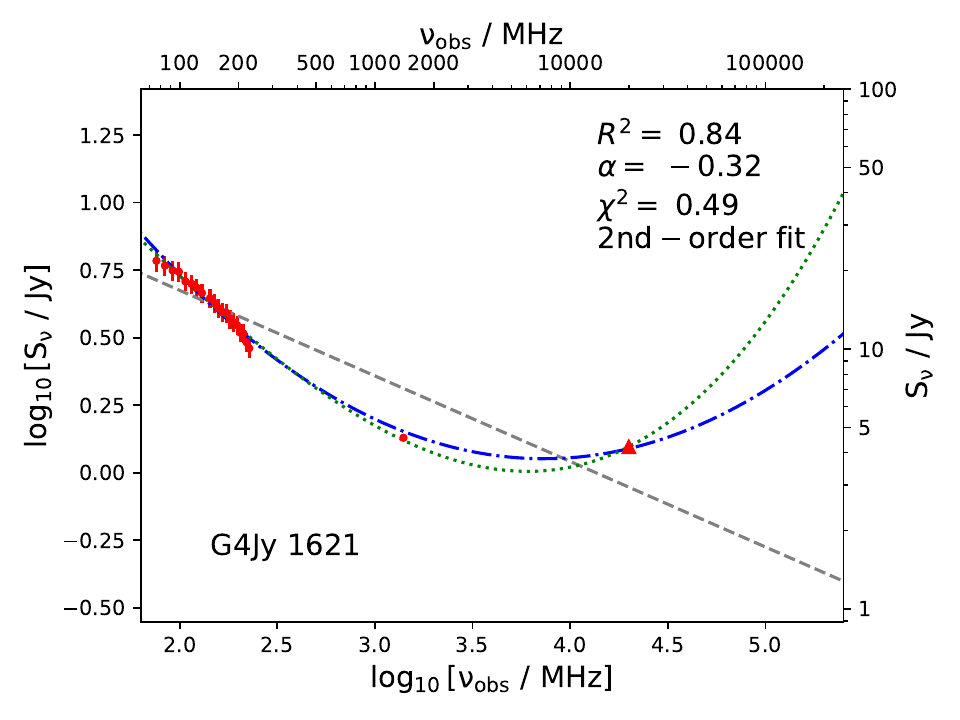} }
\subfigure[G4Jy 1665 (flat-spectrum radio quasar)]{
\includegraphics[scale=0.52]{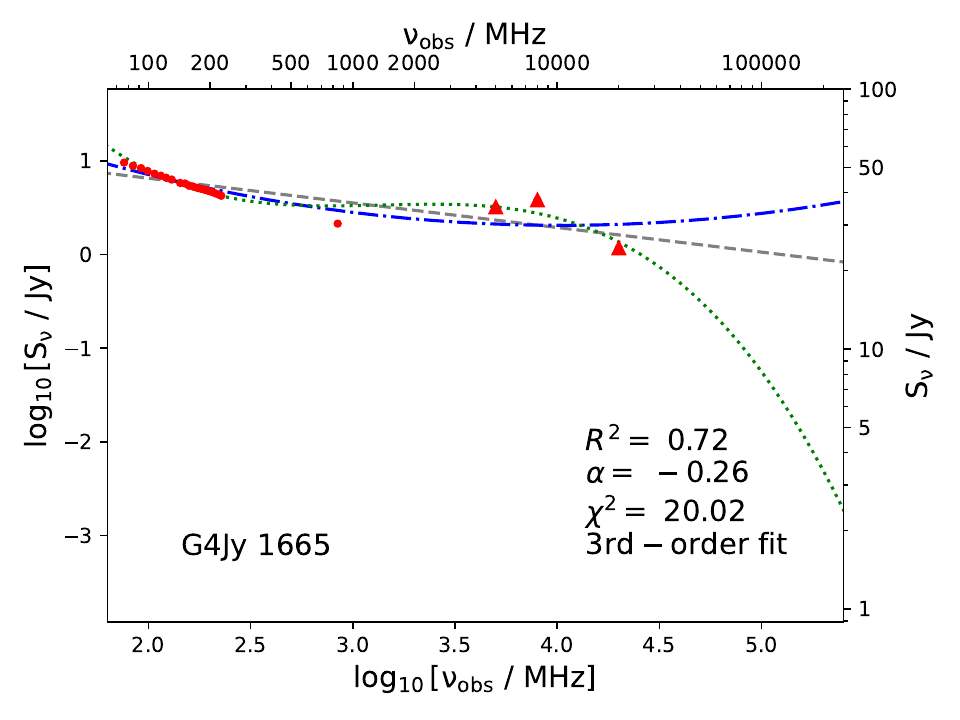} }
\subfigure[G4Jy 1802 (blazar of unknown type)]{
\includegraphics[scale=0.52]{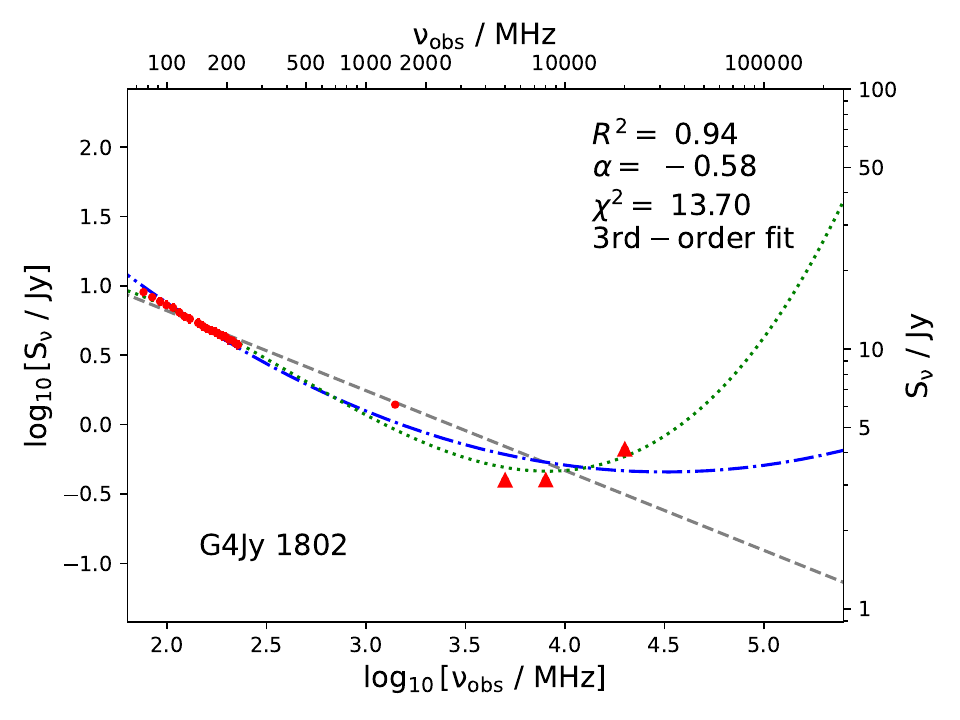} }
\caption{ $Continued$ --  The radio spectra of G4Jy-ATCA sources that appear in the BLAZE blazar catalogue \citep{Haemmerich2026}. A first-order polynomial fit to the data is represented by a grey dashed line, a second-order polynomial fit is represented by a blue dash-dotted line, and a third-order polynomial fit is represented by a green dotted line. Red dots represent G4Jy and SUMSS/NVSS flux-densities, whilst red triangles indicate that the ATCA flux-densities are lower limits. }
\label{fig:blazars3}
\end{figure*}

We visually identify 27 G4Jy-ATCA radio spectra that are particularly `concave' in their shape \citep[with an `upturn' at higher frequencies;][]{Massaro2013,dAntonio2019} -- see Fig.~\ref{fig:concave}. These are listed in Table~\ref{tab:blazars}, and we find that 70 per cent of these sources appear in the BLAZE blazar catalogue \citep{Haemmerich2026}. We note that these radio spectra (Figs.~\ref{fig:blazars1}--\ref{fig:blazars3}) need to be treated with caution given the time variability of the radio emission at high frequencies.   

\begin{table}
	\centering
	\caption{G4Jy-ATCA sources that are visually determined as having particularly concave radio spectra (Figures~\ref{fig:blazars1}--\ref{fig:blazars3}). Their blazar classification and alternative name are as adopted from the BLAZE/BlazEr1 catalogue \citep{Haemmerich2026}. `fsrq' = flat-spectrum radio quasar, `bll' = BL Lac, `bcu' = blazar of unknown type, and a suffix of `c' indicates that the source is a blazar $candidate$. The alternative-name prefix of `4FGL' refers to the fourth $Fermi$-LAT source catalogue \citep{Abdollahi2022}, `5BZQ' and `5BZU' refer to the 5th Roma BZCAT \citep{Massaro2015}, `ABC' refers to the ABC catalogue \citep{Paggi2020}, and the prefix `WIBRaLS2' refers to a blazar catalogue of \citet{DAbrusco2019}.  }
	\label{tab:blazars}
	\begin{tabular}{lccr} 
		\hline
		G4Jy name & Alternative name & Blazar classification\\
		\hline
		G4Jy 28 & -- & -- \\
		G4Jy 86 & -- & -- \\
		G4Jy 134 & 4FGL\_J0116.0$-$1136 & fsrq \\
        G4Jy 229 & 4FGL\_J0210.7$-$5101 & fsrq \\
        G4Jy 260 & -- & -- \\
        G4Jy 292 & 4FGL\_J0245.5$-$4502 & fsrq \\
        G4Jy 344 & 4FGL\_J0317.8$-$4414 & fsrq \\
        G4Jy 415 & 5BZU\_J0407$-$1211 &  bcu \\
        G4Jy 466 & -- & -- \\
        G4Jy 483 & 5BZQ\_J0437$-$2954 & fsrq \\
        G4Jy 509 & 4FGL\_J0455.7$-$4617 & fsrq \\
        G4Jy 518 & 5BZQ\_J0506$-$6109 & fsrq \\
        G4Jy 871 & 4FGL\_J1048.0$-$1912 & fsrq \\
        G4Jy 891 & -- & -- \\
        G4Jy 1071 & -- & -- \\
        G4Jy 1074 & -- & -- \\
        G4Jy 1080 & ABC\_J1336$-$3357 & bcuc \\
        G4Jy 1168 & 4FGL\_J1427.9$-$4206 & fsrq \\
        G4Jy 1202 & 4FGL\_J1454.4$-$3744 & fsrq \\
        G4Jy 1360 & 4FGL\_J1643.0$-$7714 & bll \\
        G4Jy 1521 & WIBRaLS2\_J191739.62$-$453030.7 & bllc \\
        G4Jy 1533 & 4FGL\_J1924.8$-$2914 & fsrq \\
        G4Jy 1621 & 4FGL\_J2025.6$-$0735 & fsrq \\
        G4Jy 1665 & 4FGL\_J2056.2$-$4714 & fsrq \\
        G4Jy 1732 & -- & -- \\
        G4Jy 1777 & 4FGL\_J2225.7$-$0457 & fsrq \\
        G4Jy 1802 & 5BZU\_J2303$-$1841 & bcu \\
		\hline
	\end{tabular}
\end{table}

\section{New columns of the G4Jy catalogue}
\label{app:catalogue_columns}

Descriptions of newly-added/newly-named columns in the updated G4Jy catalogue, along with 17th-row entries, can be found in Table~\ref{tab:catalogue_columns}. This includes the addition of a `classification' column, where the G4Jy source is listed as an AGN if it satisfies at least one of the following criteria: (i) has a `double', `triple' or `complex' radio morphology; (ii) has a 151-MHz radio luminosity that exceeds 10$^{24}$\,W\,Hz$^{-1}$ \citep[noting that star-forming galaxies have lower luminosities;][]{Gurkan2018}; or (iii) has a blazar classification (as indicated via the BLAZE\_CLASS column). Exceptions to this logic are G4Jy~77 \citep[a radio phoenix;][]{Raja2024}, G4Jy~86 (a star-forming galaxy), G4Jy~288 (aka NGC 1068, a known AGN), G4Jy~571 (the Flame Nebula), G4Jy~1081 (a star-forming galaxy), and G4Jy~1605 (a cluster relic). 

\begin{landscape}
\begin{table}
\centering
\caption{Column numbers, names, units, descriptions, and 17th-row entries for 124 of the 253 columns in the updated G4Jy catalogue (Appendix~\ref{app:catalogue_columns}). The 17th row is chosen to minimise the number of empty examples. `AT20G' refers to the AT20G catalogue \citep{Murphy2010}, whilst `BLAZE' and `BLAZER1' refer to the blazar catalogues of \citet{Haemmerich2026}. See appendix~E of \citet{White2020b} and appendix~A of \citet{White2026b} for the remaining G4Jy-catalogue columns.}
\begin{tabular}{@{}ccccr@{}} 

\hline
Column no. & Column name & Units & Description & 17th-row entry  \\
\hline

3 & classification & -- & Source-type classification & AGN \\
24 & alpha\_mid & -- & Spectral index between 151\,MHz and 1367.5\,MHz & $-$0.809757 \\
25 & err\_alpha\_mid & -- & Error on the spectral index between 151\,MHz and 1367.5\,MHz & 0.041011 \\
26 & alpha\_low & -- & Spectral index between 76\,MHz and 227\,MHz & $-$0.670262 \\
27 & err\_alpha\_low & -- & Error on the spectral index between 76\,MHz and 227\,MHz & 0.007851 \\

33 & syserr\_total\_int\_flux\_076 & Jy & Systematic error at 72-80 MHz due to flux-density scale & 0.970512 \\
36 & syserr\_total\_int\_flux\_084 & Jy & Systematic error at 80-88 MHz due to flux-density scale & 0.922980 \\
39 & syserr\_total\_int\_flux\_092 & Jy & Systematic error at 88-95 MHz due to flux-density scale & 0.880273 \\
42 & syserr\_total\_int\_flux\_099 & Jy & Systematic error at 95-103 MHz due to flux-density scale & 0.817363 \\
45 & syserr\_total\_int\_flux\_107 & Jy & Systematic error at 103-111 MHz due to flux-density scale & 0.795469 \\
48 & syserr\_total\_int\_flux\_115 & Jy & Systematic error at 111-118 MHz due to flux-density scale & 0.755457 \\
51 & syserr\_total\_int\_flux\_122 & Jy & Systematic error at 118-126 MHz due to flux-density scale & 0.726722 \\
54 & syserr\_total\_int\_flux\_130 & Jy & Systematic error at 126-134 MHz due to flux-density scale & 0.691611 \\
57 & syserr\_total\_int\_flux\_143 & Jy & Systematic error at 139-147 MHz due to flux-density scale & 0.661601 \\
60 & syserr\_total\_int\_flux\_151 & Jy & Systematic error at 147-154 MHz due to flux-density scale & 0.636181 \\
63 & syserr\_total\_int\_flux\_158 & Jy & Systematic error at 154-162 MHz due to flux-density scale & 0.615912 \\
66 & syserr\_total\_int\_flux\_166 & Jy & Systematic error at 162-170 MHz due to flux-density scale & 0.597573 \\
69 & syserr\_total\_int\_flux\_174 & Jy & Systematic error at 170-177 MHz due to flux-density scale & 0.584969 \\
72 & syserr\_total\_int\_flux\_181 & Jy & Systematic error at 177-185 MHz due to flux-density scale & 0.566740 \\
75 & syserr\_total\_int\_flux\_189 & Jy & Systematic error at 185-193 MHz due to flux-density scale & 0.549959 \\
78 & syserr\_total\_int\_flux\_197 & Jy & Systematic error at 193-200 MHz due to flux-density scale & 0.529204 \\
81 & syserr\_total\_int\_flux\_204 & Jy & Systematic error at 200-208 MHz due to flux-density scale & 0.505618 \\
84 & syserr\_total\_int\_flux\_212 & Jy & Systematic error at 208-216 MHz due to flux-density scale & 0.491888 \\
87 & syserr\_total\_int\_flux\_220 & Jy & Systematic error at 216-223 MHz due to flux-density scale & 0.477011 \\
90 & syserr\_total\_int\_flux\_227 & Jy & Systematic error at 223-231 MHz due to flux-density scale & 0.466105 \\

155 & RACSlow\_localrms & Jy\,beam$^{-1}$ & Local rms in the RACS-low image & 0.000826 \\
156 & RACSlow\_n\_cmp & -- & Number of associated RACS-low Gaussian components & 1 \\
157 & RACS\_low\_flag & -- & Flags an under- (-1) or over-estimated (+1) RACS-low flux-density & 0 \\
158 & S\_RACSlow\_int & Jy & Integrated flux-density in RACS-low at 887.5\,MHz  & 2.086022 \\
159 & err\_S\_RACSlow\_int & Jy & Error in the integrated RACS-low flux-density  & 0.146531 \\
160 & RACSmid\_localrms & Jy\,beam$^{-1}$ & Local rms in the RACS-mid image & 0.000357 \\
161 & RACSmid\_n\_cmp & -- & Number of associated RACS-mid Gaussian components & 4 \\
162 & RACS\_mid\_flag & -- & Flags an under- (-1) or over-estimated (+1) RACS-mid flux-density & 0 \\
163 & S\_RACSmid\_int & Jy & Integrated flux-density in RACS-mid at 1367.5\,MHz  & 1.335346 \\
164 & err\_S\_RACSmid\_int & Jy & Error in the integrated RACS-mid flux-density  & 0.056138 \\
165 & RACShigh\_localrms & Jy\,beam$^{-1}$ & Local rms in the RACS-high image & 0.000336 \\
166 & RACShigh\_n\_cmp & -- & Number of associated RACS-high Gaussian components  & 1 \\
167 & RACS\_high\_flag & -- & Flags an under- (-1) or over-estimated (+1) RACS-high flux-density  & 0 \\
168 & S\_RACShigh\_int & Jy & Integrated flux-density in RACS-high at 1655.5\,MHz  & 1.098967 \\
169 & err\_S\_RACShigh\_int & Jy & Error in the integrated RACS-high flux-density  & 0.109928 \\
170 & polynomial\_constant & -- & Constant for polynomial fit of 72-MHz to 1.4-GHz data & 2.657688 \\
171 & first\_polynomial\_coeff & -- & First polynomial coefficient for fitting 72-MHz to 1.4-GHz data & $-$0.806632 \\

\hline
\label{tab:catalogue_columns}
\end{tabular}
\end{table}
\end{landscape}

\setcounter{table}{0}

\begin{landscape}
\begin{table}
\centering
\caption{ {\it Continued} -- Column numbers, names, units, and descriptions for 124 of the 253 columns in the updated G4Jy catalogue (Appendix~\ref{app:catalogue_columns}). `AT20G' refers to the AT20G catalogue \citep{Murphy2010}, whilst `BLAZE' and `BLAZER1' refer to the blazar catalogues of \citet{Haemmerich2026}. See appendix~E of \citet{White2020b} and appendix~A of \citet{White2026b} for the remaining G4Jy-catalogue columns.}
\begin{tabular}{@{}ccccr@{}} 

\hline
Column no. & Column name & Units & Description & 17th-row entry  \\
\hline

172 & second\_polynomial\_coeff & -- & Second polynomial coefficient for fitting 72-MHz to 1.4-GHz data & -- \\
173 & third\_polynomial\_coeff & -- & Third polynomial coefficient for fitting 72-MHz to 1.4-GHz data & -- \\
174 & polynomial\_reduced\_chi2 & -- & Reduced $\chi^{2}$ value for polynomial fitting & 0.452083 \\
175 & restframe\_polynomial\_constant & -- & Constant for rest-frame polynomial fit of 72-MHz to 1.4-GHz data & -- \\
176 & restframe\_first\_polynomial\_coeff & -- & First polynomial coefficient for fitting 72-MHz to 1.4-GHz data in the rest-frame & -- \\
177 & restframe\_second\_polynomial\_coeff & -- & Second polynomial coefficient for fitting 72-MHz to 1.4-GHz data in the rest-frame & -- \\
178 & restframe\_third\_polynomial\_coeff & -- & Third polynomial coefficient for fitting 72-MHz to 1.4-GHz data in the rest-frame & -- \\
179 & restframe\_polynomial\_reduced\_chi2 & -- & Reduced $\chi^{2}$ value for rest-frame polynomial fitting & -- \\
180 & Name\_AT20G & -- & Name in the AT20G catalogue & AT20GJ000829$-$055845 \\
181 & RA\_hms\_AT20G & h:m:s & Right Ascension in the AT20G catalogue, in sexagesimal units & 00:08:29.33 \\
182 & Dec\_dms\_AT20G & d:m:s & Declination in the AT20G catalogue, in sexagesimal units & $-$05:58:45.2 \\
183 & RA\_deg\_AT20G & deg & Right Ascension in the AT20G catalogue, in degrees & 2.122208 \\
184 & Dec\_deg\_AT20G & deg & Declination in the AT20G catalogue, in degrees & $-$5.979222 \\
185 & S5\_AT20G\_peak & mJy & Peak flux-density at 5\,GHz in the AT20G catalogue & -- \\
186 & err\_S5\_AT20G\_peak & mJy & Error in the peak flux-density at 5\,GHz in the AT20G catalogue & -- \\
187 & S8\_AT20G\_peak & mJy & Peak flux-density at 8\,GHz in the AT20G catalogue & -- \\
188 & err\_S8\_AT20G\_peak & mJy & Error in the peak flux-density at 8\,GHz in the AT20G catalogue & -- \\
189 & S20\_AT20G\_peak & mJy & Peak flux-density at 20\,GHz in the AT20G catalogue & 66.0 \\
190 & err\_S20\_AT20G\_peak & mJy & Error in the peak flux-density at 20\,GHz in the AT20G catalogue & 4\\
191 & Epoch\_AT20G & -- & Epochs of the observations at 20, 8, 5\,GHz & 6 \\
192 & Quality\_AT20G & -- & Quality flag (g = good or p = poor) in the AT20G catalogue & g \\
193 & Flags\_AT20G & -- &  Other flags in the AT20G catalogue & -- \\
194 & P20\_mJy\_beam\_AT20G & mJy & Polarised flux density (or limit) at 20\,GHz & $<7$ \\

195 & e\_P20\_AT20G & mJy & Error in the polarised flux density at 20\,GHz & -- \\
196 & m20\_AT20G & \% & Fractional polarisation at 20\,GHz & $<10.5$\\
197 & PA20\_AT20G & deg & Position angle at 20\,GHz & -- \\
198 & P8\_mJy\_beam\_AT20G & mJy & Polarised flux density (or limit) at 8\,GHz & -- \\
199 & e\_P8\_AT20G & mJy & Error in the polarised flux density at 8\,GHz & -- \\
200 & m8\_AT20G & \% & Fractional polarisation at 8\,GHz & -- \\
201 & PA8\_AT20G & deg & Position angle at 8\,GHz & -- \\
202 & P5\_mJy\_beam\_AT20G & mJy & Polarised flux density (or limit) at 5\,GHz & -- \\
203 & e\_P5\_AT20G & mJy & Error in the polarised flux density at 5\,GHz & -- \\
204 & m5\_AT20G & \% & Fractional polarisation at 5\,GHz & -- \\
205 & PA5\_AT20G & deg & Position angle at 5\,GHz & -- \\
206 & SNR\_20GHz & -- & Signal-to-noise ratio at 20\,GHz & 16.5 \\
207 & S5\_ATCAfollowup\_int & mJy & Integrated flux-density at 5\,GHz from ATCA follow-up & -- \\
208 & err\_S5\_ATCAfollowup\_int & mJy & Error in the integrated flux-density at 5\,GHz, from ATCA follow-up & -- \\
209 & flag5\_ATCAfollowup\_int & -- & Flag to indicate unreliability of the 5\,GHz data & -- \\
210 & DR5\_ATCAfollowup & -- & Dynamic range at 5\,GHz, from ATCA follow-up & -- \\
211 & ncmp5\_ATCAfollowup & -- & Number of 5\,GHz components, from ATCA follow-up & -- \\
212 & S9\_ATCAfollowup\_int & mJy & Integrated flux-density at 9\,GHz from ATCA follow-up & -- \\

\hline
\label{tab:catalogue_columns}
\end{tabular}
\end{table}
\end{landscape}

\setcounter{table}{0}

\begin{landscape}
\begin{table}
\centering
\caption{ {\it Continued} -- Column numbers, names, units, and descriptions for 124 of the 253 columns in the updated G4Jy catalogue (Appendix~\ref{app:catalogue_columns}). `AT20G' refers to the AT20G catalogue \citep{Murphy2010}, whilst `BLAZE' and `BLAZER1' refer to the blazar catalogues of \citet{Haemmerich2026}. See appendix~E of \citet{White2020b} and appendix~A of \citet{White2026b} for the remaining G4Jy-catalogue columns.}
\begin{tabular}{@{}ccccr@{}} 

\hline
Column no. & Column name & Units & Description & 17th-row entry  \\
\hline

213 & err\_S9\_ATCAfollowup\_int & mJy & Error in the integrated flux-density at 9\,GHz, from ATCA follow-up & -- \\
214 & flag9\_ATCAfollowup\_int & -- & Flag to indicate unreliability of the 9\,GHz data & -- \\
215 & DR9\_ATCAfollowup & -- & Dynamic range at 9\,GHz, from ATCA follow-up & -- \\
216 & ncmp9\_ATCAfollowup & -- & Number of 9\,GHz components, from ATCA follow-up & -- \\
217 & S18\_ATCAfollowup\_int & mJy & Integrated flux-density at 18\,GHz from ATCA follow-up & --\\
218 & err\_S18\_ATCAfollowup\_int & mJy & Error in the integrated flux-density at 18\,GHz, from ATCA follow-up & -- \\
219 & flag18\_ATCAfollowup\_int & -- & Flag to indicate unreliability of the 18\,GHz data & -- \\
220 & DR18\_ATCAfollowup & -- & Dynamic range at 18\,GHz, from ATCA follow-up & -- \\
221 & ncmp18\_ATCAfollowup & -- & Number of 18\,GHz components, from ATCA follow-up & -- \\
222 & polynomial\_constant\_ATCA & -- & Constant for polynomial fit of 72-MHz to 20-GHz data & 1.611203 \\
223 & first\_polynomial\_coeff\_ATCA & -- & First polynomial coefficient for fitting 72-MHz to 20-GHz data & 0.003040   \\
224 & second\_polynomial\_coeff\_ATCA & -- & Second polynomial coefficient for fitting 72-MHz to 20-GHz data & $-$0.151617 \\
225 & third\_polynomial\_coeff\_ATCA & -- & Third polynomial coefficient for fitting 72-MHz to 20-GHz data & -- \\
226 & polynomial\_reduced\_chi2\_ATCA & -- & Reduced $\chi^{2}$ value for polynomial fitting &  0.033276 \\
227 & alpha\_high & -- & Spectral index between 1367.5\,MHz and 18/20\,GHz & $-$1.120974\\
228 & err\_alpha\_high & -- & Error on the spectral index between 1367.5\,MHz and 18/20\,GHz & 0.015093 \\
229 & BLAZE\_SEP & arcsec & Angular separation between G4Jy source and BLAZE counterpart & -- \\
230 & BLAZE\_SRC\_NAME & -- & Name of the source in the BLAZE catalogue & -- \\
231 & BLAZE\_CLASS & -- & Blazar classification in the BLAZE catalogue & -- \\
232 & BLAZER1\_SEP & arcsec & Angular separation between G4Jy source and BLAZER1 counterpart & -- \\
233 & BLAZER1\_SRC\_NAME & -- & Name of the source in the BLAZER1 catalogue & -- \\
234 & BLAZER1\_CLASS & -- & Blazar classification in the BLAZER1 catalogue & -- \\
235 & redshift\_for\_models & -- & Redshift used for modelling sources with RAiSE & 0.226 \\
236 & log10\_jet\_power\_12d5\_low & W & Lower limit of the jet power, assuming a halo mass of $10^{12.5} \Msol$ & 38.616359\\
237 & log10\_jet\_power\_12d5\_med & W & Median of the jet power, assuming a halo mass of $10^{12.5} \Msol$ & 38.652865 \\
238 & log10\_jet\_power\_12d5\_upp & W & Upper limit of the jet power, assuming a halo mass of $10^{12.5} \Msol$ & 38.689370 \\

239 & log10\_jet\_power\_13d5\_low & W & Lower limit of the jet power, assuming a halo mass of $10^{13.5} \Msol$ & 38.909640 \\
240 & log10\_jet\_power\_13d5\_med & W & Median of the jet power, assuming a halo mass of $10^{13.5} \Msol$ & 38.946214\\
241 & log10\_jet\_power\_13d5\_upp & W & Upper limit of the jet power, assuming a halo mass of $10^{13.5} \Msol$ & 38.982786 \\
242 & log10\_jet\_power\_14d5\_low & W & Lower limit of the jet power, assuming a halo mass of $10^{14.5} \Msol$ & 39.078504 \\
243 & log10\_jet\_power\_14d5\_med & W & Median of the jet power, assuming a halo mass of $10^{14.5} \Msol$ & 39.137201 \\
244 & log10\_jet\_power\_14d5\_upp & W & Upper limit of the jet power, assuming a halo mass of $10^{14.5} \Msol$ & 39.179904 \\
245 & log10\_source\_age\_12d5\_low & yr & Lower limit of the source age, assuming a halo mass of $10^{12.5} \Msol$ & 6.216359\\
246 & log10\_source\_age\_12d5\_med & yr & Median of the source age, assuming a halo mass of $10^{12.5} \Msol$ & 6.252865 \\
247 & log10\_source\_age\_12d5\_upp & yr & Upper limit of the source age, assuming a halo mass of $10^{12.5} \Msol$ & 6.289370 \\
248 & log10\_source\_age\_13d5\_low & yr & Lower limit of the source age, assuming a halo mass of $10^{13.5} \Msol$ & 6.509640 \\
249 & log10\_source\_age\_13d5\_med & yr & Median of the source age, assuming a halo mass of $10^{13.5} \Msol$ & 6.546214 \\
250 & log10\_source\_age\_13d5\_upp & yr & Upper limit of the source age, assuming a halo mass of $10^{13.5} \Msol$ & 6.582789 \\
251 & log10\_source\_age\_14d5\_low & yr & Lower limit of the source age, assuming a halo mass of $10^{14.5} \Msol$ & 6.578509\\
252 & log10\_source\_age\_14d5\_med & yr & Median of the source age, assuming a halo mass of $10^{14.5} \Msol$ & 6.737230 \\
253 & log10\_source\_age\_14d5\_upp & yr & Upper limit of the source age, assuming a halo mass of $10^{14.5} \Msol$ & 6.779966 \\

\hline
\label{tab:catalogue_columns}
\end{tabular}
\end{table}
\end{landscape}


\bsp	
\label{lastpage}
\end{document}